\documentclass[fleqn,usenatbib]{mnras}

\usepackage{lmodern}
\usepackage{newtxtext,newtxmath}
\usepackage{multirow}
\usepackage[T1]{fontenc}

\DeclareRobustCommand{\VAN}[3]{#2}
\let\VANthebibliography\thebibliography
\def\thebibliography{\DeclareRobustCommand{\VAN}[3]{##3}\VANthebibliography}

\usepackage{graphicx}
\usepackage{amsmath}
\usepackage{physics}
\usepackage[dvipsnames,table]{xcolor}
\usepackage{ulem}

\title[Efficient Star Formation at Cosmic Dawn]{On the Origin of the High Star-Formation Efficiency in Massive Galaxies at Cosmic Dawn}

\author[Z. L. Andalman et al.]{
Zachary L. Andalman$^{1}$\thanks{E-mail: zack.andalman@princeton.edu}, Romain Teyssier$^{1}$, and Avishai Dekel$^{2}$
\\
$^{1}$Department of Astrophysical Sciences, Princeton University, 4 Ivy Lane, 08540, Princeton, NJ, USA\\
$^{2}$Racah Institute of Physics, The Hebrew University of Jerusalem, Jerusalem, 91904, Israel
}

\date{Accepted XXX. Received YYY; in original form ZZZ}

\pubyear{2024}

\begin{document}
\label{firstpage}
\pagerange{\pageref{firstpage}--\pageref{lastpage}}
\maketitle

\begin{abstract}
Motivated by the early excess of bright galaxies seen by \textit{JWST}, we run zoom-in cosmological simulations of a massive galaxy at Cosmic Dawn, in a halo of $10^{11} M_\odot$ at $z = 9$, using the hydro-gravitational code \texttt{RAMSES} at an effective resolution $\sim 10~{\rm pc}$. We investigate physical mechanisms that enhance the star-formation efficiencies (SFEs) at the high gas densities of the star-forming regions in this galaxy ($\sim 3\times 10^3~{\rm cm^{-3}}$, $\sim 10^4~M_\odot/{\rm pc^2}$). Our fiducial star formation recipe uses a physically-motivated, turbulence-based, multi-freefall model, avoiding \textit{ad hoc} extrapolation from lower redshifts. By $z = 9$, our simulated galaxy is a clumpy, thick, rotating disc with a high stellar mass $\sim 3\times 10^9~M_\odot$ and high star formation rate $\sim 50~M_\odot/{\rm yr}$. The high gas density makes supernova (SN) feedback less efficient, producing a high local SFE $\gtrsim 10\%$. The global SFE is set by feedback-driven outflows and only weakly correlated with the local SFE. Photoionization heating makes SN feedback more efficient, but the integrated SFE always remains high. Intense accretion at Cosmic Dawn seeds turbulence which reduces local SFE, but this only weakly affects the global SFE. The star formation histories of our simulated galaxies are similar to observed massive galaxies at Cosmic Dawn, despite our limited resolution. We set the stage for future simulations which treat radiation self-consistently and use a higher effective resolution $\sim 1~{\rm pc}$ that captures the physics of star-forming clouds.

\end{abstract}

\begin{keywords}
galaxies: high-redshift -- software: simulations -- galaxies: star formation
\end{keywords}

\section{Introduction}
\label{sec:intro}
The James Webb Space Telescope (\textit{JWST}) has opened a new frontier for empirical constraints on galaxy formation models at Cosmic Dawn ($z \geq 9$), an epoch starting with the formation of the first stars and ending with the reionization of the inter-galactic medium (IGM). At these redshifts, \textit{JWST}'s \texttt{NIRCam} and \texttt{NIRSpec} instruments probe the rest-frame ultraviolet (UV), where they provide estimates for stellar mass and star formation rate (SFR). 

So far, \textit{JWST} has found a growing list of galaxies at Cosmic Dawn with high stellar masses $M_* \gtrsim 10^8~M_\odot$ and/or high SFRs $\dot{M}_* \gtrsim 1~M_\odot/{\rm yr}$ \citep[e.g.][]{Finkelstein+2023, Labbe+2023, Mason+2023, Carniani+2024}. These massive Dawn galaxies (MDGs) constitute an excess in the abundance of the (UV-)brightest galaxies relative to expectations of standard galaxy formation models \citep{Boylan-Kolchin+2023}. \citet{Shen+2023} show that to reproduce the UV luminosity function (UVLF) inferred from MDGs, one requires that $\sim 30\%$ of the baryonic matter accreting onto a dark matter (DM) halo is converted into stars. This is far higher than the canonical value of $\lesssim 5\%$ inferred for galaxies at lower redshift based on abundance matching of stellar masses and $\Lambda {\rm CDM}$ haloes \citep{RodriguezPuebla+2017, Moster+2018, Behroozi+2019}.

The explanations for this tension fall into four categories. First, star formation might be intrinsically more efficient at Cosmic Dawn \citep{Dekel+2023}. Second, a low mass-to-light ratio at Cosmic Dawn might cause models calibrated to the lower-redshift universe to overestimate stellar masses. These explanations include the presence of active galactic nuclei (AGN) \citep{Wang+2024}, a top-heavy stellar initial mass function (IMF) \citep{Sharda&Krumholz2022, Cameron+2023}, a lack of dust along the line of sight \citep{Mason+2023, Ferrara+2023}, and differences in dust properties. Third, observations could be biased by bursty star formation at Cosmic Dawn \citep{Sun+2023, Shen+2023}. In this scenario, the bright end of the UVLF is dominated by less intrinsically bright galaxies which have undergone a recent burst. Finally, a modified $\Lambda$-CDM cosmology can produce a greater abundance of massive haloes at Cosmic Dawn. These explanations include ``Early Dark Energy'' \citep{Klypin+2021, Shen+2024} or massive primordial black hole seeds \citep{Liu&Bromm2022}. 

Some combination of effects may contribute to the tension. In this work, we explore the possibility that star formation is intrinsically more efficient at Cosmic Dawn due to differences in the properties of the interstellar medium (ISM). The mean baryon density of the Universe scales as $(1 + z)^{-3}$ due to cosmic expansion, suggesting that massive dark matter haloes at Cosmic Dawn may host galaxies with high gas densities.

\begin{figure}
    \centering
    \includegraphics[width=\linewidth]{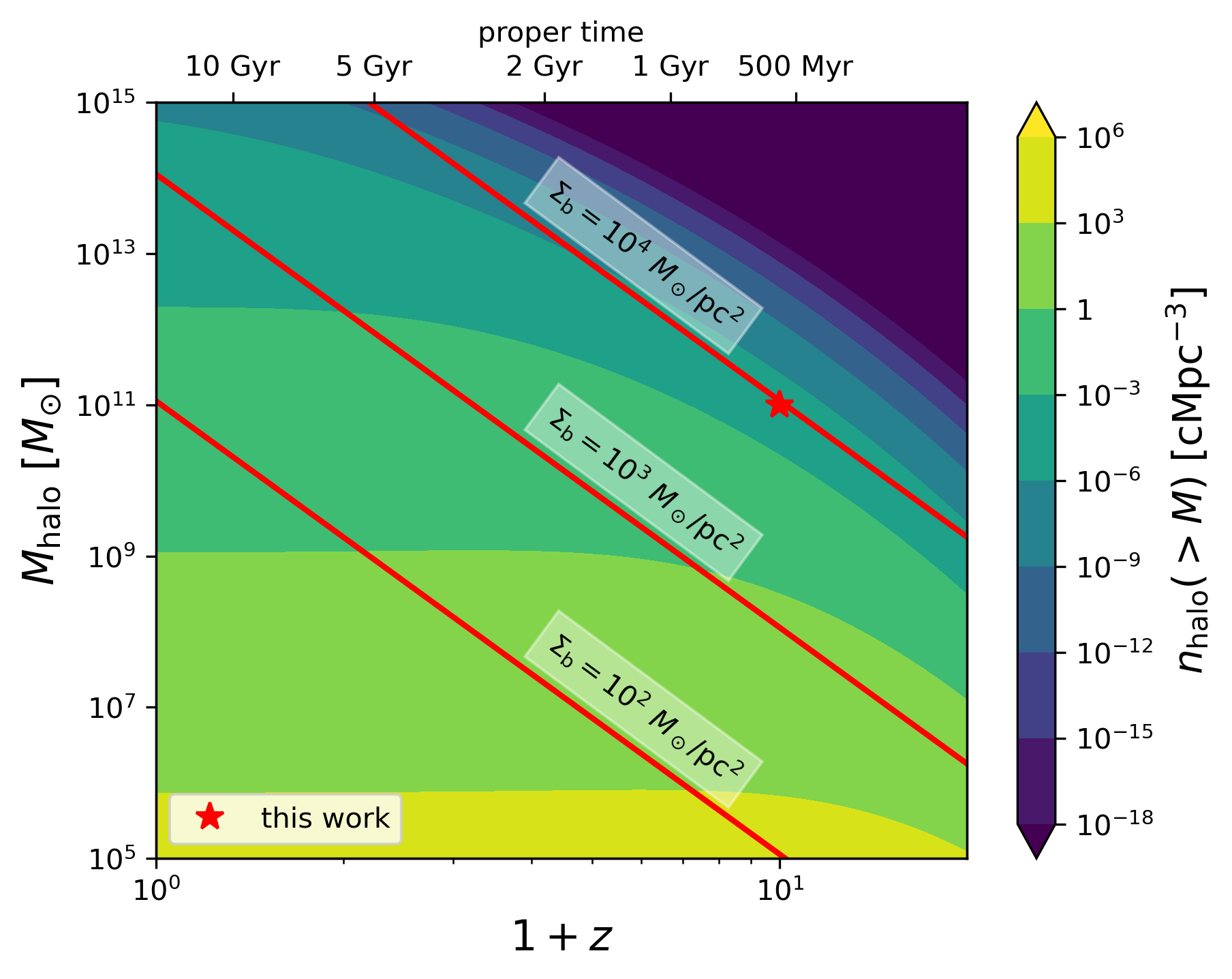}
    \caption{The comoving number density of dark matter haloes as a function of halo mass and redshift. A simple estimate for baryon surface density (App.~\ref{sec:surf_dens}) is used to plot lines of constant baryon surface density in red. The most massive galaxies at Cosmic Dawn seen by \textit{JWST} likely have comoving number densities $\sim 10^{-6}~{\rm cMpc}^{-3}$. Our simulation (red star) probes galaxies with similar number densities. The high baryon surface densities $\sim 10^4~M_\odot/{\rm pc^2}$ and number densities $\sim 3\times 10^{3}~{\rm cm^{-3}}$ create a unique environment for star formation which is not realized in the lower redshift Universe.}
    \label{fig:haloregime}
\end{figure}

At high redshift, massive haloes become increasingly rare. Typical comoving volumes of \textit{JWST} galaxy surveys are $\sim 10^5$ to $10^6~{\rm cMpc^3}$. For example, in the Cosmic Evolution Early Release Science \citep[CEERS][ERS-1345]{Finkelstein+2023b}, the effective survey volume at UV magnitude $\le -20$ is $\simeq 160000~{\rm cMpc^3}$. Thus, the haloes of observed galaxies should have comoving number densities $\gtrsim 10^{-6}~{\rm cMpc}^{-3}$. In Figure~\ref{fig:haloregime}, we plot the comoving number density of dark matter haloes as a function of halo mass and redshift, assuming cosmological parameters from the Plank 2018 results \citep{Plank+2020}. We estimate the number density by integrating the Press-Schechter halo mass function \citep{Press&Schechter1974} using the Python package \textsc{COLOSSUS} \citep{Diemer2018}. We also plot an estimate for the typical baryon surface density of the galaxy hosted by the halo. The calculation is described in Appendix~\ref{sec:surf_dens}.

Haloes at redshifts $z\sim 9$ and masses $\sim 10^{11}~M_\odot$ (marked by the red star) are interesting objects because they have number densities $\sim2 \times 10^{-5}~{\rm cMpc^{-3}}$, so a small sample of such objects should exist within current surveys. Simultaneously, they are expected to host galaxies with high baryon surface densities $\sim 10^4~M_\odot/{\rm pc^2}$. For comparison, a Milky-Way-like galaxy at $z=0$ with halo mass $10^{12}~M_\odot$ has number density $\sim 2 \times 10^{-3}~{\rm cMpc^{-3}}$ and baryon surface density $\sim 3\times 10^2~M_\odot/{\rm pc^2}$. We expect the difference in surface densities to be even greater for the gas component, because at low redshift, massive galaxies tend to be quenched with most of their baryons in stars. We can also estimate the typical Hydrogen number density by assuming a disc scale height (App.~\ref{sec:surf_dens}). For a scale height $(H/R)_{\rm gal} = 0.1$ and 50\% of the baryon mass in gas, the number density is $\sim 3\times 10^{3}~{\rm cm^{-3}}$.

Such high densities create a unique environment for star formation which is not realized in the lower redshift Universe, except in nuclear regions of some galaxies, where the typical result is a starburst \citep[e.g.][]{Naab&Ostriker2017}. MDGs may have starburst-like SFR densities throughout an extended region, rather than just in their center. In their feedback-free starburst (FFB) model, \citet{Dekel+2023} argue that the high density conditions of Cosmic Dawn suppress the stellar feedback processes which usually regulate star formation.

The physical conditions in MDGs differ from present day galaxies in other ways too. The rapid accretion of cold streams in massive galaxies at high redshift is expected to seed strong turbulence \citep{Dekel+2009, Ginzburg+2022}. This is empirically supported by the high velocity dispersions observed in galaxies at $z\gtrsim 6$ \citep{deGraaff+2024b}. Turbulent pressure supports molecular clouds against gravitational collapse, suppressing local star formation. Turbulence also reduces the clustering of supernovae (SNe), making SN feedback less effective \citep{Gentry+2017}. In addition to turbulence, the low metal content of the ISM and the higher CMB temperature at Cosmic Dawn may suppress star formation by making cooling less efficient.

Galaxy formation during the epoch of reionization is often studied with big box cosmological simulations such as \textsc{MassiveBlack} \citep{DiMatteo2012}, \textsc{CROC} \citep{Gnedin2014}, \textsc{Eagle} \citep{Schaye+2015, Crain+2015}, \textsc{BlueTides} \citep{Feng+2016}, \textsc{Aurora} \citep{Pawlik+2017}, \textsc{Illustris-Tng} \citep{Naiman+2018, Nelson+2018, Marinacci+2018, Springel+2018, Pillepich+2018}, \textsc{Sphinx} \citep{Rosdahl+2018}, \textsc{SIMBA} \citep{Dave+2019}, \textsc{Cosmic Dawn II} \citep{Ocvirk+2020}, and \textsc{Thesan} \citep{Kannan+2022}. However, it is challenging to simultaneously resolve the structure of the ISM at scales $\sim 10~{\rm pc}$ and use a large enough volume $\simeq (100 h_0^{-1}{\rm cMpc})^3$ to sample rare massive haloes.

Previous work has gotten around this by using computationally cheap dark-matter-only simulations to select massive haloes as sources for zoom-in simulations with full hydrodynamics \citep{Katz+1993, Tormen+1997}, including \textsc{Renaissance} \citep{OShea2015}, \textsc{FirstLight} \citep{Ceverino+2017}, \textsc{Fire-2} \citep{Ma+2018}, \textsc{Flares} \citep{Lovell+2021, Vijayan+2021}, and \textsc{serra} \citep{Pallottini+2022}. These works generally find that massive galaxies at high redshift are associated with efficient and bursty star formation. Initial work focused on forward modeling of observations and population-level properties like the UVLF rather than the details of star formation in individual galaxies. However, recent work has investigated the factors that contribute to efficient star formation in detail. \citet{Bassini+2023} ran \textsc{Fire-2} simulations and found that high gas volume and surface densities naturally led to efficient star formation. Similarly, \citet{Ceverino+2024} analyzed \textsc{FirstLight} simulations and found that star formation efficiency (SFE) is correlated with redshift and halo mass, reaching up to $\sim 10\%$ for $10^{11}~M_\odot$ haloes at $z=9$ (their Fig.~7).

In this work, we also use cosmological zoom-in simulations with the principal aim of exploring the physics of star formation and feedback in the most extreme MDGs. We simulate a halo which reaches $10^{11}~M_\odot$ at $z = 9$. We vary our treatment of turbulence and stellar feedback, allowing us to connect star formation on a global scale with ISM physics on a local scale. Our simulations differs from previous works in their treatment of star formation. Most high-redshift galaxy formation simulations use a star formation recipe based on the local Kennicutt-Schmidt Law \citep{Kennicutt1998}, where the local SFR density is $\epsilon_{\rm ff} \rho / \tau_{\rm ff}$ and $\epsilon_{\rm ff}$ is the local star formation efficiency (SFE) per freefall time $\tau_{\rm ff}$. $\epsilon_{\rm ff}$ is typically set to a constant value on the order of a few percent, consistent with observations of the local Universe \citep[e.g.][]{Hopkins+2014}. In contrast, we use a multi-freefall model \citep{Federrath&Klessen2012}, a physically-motivated model which predicts the SFR using a model of subgrid turbulence. This avoids \textit{ad hoc} extrapolation from lower redshifts.

This work is only the first step towards understanding the physics of star formation in MDGs. Our treatment of stellar feedback is crude due to the limitations of our effective resolution $\sim 10~{\rm pc}$ in a pure hydrodynamics simulation. Therefore, our quantitative results should be taken with a grain of salt. Future work will treat radiation self-consistently and use a higher effective resolution $\sim 1~{\rm pc}$ that captures the physics of star-forming clouds.

In Section~\ref{sec:methods}, we explain our numerical methods and our suite of models. In Section~\ref{sec:results}, we analyze our results. In Section~\ref{sec:discussion}, we interpret our results with toy models and discuss caveats and observational implications. We conclude in Section~\ref{sec:conclusion}. Tangential discussions are provided in the appendices.

\section{Numerical Methods}
\label{sec:methods}
We use the adaptive mesh refinement (AMR) code \texttt{RAMSES} \citep{Teyssier2002} to model the interaction of dark matter, star clusters, and baryonic gas in a cosmological context. We start with a discussion of the relevant technical details of \texttt{RAMSES} and our initial conditions (Sec.~\ref{sec:ramses}). We follow with the details of our recipes for subgrid turbulence (Sec.~\ref{sec:turb}), star formation (Sec.~\ref{sec:stars}), SN feedback (Sec.~\ref{sec:sne}), and early feedback (Sec.~\ref{sec:earlyfbk}). Finally, we describe our suite of simulations (Sec.~\ref{sec:models}).

\subsection{RAMSES}
\label{sec:ramses}

\texttt{RAMSES} evolves gas quantities in an Eulerian sense on a discretized grid of cubical cells, which can be adaptively refined to increase the spatial resolution locally according to some adopted Adaptive Mesh Refinement (AMR) criterion. Dark matter and star cluster particles are evolved in a Lagrangian sense. Dark matter particles and gas cells are coupled via the Poisson equation while star cluster particles and gas cells are coupled via recipes for star formation, early feedback, and SN feedback. We adopt standard cosmological parameters $h_0 = 0.7$, $\Omega_{\rm m,0} = 0.276$, $\Omega_{\rm b, 0} = 0.049$, $\Omega_{\Lambda, 0} = 0.724$, and zero curvature, consistent with current constraints.

Each particle is defined by a unique identification number (ID), a position $\vb{x}$ and a velocity $\vb{v}$. Star cluster particles carry additional metadata including their mass $m_*$, metallicity $Z$, and time of birth $t_{\rm birth}$. We evolve density $\rho$, pressure $P$, velocity $\vb{v}$, and metallicity $Z$ on the grid according to the Euler equations using the MUSCL-Hancock scheme, a second-order Godunov method, and the Harten-Lax-van Leer-Contact Leer (HLLC) Riemann solver \citep{Harten+1983}. We use a fine timestep controlled by several stability conditions \citep[see][]{Teyssier2002} and whose size is roughly $\simeq 500~{\rm yr}$.

We assume an ideal gas equation of state (EOS) appropriate for a rarefied plasma with $\gamma=5/3$. The specific turbulent kinetic energy (TKE) $\varepsilon_{\rm turb}$ and a refinement mask $M$ are advected passively with the flow. The trajectories of dark matter and star cluster particles are computed using a particle-mesh solver. The Poisson equation is solved using a multi-grid method \citep{Guillet&Teyssier2011} with Dirichlet boundary conditions at refinement level boundaries. 

The grid resolution can be defined in terms of a refinement level $\ell$ as $\Delta x = L / 2^{\ell}$, where $L$ is the size of the grid. Similarly, we can define the mass resolution of dark matter particles as
\begin{equation}
	m_{\rm dm} = \Omega_{\rm m,0} \rho_{\rm crit,0} (L / 2^{\ell})^3
	\label{eq:mass_dm}
\end{equation}
where $\rho_{\rm crit,0} = 3 H_0^2 / (8 \pi G)$ is the critical density at $z=0$.

We generate initial conditions for a low-resolution ($\ell = 9$) dark matter only simulation in a comoving volume $(100 h_0^{-1} {\rm cMpc})^3$ at $z=100$ using \texttt{MUSIC} \citep{Hahn&Abel2011}. We run the simulation until $z=9$ and identify a candidate halo with a mass $\simeq 10^{11}~M_\odot$. We trace the dark matter particles of the candidate halo back to their positions at $z=100$ and define the convex hull created by these particles as the zoom region. 

We generate new initial conditions using \texttt{MUSIC} including dark matter and baryons. The simulation box is translated such that the candidate halo will form near the center of the box at $z=9$. Within the zoom region, we use a resolution $\ell = 14$. Outside the zoom region, we progressively degrade the resolution until $\ell = 7$, enforcing a separation of 5 cells between level boundaries (from $\ell = 7$ to $\ell = 14$). We refer to this initial grid as the coarse grid. We track the Lagrangian volume of the halo throughout the simulation using the refinement mask, which we set equal to unity inside the zoom region and zero otherwise.

The dark matter mass resolution in the zoom region is $m_{\rm dm,min} \simeq 2.48\times 10^4~M_\odot$. The corresponding baryon mass resolution is $m_{\rm b,min} = f_{\rm b} m_{\rm dm,min} \simeq 4400~M_\odot$, where $f_{\rm b} = \Omega_{\rm b,0} / \Omega_{\rm m,0} \simeq 0.17$ is the baryon fraction. As a \textit{post hoc} confirmation of our approach, we have verified that all dark matter particles within a $50~{\rm kpc}$ sphere of the galaxy have mass $m_{\rm dm}$ throughout the duration of our simulations.

As the simulation runs, we refine the initial coarse grid based on a quasi-Lagrangian approach. A cell is refined if
\begin{enumerate}
	\item The refinement mask $M > 0.1$
\end{enumerate}
and at least one of the following conditions is met:
\begin{enumerate}\addtocounter{enumi}{1}
	\item The dark matter mass in a cell exceeds $8 m_{\rm dm}$
	\item There baryonic (gas + star) mass in a cell exceeds $8 m_{\rm b}$
	\item There are more than 5 Jeans lengths per cell.
\end{enumerate}
where the Jeans length is $\lambda_{\rm J} = (\pi c_{\rm s}^2 / G \rho)^{1/2}$ and $c_{\rm s} = \sqrt{\gamma P / \rho}$ is the sound speed.

Criterion (i) ensures that only cells which exchange gas with the initial zoom-in region are refined. Criterion (iv) ensures that we satisfy the \citet{Truelove+1997} criterion $\Delta x / \lambda_{\rm J} < 0.25$ in cells that are not maximally refined, preventing artificial fragmentation. 

Artificial fragmentation can still occur when maximally refined cells violate the Truelove criterion. However, these cells only persist for one freefall time due to rapid star formation, limiting the development of artificial fragmentation. In our fiducial simulation at $z=9$, only 6\% of maximally refined cells violate the Truelove criterion. In future simulations, we may use a pressure floor to eliminate artificial fragmentation altogether, following previous works \citep{Robertson&Kravtsov2008, Agertz+2009}.

We increment the maximum refinement level $\ell_{\rm max}$ each time the physical box size doubles due to cosmic expansion, reaching $\ell_{\rm max} = 20$ at expansion factor $a\in [0.05, 0.1]$. This approach maintains a constant physical effective resolution $\Delta x_{\rm min} = L/2^{\ell_{\rm max}} \simeq 10~{\rm pc}$ throughout the duration of the simulation. 

We do not directly model Pop III star formation. Instead, we enrich the ISM to $10^{-3} Z_\odot$ in the initial conditions. Previous simulation work shows that a single pair instability SN can enrich the host halo to at least this metallicity \citep{Wise+2012}. Furthermore, this is above the critical metallicity $\sim 10^{-3.5} Z_\odot$ at which Pop III stars cease to form due to fine-structure Carbon and Oxygen line cooling \citep{Bromm&Loeb2003}, so our initial conditions are consistent with no Pop III stars.

Gas cooling and heating is implemented using equilibrium chemistry for Hydrogen and Helium \citep{Katz+1996}, with a metallicity-dependence given by \citet{Sutherland&Dopita1993} and cosmological abundances ($X=0.76$, $Y=0.24$). In diffuse gas $n_{\rm H} \le 0.01~{\rm cm^{-3}}$ \citep{Aubert&Teyssier2010}, we account for heating from the extragalactic UV background assuming a uniform radiation field which evolves with time according to \citet{Haardt&Madau1996}. This likely underestimates the UV background in MDGs due to clustering effects, but the error is acceptable given the simplicity of our model.

We floor the temperature at the cosmic microwave background (CMB) temperature to account for heating by CMB photons. This effect is important at Cosmic Dawn, when the CMB temperature is higher than the temperatures of star-forming clouds at the present day. When we stop our simulations at $z=9$, the CMB temperature is $T_{\rm CMB} \simeq 27.25~{\rm K}$. 

We run our simulations on the Stellar cluster at Princeton University. Each simulation was run on 4 nodes with 96 cores per code, making 384 cores total. The cores on stellar are 2.9 GHz Intel Cascade Lake processors. Each simulation took approximately 2 weeks of runtime to reach completion.

\subsection{Subgrid turbulence model}
\label{sec:turb}

We use a Large Eddy Simulation (LES) method for subgrid turbulence as implemented in \textsc{RAMSES} by \citet{Kretschmer&Teyssier2020}, which we describe here for clarity. To describe the model, we decompose the density field into a bulk component $\overline{\rho}$ and a turbulent component $\rho'$:
\begin{equation}
	\rho = \overline{\rho} + \rho'
\end{equation}
The bulk component is defined as the volume-average density field smoothed on the scale of the cell $\Delta x$. We also decompose the subgrid temperature $T$ and velocity $\vb{v}$ fields into bulk components $\widetilde{T}$, $\widetilde{\vb{v}}$ and turbulent components $T''$, $\vb{v}''$:
\begin{equation}
	T = \widetilde{T} + T'', \quad \vb{v} = \widetilde{\vb{v}} + \vb{v}'' \quad {\rm with} \quad \widetilde{T} = \frac{\overline{\rho T}}{\overline{\rho}}, \quad \widetilde{\vb{v}} = \frac{\overline{\rho \vb{v}}}{\overline{\rho}}
\end{equation}
The bulk components are defined as the mass-weighted volume-average fields, also known as Favre-averaged fields. The TKE is defined as
\begin{equation}
    e_{\rm turb} = \frac{1}{2} \overline{ \rho v''^2 } = \frac{1}{2}\overline{\rho} \sigma_{\rm 3D}^2 = \frac{3}{2}\overline{\rho} \sigma_{\rm 1D}^2
\end{equation}
where $\sigma_{\rm 3D}$ (resp. $\sigma_{\rm 1D}$) is the three-dimensional (resp. one-dimensional) turbulent velocity dispersion. Using these definitions, one can derive equations for the TKE and bulk fluid components \citep{Schmidt&Federrath2011, Semenov+2017}

The bulk fluid equations include new terms for turbulent diffusion. However, the turbulent diffusion is generally small compared to the diffusion introduced by the numerical scheme. We maintain the original fluid equations rather than adding a diffusive term to an already too-diffusive scheme, following previous work \citep{Schmidt+2006}. The TKE equation reads
\begin{equation}
	\pdv{t} e_{\rm turb} + \pdv{x_j} (e_{\rm turb} \widetilde{v}_j ) + P_{\rm turb} \pdv{\widetilde{v}_j}{x_j} = \mathcal{C}_{\rm turb} - \mathcal{D}_{\rm turb}
	\label{eq:turb}
\end{equation}
where $P_{\rm turb} = (2/3) e_{\rm turb} = \overline{\rho} \sigma_{\rm 1D}^2$ is the turbulent pressure and $\mathcal{C}_{\rm turb}$ and $\mathcal{D}_{\rm turb}$ are creation and destruction source terms respectively. 

In the LES framework, we assume that resolved and unresolved turbulence are coupled by eddies at the resolution scale. Defining the turbulent viscosity   $\mu_{\rm turb} = \overline{\rho} \Delta x \sigma_{\rm 1D}$, the turbulence creation term is
\begin{equation}
	\mathcal{C}_{\rm turb} = 2 \mu_{\rm turb} \sum_{ij} \left[ \frac{1}{2} \left( \pdv{\widetilde{v_i}}{x_j} + \pdv{\widetilde{v_j}}{x_i} \right) - \frac{1}{3} ( \div{\widetilde{\vb{v}}} ) \delta_{ij} \right] = \frac{1}{2} \mu_{\rm turb} \abs{ \widetilde{S}_{ij} }^2
	\label{eq:creation}
\end{equation}
where $\widetilde{S}$ is the bulk viscous stress tensor. Defining the turbulent dissipation timescale $\tau_{\rm diss} = \Delta x / \sigma_{\rm 1D}$, the turbulence destruction term is
\begin{equation}
	\mathcal{D}_{\rm turb} = \frac{\varepsilon_{\rm turb}}{\tau_{\rm diss}}
\end{equation}

One caveat of the LES model is that gravitational forces do not enter into the turbulence creation term (Eq.~\ref{eq:creation}). In a stable Keplerian disc, gravitational forces can stabilize shear flows against the Kelvin-Helmholtz instability, so the model overestimates the TKE source term. In a gravitationally unstable disc, gravitational forces can source additional turbulence, so the model underestimates the TKE source term.

\subsection{Star formation recipe}
\label{sec:stars}

With slight modification (Sec.~\ref{sec:b_turb}), we model star formation using the multi-freefall model of \citet{Federrath&Klessen2012} as implemented in \textsc{RAMSES} by \citet{Kretschmer&Teyssier2020}, which we describe here for clarity.

\subsubsection{Subgrid density PDF}

Empirically, star formation is described by the Kennicutt-Schmidt (KS) law \citep{Kennicutt1998}, which relates the SFR surface density and gas surface density by $\Sigma_{\rm SFR} \propto \Sigma_{\rm gas}^{1.5}$. The KS relation holds at scales as small as $1~{\rm kpc}$ \citep{Kennicutt+2007}, supporting the idea that star formation can be modeled locally as volumetric Schmidt law \citep{Schmidt1959}
\begin{equation}
	\dot{\rho}_* = \epsilon_{\rm ff} \frac{\rho}{\tau_{\rm ff}} \quad {\rm with} \quad \tau_{\rm ff} = \sqrt{\frac{3\pi}{32 G \rho}}
	\label{eq:KSlaw}
\end{equation}
where $\dot{\rho}_*$ is the local SFR density and $\epsilon_{\rm ff}$ is the SFE per local freefall time $\tau_{\rm ff}$. This model naturally explains the power law index of $1.5$ in the empirical KS law. 

Equation~\ref{eq:KSlaw} is often combined with a fixed density threshold to obtain a simple star formation recipe which can be used in cosmological simulations \citep[e.g.][]{Rasera&Teyssier2006}. The parameter $\epsilon_{\rm ff}$ is chosen to match the observed KS law in nearby resolved galaxies, typically resulting in a value $\lesssim 5\%$. However, the SFE in the local Universe cannot necessarily be extrapolated to Cosmic Dawn. This motivates a prescriptive star formation recipe which does not rely on empirical calibration.

One approach is to consider the structure of the density field on unresolved scales $\le \Delta x_{\rm min} \simeq 10~{\rm pc}$. We extrapolate the turbulence spectrum to unresolved scales assuming Burgers turbulence:
\begin{equation}
	\sigma(\ell) = \sigma_{\rm 1D} \left( \frac{\ell}{\Delta x} \right)^{1/2}
	\label{eq:burger}
\end{equation}
where $\ell$ is the spatial scale.

The turbulence transitions from supersonic to subsonic at the sonic scale $\ell_{\rm s}$ where the turbulent velocity dispersion (Eq.~\ref{eq:burger}) is equal to the sound speed. Defining the turbulent Mach number $\mathcal{M}_{\rm turb} = \sigma_{\rm 1D}/c_{\rm s}$, the sonic scale is $\ell_s = \Delta x / \mathcal{M}_{\rm turb}^2$. Below the sonic scale, density fluctuations are weak, and a gas cloud can be treated as a quasi-homogeneous region. 

We assume that each parcel of gas which is gravitationally unstable on the sonic scale will eventually collapse and form stars. On intermediate scales between the sonic scale and the resolution scale, density fluctuations can be significant. The probability distribution function (PDF) of density in a supersonic turbulent medium is well-described by a log-normal distribution \citep{VazquezSemadeni1994, Kritsuk+2007}:
\begin{equation}
	p_V(s) =\dv{V}{s} = \frac{1}{\sqrt{2\pi \sigma_s^2}} \exp ( -\frac{(s - \overline{s})^2}{2 \sigma_s^2} )
	\label{eq:mffpdf}
\end{equation}
with normalization condition
\begin{equation}
	\int_{-\infty}^\infty p_V(s) \dd s = 1 \quad {\rm and} \quad \int_{-\infty}^\infty \rho p_V(s) \dd s = \overline{\rho}
\end{equation}
where $s = \ln(\rho / \overline{\rho})$ is the logarithmic density, $\sigma_s$ is its standard deviation, and $\overline{s} = -1/2\sigma_s^2$ is its mean.

\citet{Padoan&Nordlund2011} fit a simple analytic form to $\sigma_s$ using non-magnetized, isothermal turbulence simulations forced by an Ornstein-Uhlenbeck process:
\begin{equation}
	\sigma_s^2 = \ln ( 1 + b^2 \mathcal{M}_{\rm turb}^2 )
	\label{eq:sigs}
\end{equation}
The parameter $b$ describes the turbulence forcing in the simulations. For purely solenoidal (divergence-free) forcing, $b = 1/3$. For purely compressive (curl-free) forcing, $b=1$. The column density PDF is also well-described by a lognormal distribution, although the dispersion is generally smaller because fluctuations are averaged out by integration along the line of sight (App.~\ref{sec:coldenspdf}).

\subsubsection{Turbulence forcing parameter}
\label{sec:b_turb}

In most of our simulations, we set the forcing parameter to a constant value. However, we run one experimental simulation where we derive the forcing parameter from the local velocity field, a novel addition to \textsc{RAMSES} in this work. Using a Helmholtz decomposition, the velocity field can be represented as a curl-free term $\vb{v}_{\rm comp} = -\grad{\phi}$ and divergence-free term $\vb{v}_{\rm sol} = \curl{\vb{A}}$.
\begin{equation}
	\vb{v} = \vb{v}_{\rm comp} + \vb{v}_{\rm sol} = -\grad{\phi} + \curl{\vb{A}}
	\label{eq:helm}
\end{equation}

Computing the Helmholtz decomposition would require a computationally expensive iterative solver on the global velocity field. Instead, we use a heuristic approach motivated by the following. Taking the divergence of Equation~\ref{eq:helm} reveals that the curl-free component of the velocity field depends only on the divergence. Similarly, taking the curl reveals that the divergence-free component of the velocity field depends only on the curl. We make a simple approximation that the magnitudes of the curl-free and divergence-free components of the velocity field are proportional to the divergence and curl respectively. This gives the following expression for ratio of power in compressive to solenoidal forcing modes:
\begin{equation}
	\psi = \frac{P_{\rm comp}}{P_{\rm sol}} = \frac{(\div{\vb{v}})^2}{\norm{\curl{\vb{v}}}^2}
	\label{eq:power}
\end{equation}
See \citet{Ginzburg+2025} for a more rigorous justification of this expression. \citet{Federrath+2010} fit a simple analytic form to $b_{\rm turb}$ using detailed turbulence simulations (their Eq.~23): 
\begin{equation}
	b_{\rm turb} \simeq \frac{1}{3} + \frac{2}{3} \left( \frac{\psi}{\psi + 1} \right)^3
	\label{eq:bturb}
\end{equation}
We use this formula to compute a local turbulence forcing for each cell in the simulation.

\subsubsection{Star formation efficiency}

Assuming that star-forming cores are homogeneous spheres with diameter $\ell_{\rm s}$ \citep{Krumholz&McKee2005}, the gravitational stability condition is $\alpha_{\rm vir, core} \ge 1$, where $\alpha_{\rm vir, core}$ is the virial parameter of the core given by
\begin{equation}
	\alpha_{\rm vir, core} = \frac{2 E_{\rm kin}}{\abs{E_{\rm grav}}} = \frac{15}{\pi} \frac{c_{\rm s}^2 + \sigma(\ell_{\rm s})^2}{G \rho \ell_{\rm s}^2}
	\label{eq:alphavir_first}
\end{equation}

The gravitational stability condition can alternatively be expressed as a condition on the density $\rho \le \rho_{\rm crit}$, where
\begin{equation}
	\rho_{\rm crit} = \frac{15}{\pi} \frac{2 c_{\rm s}^2 \mathcal{M}_{\rm turb}^4}{G \Delta x} = \alpha_{\rm vir} \overline{\rho} \frac{ 2 \mathcal{M}_{\rm turb}^4 }{1 + \mathcal{M}_{\rm turb}^2}
\end{equation}
and
\begin{equation}
	\alpha_{\rm vir} = \frac{15}{\pi} \frac{c_{\rm s}^2 + \sigma_{\rm 1D}^2}{G \overline{\rho} \Delta x^2} = \frac{15}{\pi} \frac{c_{\rm s}^2}{G \overline{\rho} \Delta x^2} (1 + \mathcal{M}_{\rm turb}^2)
	\label{eq:alphavir}
\end{equation}
is the virial parameter of the entire cell. The critical logarithmic density is
\begin{equation}
	s_{\rm crit} = \ln \left[ \alpha_{\rm vir} \frac{2 \mathcal{M}_{\rm turb}^4}{1 + \mathcal{M}_{\rm turb}^2} \right]
\end{equation}
$s_{\rm crit}$ depends on the cell size. At higher resolutions, a larger density is required to become gravitationally unstable, but gas will naturally reach those higher densities as it collapses down to the smaller cell size.

The model breaks down when the entire cell is subsonic $\mathcal{M}_{\rm turb} \lesssim 1$ \citep{Federrath&Klessen2012} and the sonic scale is larger than the resolution scale. In this case, the density PDF should be interpreted as the probability distribution for the density across the entire cell and the gravitational stability condition (Eq.~\ref{eq:alphavir_first}) should use the resolution scale rather than the sonic scale. We smoothly interpolate between both stability conditions by defining a modified critical logarithmic density \citep{Kretschmer&Teyssier2020}:
\begin{equation}
	s_{\rm crit} = \ln \left[ \alpha_{\rm vir} \left( 1 + \frac{2 \mathcal{M}_{\rm turb}^4}{1 + \mathcal{M}_{\rm turb}^2} \right) \right]
	\label{eq:scrit}
\end{equation}

If each unstable gas parcel collapses in one freefall time and converts all its mass into stars, then the local SFR is given by
\begin{equation}
	\dot{\rho}_* = \int_{s_{\rm crit}}^\infty \frac{\rho}{\tau_{\rm ff}(\rho)} p(s) \dd s = \epsilon_{\rm ff} \frac{\overline{\rho}}{\tau_{\rm ff}(\overline{\rho})}
\end{equation}
where $\epsilon_{\rm ff}$ is the local SFE per freefall time given by
\begin{equation}\begin{split}
	\epsilon_{\rm ff} =\ & \int_{s_{\rm crit}}^\infty \frac{\tau_{\rm ff}(\overline{\rho})}{\tau_{\rm ff}(\rho)} \frac{\rho}{\overline{\rho}} p(s) \dd s\\
	=\ & \frac{1}{2} \exp( \frac{3}{8} \sigma_s^2 ) \left[ 1 + {\rm erf} \left( \frac{\sigma_s^2 - s_{\rm crit}}{\sqrt{2 \sigma_s^2}} \right) \right]
	\label{eq:mff}
\end{split}\end{equation}
For notational convenience, we drop the overlines on the bulk fluid components for the remainder of the paper.

This is the multi-freefall (MFF) model of \citet{Federrath&Klessen2012}, validated by detailed turbulence simulations in the same work. This recipe is an important first step towards predictive models at Cosmic Dawn. However, there are several caveats. The model neglects deviations from a lognormal density PDF, the spatial correlation of the density field, and the timescale required to replenish the density PDF after a star has formed. Some of these issues are addressed by the recent turbulent support (TS) model of \citet{Hennebelle+2024} (Sec.~\ref{sec:beyond_mff}). The MFF model implies that as the turbulence forcing becomes more solenoidal, the subgrid density PDF becomes more narrow (Eq.~\ref{eq:sigs}), generally reducing the local SFE for the same gas conditions and TKE (Eq.~\ref{eq:mff}). We leave further discussion to Section~\ref{sec:turbdiscuss}.

The expected number of star cluster particles to form in one timestep $\Delta t$ is
\begin{equation}
	\expval{N_{\rm cl}} = \frac{\dot{\rho}_* \Delta x^3 \Delta t}{M_{\rm cl}}
\end{equation}
where $M_{\rm cl}$ is the star cluster particle mass. In our simulations, the fiducial value $M_{\rm cl} = M_{\rm b,min} = 4400~M_\odot$ is the minimum baryon mass resolution on the coarse grid defined in Section~\ref{sec:ramses}. Each timestep, we form a number of star cluster particles sampled from a Poisson distribution with a Poisson parameter $\expval{N_{\rm cl}}$. This procedure allows us to model star formation as discrete events distributed in time, independent of the adopted star cluster particle mass. 

Although our star formation recipe is independent of the adopted star cluster particle mass, our recipe for photoionization feedback is not. By adjusting the star particle mass, we adjust the efficiency of photoionization feedback (Sec.~\ref{sec:earlyfbk}). We exploit this effect in our early feedback series (Sec.~\ref{sec:models}).

\subsection{Supernovae recipe}
\label{sec:sne}

We use a SN feedback model based on the detailed turbulence simulations of \citet{Martizzi+2015} as implemented in \textsc{RAMSES} by \citet{Kretschmer&Teyssier2020}, which we describe here for clarity. SN explosions are one of the key feedback processes which regulate star formation. As the SFR increases, the SN rate increases commensurately. SNe inject energy and momentum into the ISM, depleting the reservoir of cold gas available to form stars. 

Our simulation duration $\sim 550~{\rm Myr}$ is shorter than the timescales required for Type Ia SNe $\gtrsim 1~{\rm Gyr}$, associated with the explosion of a white dwarf in a close binary system \citep{Tinsley1979}. We only model Type II SNe, associated with the explosions of massive stars $m_* \gtrsim 8~M_\odot$.

Type II SNe are expected to occur during a window of time from $\tau_{\rm start} \simeq 3 {\rm Myr}$ to $\tau_{\rm end} \simeq 20~{\rm Myr}$ after the birth of a massive star. These values for $\tau_{\rm start}$ and $\tau_{\rm end}$ are consistent with previous work using the population synthesis code \textsc{Starburst99} \citep{Leitherer+1999, Kimm+2015}. 

Assuming that it samples the IMF, a star cluster particle is expected to produce a number of SNe given by
\begin{equation}
	N_{\rm SNe} = \frac{\chi M_{\rm cl}}{m_{\rm big}}
\end{equation}
where $M_{\rm cl}$ is the star cluster particle mass, $m_{\rm big}$ is the typical mass of a supernova progenitor, and $\chi$ is the mass fraction of supernova progenitors within the stellar population. $m_{\rm big}$ and $\chi$ are related to the IMF $\xi(m)$ via
\begin{align}
	m_{\rm big} = \frac{\int_{m_{\rm min}}^{m_{\rm max}} m \xi(m) \dd m }{\int_{m_{\rm min}}^{m_{\rm max}} \xi(m) \dd m}, \quad \chi = \frac{\int_{m_{\rm min}}^{m_{\rm max}} m \xi(m) \dd m }{\int_{0}^{m_{\rm max}} m \xi(m) \dd m}
    \label{eq:mbig}
\end{align}
where $m_{\rm min}$ is the minimum mass of a SN progenitor and $m_{\rm max}$ is the maximum stellar mass.

The IMF only enters into our modeling through our choice of $m_{\rm big}$ and $\chi$. Using a Chabrier IMF \citep{Chabrier2003} with $m_{\rm min} = 8~M_\odot$ and $m_{\rm max} = 100~M_\odot$, one finds $m_{\rm big} \simeq 19~M_\odot$ and $\chi \simeq 0.21$. Following \citet{Kretschmer&Teyssier2020}, we adopt $m_{\rm big} = 10~M_\odot$ and $\chi = 0.2$. Our smaller value of $m_{\rm big}$ results in approximately twice as many SNe than predicted by a Chabrier IMF. This discrepancy was unintentional. However, we note that the value of $m_{\rm big}$ is uncertain due to the sensitivity of Equation~\ref{eq:mbig} to the exact choices of $m_{\rm min}$, $m_{\rm max}$, and the functional form of the IMF.

In each simulation timestep, we compute the SN rate assuming that the $N_{\rm SNe}$ events are uniformally distributed in the interval between $\tau_{\rm start}$ and $\tau_{\rm end}$:
\begin{equation}
	\dot{N}_{\rm SNe} = \frac{N_{\rm SNe}}{\tau_{\rm end} - \tau_{\rm start}}
\end{equation}
although physically, the SN distribution should be skewed towards later times (Sec.~\ref{sec:snecav}). 

The expected number of SN events per timestep is $\expval{N_{\rm SNe}} = \dot{N}_{\rm SNe} \Delta t$. Each timestep, we trigger a number of SNe sampled from a Poisson distribution with a Poisson parameter $\expval{N_{\rm SNe}}$, similar to \citet{Hopkins+2018}. This procedure allows us to model SNe as discrete events distributed in time, independent of the adopted star cluster particle mass. 

The exploding SN blastwave interacts with the surrounding medium in 4 stages. Initially, the SN ejecta expands freely. Once the ejecta has swept up a mass in the ISM comparable to its own, a shock forms and expands following the Sedov-Taylor solution, approximately conserving energy. Eventually, radiative cooling behind the shock front causes the expansion to deviate from the Sedov-Taylor solution, and the expansion proceeds approximately conserving momentum. Finally, the blast wave slows due to radiative losses and accumulated material, fading into the ISM. The transition radius between the second Sedov-Taylor phase and the third ``snowplow'' phase is called the cooling radius.

Ideally, one should resolve the energy-conserving Sedov-Taylor phase and model a SN explosion by injecting a thermal energy $E_{\rm SN} \simeq 10^{51}~{\rm erg}$ into the surrounding gas. However, at high densities, the cooling radius is smaller than the resolution scale. In this case, injecting energy into the surrounding gas would result in the energy from the SN being artificially radiated away before it could accelerate gas in the snowplow phase. To compensate for this effect, we inject momentum in addition to thermal energy when the cooling radius is not resolved. This momentum feedback method is common practice in galaxy simulations \citep{Hopkins+2011, Hopkins+2018, Kim&Ostriker2015, Kimm+2015}.

\citet{Martizzi+2015} fit power laws in density and metallicity to the cooling radius $R_{\rm cool}$ and the SN terminal momentum $P_{\rm SN}$ using high-resolution simulations of individual SN explosions:
\begin{align}
	R_{\rm cool} \simeq\ & 3.0\left( \frac{Z}{Z_\odot} \right)^{-0.082} \left( \frac{n_{\rm H}}{100~{\rm cm^{-2}}} \right)^{-0.42}~{\rm pc} \label{eq:Rcool}\\
	P_{\rm SN} \simeq\ & 1420 \left( \frac{Z}{Z_\odot} \right)^{-0.137} \left( \frac{n_{\rm H}}{100~{\rm cm^{-2}}} \right)^{-0.16}~M_\odot {\rm km / s} \label{eq:pSN}
\end{align}
We use these equations to estimate $R_{\rm cool}$ and $P_{\rm SN}$ and inject momentum according to
\begin{equation}
	P = P_{\rm SN} N_{\rm SN} \begin{cases} 1 & R_{\rm cool} < \Delta x_{\rm min}\\ (\Delta x_{\rm min}/R_{\rm cool})^{3/2} & \Delta x_{\rm min} < R_{\rm cool} < 4 \Delta x_{\rm min}\\ 0 & R_{\rm cool} > \Delta x_{\rm min} \end{cases}
\end{equation}

Our approach is only an approximation of the full model recommended by \citet{Martizzi+2015}, which incorporates additional physical scales into the calculation of injected momentum and energy. \citet{Martizzi+2015} also provide versions of Equations~\ref{eq:Rcool} and \ref{eq:pSN} for SN which occur in an inhomogeneous medium with $\mathcal{M}_{\rm turb} = 30$. In an inhomogeneous medium, the cooling radius is larger by a factor of a few because the blast wave preferentially moves through low-density channels, but the terminal energy and momentum deposited in the ISM by the SN remnant do not change significantly. Ideally, one should use a model which depends continuously on the turbulent Mach number. 

We inject momentum onto the grid by dividing momentum flux $P \Delta x^{-2} \Delta t^{-1}$ equally between the six faces of the cell and directly adding it to the thermal pressure in the Riemann solver \citep{Agertz+2013}. We remove the associated $P \dd V$ work in the internal energy equation to avoid spurious heating \citep{Kretschmer&Teyssier2020}. We inject thermal energy onto the grid by adding energy $N_{\rm SN} E_{\rm SN}$ to the thermal energy of the cell hosting the star particle. We assume a SN metal yield $y=0.1$, so each SN event deposits a mass $y N_{\rm SN} m_{\rm big} = N_{\rm SN} M_\odot$ of metals into the ISM.

Momentum feedback approximates the long-term effect of SNe on the surrounding ISM, but it fails to capture heating on short timescales in the unresolved Sedov-Taylor phase. Our results concerning SN feedback should be considered with caution, despite our physically-motivated subgrid prescription. We leave further discussion to Section~\ref{sec:snecav}.

\subsection{Early feedback recipe}
\label{sec:earlyfbk}

In addition to SN feedback, stellar feedback includes thermal pressure from photoionized gas, stellar winds, and radiation pressure. We call these early feedback processes because they begin immediately after a massive star forms, in contrast to SN feedback. Out of the early feedback processes, we only model thermal pressure from photoionized gas. We discuss the effects of other early feedback processes in Section~\ref{sec:otherfbk}.

In each cell containing a star cluster particle younger than $\tau_{\rm end} = 20~{\rm Myr}$, we set the temperature $T_{\rm phot} = 20000~{\rm K}$. This forces the gas into a fast cooling regime where it approaches the photoionization temperature within a few timesteps. In practice, we achieve this by injecting a thermal energy associated with a sound speed 
\begin{equation}
	c_{\rm s} = \sqrt{\frac{\gamma k_{\rm B} T_{\rm phot}}{\mu m_{\rm p}}} \simeq 22~{\rm km/s}
\end{equation}
where we have assumed a $\gamma$ = 5/3 adiabatic EOS and mean molecular weight $\mu = 0.6$.

Our early feedback recipe depends on the choice of $M_{\rm cl}$ because each star particle has the same effect on its parent cell, regardless of its mass. As the adopted value of $M_{\rm cl}$ decreases, less stellar mass is required to photoionize the parent cell, representing increasingly efficient photoionization feedback. We select a fiducial value of $M_{\rm cl}$ such that the H II region volume equals the volume of the most refined cell under typical star-forming conditions. In this case, a star particle is formed exactly when its H II region becomes resolved.

Let $\expval{Q/m_*}$ be the average rate of Hydrogen-ionizing photons emitted per stellar mass. This can be derived from the IMF $\xi(m)$ if the rate of ionizing photon emission $Q(m)$ is known as a function mass:
\begin{equation}
	\expval{\frac{Q}{m_*}} = \frac{\int Q(m) \xi(m) \dd m}{\int m \xi(m) \dd m}
\end{equation}

Using data for main sequence stars given by Table~15.1 of \citet{Draine2011}, we fit $Q(m)$ to a simple analytic form for masses $\ge 15.6~M_\odot$ (App.~\ref{sec:QvsM}):
\begin{equation}
	Q(m) \simeq \left[ 8.37\times 10^{45} \left(\frac{m}{M_\odot}\right)^{2.12} - 2.33\times 10^{48}\right]~{\rm cm^{-2}s^{-1}}
\end{equation}
Assuming a Chabrier IMF \citep{Chabrier2003}, we estimate $\expval{Q / m_*} \simeq 5.85 \times 10^{46}~{\rm cm^{-2} s^{-1}}M_\odot^{-1}$. Higher mass stars contribute more ionizing photons, but have a lower abundance. In our calculation, most of the contribution to the ionizing photons comes from stars $\simeq 35~M_\odot$. Star clusters which are too small to sample the IMF up to this mass may have lower ionization rates.

Equating the rates of photoionization and radiative recombination, we find the classic result from \citet{Stromgren1939} that the star cluster creates a Str{\"o}mgren sphere of ionized gas with radius
\begin{equation}
	R_{\rm S} = \left( \frac{3 M_{\rm cl}}{4 \pi \alpha_B n_{\rm H}^2}  \expval{ \frac{Q}{m_*} }  \right)^{1/3}
\end{equation}
where $\alpha_B \simeq 2.58 \times 10^{-13}~{\rm cm^3/s}$ is the case B recombination rate at an electron temperature $10^4~{\rm K}$ \citep{Ferland+1992}. We use the on-the-spot approximation, assuming that electrons which recombine directly to the ground state do not contribute to the net ionization.

The stellar mass required to photoionize a cell is given by equating Str{\"o}mgren sphere and cell volumes and solving for the star cluster mass, which yields a similar value to our fiducial star particle mass $4400~M_\odot$:
\begin{equation}\begin{split}
	M_{\rm cl} =\ & \frac{4\pi}{3} \frac{\alpha_B n_{\rm H}^2 \Delta x_{\rm min}^3}{\expval{Q/m_*}}\\
	\simeq\ & 5000~M_\odot \left( \frac{n_{\rm H}}{100~{\rm cm^{-3}}} \right)^2 \left( \frac{\Delta x}{10~{\rm pc}} \right)^3 \left( \frac{\expval{Q/m_*}~{\rm cm^2 s}M_\odot}{5.85\times 10^{46}} \right)^{-1}
	\label{eq:Mcl}
\end{split}\end{equation}

In MDGs, stars form at significantly higher densities than the value $\sim 10^2~{\rm cm^{-2}}$ which appears in Equation~\ref{eq:Mcl}. However, the strong turbulence in MDGs lowers the effective density relevant for photoionization because photoionizing photons preferentially escape through low-density channels in the gas. We can estimate the effective density using a PDF-weighted harmonic mean
\begin{equation}
	\rho_{\rm eff} = \left[ \int \frac{1}{\rho} p_V(s) \dd s \right]^{-1} = e^{-\sigma_s^2} \overline{\rho} = \frac{\overline{\rho}}{1 + b_{\rm turb}^2 \mathcal{M}_{\rm turb}^2}
	\label{eq:rhoeff}
\end{equation}
This is equivalent to computing the density associated with the average photon mean free path, which scales as $\propto \rho^{-1}$ for constant opacity. The result comports with our intuition that the effective density should decrease with increasing turbulent Mach number. This is similar in spirit to the more heuristic definition of the effective density in Appendix~B of \citet{Faucher-Giguere+2013}. For the typical densities and turbulent Mach numbers at star-formation sites in our simulations ($n_{\rm H} \sim 10^4~{\rm cm^{-3}}$, $\mathcal{M}_{\rm turb} \sim 10$), the effective density is indeed $\sim 10^2~{\rm cm^{-3}}$.

Our crude model (single star particle mass) cannot account for the density dependence in Equation~\ref{eq:Mcl}. We run several simulations with different star particle masses to determine how this parameter effects our results (Sec.~\ref{sec:models}). In future work, we will improve upon our early feedback model.

We also do not properly account for multiple star cluster particles in a cell. If one star cluster particle photoionizes a cell, then by the Str\"{o}mgren sphere argument, multiple star particles should photoionize a region larger than a cell. In our simulations, these situations are instead handled by multiplying the injected thermal energy by the number of star cluster particles, artificially restricting the Str\"{o}mgren sphere to the cell size.

\subsection{The suite of simulations}
\label{sec:models}

\begin{table*}
    \rowcolors{1}{}{gray!15}
    \centering
    \begin{tabular}{lccccccccccc}\hline
	\rowcolor{gray!50}\multicolumn{12}{c}{Feedback series}\\
		\rowcolor{gray!30}Name & $M_{\rm cl}$ [$M_\odot$] & $b_{\rm turb}$ & $\tau_{\rm start}$ [${\rm Myr}$] & MFF? & Phot? & SN fbk? & $M_*$ [$10^9 M_\odot$] & ${\rm SFR}_{50}$ [${\rm M_\odot/{\rm yr}}$] & $\epsilon_{\rm int}$ & ${\rm med}(\epsilon_{\rm ff})$ & $\eta$\\
		\texttt{yesFbk}$^*$ & $4400$ & $1.0$ & $3.0$ & yes & yes & yes & $3.52$ & $46.7$ & 19.8\% & $13.8\%$ & $1.39$ \\
		\texttt{SNeOnly} & $880$ & $1.0$ & $3.0$ & yes & no & yes & $6.03$ & $51.5$ & 33.9\% & $20.2\%$ & $0.50$ \\
		\texttt{photOnly} & $4400$ & $1.0$ & $3.0$ & yes & yes & no & $9.16$ & $82.5$ & 51.6\% & $17.7\%$ & $0.03$ \\
		\texttt{noFbk} & $4400$ & $1.0$ & $3.0$ & yes & no & no & $9.20$ & $83.2$ & 51.8\% & $18.1\%$ & $0.03$ \\
	\rowcolor{gray!50}\multicolumn{12}{c}{Early feedback series}\\
		\rowcolor{gray!30}Name & $M_{\rm cl}$ [$M_\odot$] & $b_{\rm turb}$ & $\tau_{\rm start}$ [${\rm Myr}$] & MFF? & Phot? & SN fbk? & $M_*$ [$10^9 M_\odot$] & ${\rm SFR}_{50}$ [${\rm M_\odot/{\rm yr}}$] &$\epsilon_{\rm int}$ & ${\rm med}(\epsilon_{\rm ff})$ & $\eta$\\
		\texttt{highPhot} & $880$ & $1.0$ & $3.0$ & yes & yes & yes & $1.81$ & $17.3$ & $10.2\%$ & $7.2\%$ & $10.1$ \\
		\texttt{medPhot}$^*$ & $4400$ & $1.0$ & $3.0$ & yes & yes & yes & $3.52$ & $46.7$ & $19.8\%$ & $13.8\%$ & $1.39$ \\
		\texttt{lowPhot} & $22000$ & $1.0$ & $3.0$ & yes & yes & yes & $5.06$ & $56.0$ & $28.5\%$ & $20.7\%$ & $0.56$ \\
		\texttt{highPhotC} & $880$ & $1.0$  & $3.0$& no & yes & yes & $2.05$ & $23.0$ & $11.6\%$ & $13.8\%$ (const) & $7.42$ \\
		\texttt{medPhotC} & $4400$ & $1.0$ & $3.0$ & no & yes & yes & $3.12$ & $41.8$ & $17.6\%$ & $13.8\%$ (const) & $1.35$ \\
		\texttt{lowPhotC} & $22000$ & $1.0$ & $3.0$ & no & yes & yes & $6.30$ & $66.5$ & $35.5\%$ & $13.8\%$ (const) & $0.57$ \\
	\rowcolor{gray!50}\multicolumn{12}{c}{Supernovae feedback series}\\
		\rowcolor{gray!30}Name & $M_{\rm cl}$ [$M_\odot$] & $b_{\rm turb}$ & $\tau_{\rm start}$ [${\rm Myr}$] & MFF? & Phot? & SN fbk? & $M_*$ [$10^9 M_\odot$] & ${\rm SFR}_{50}$ [${\rm M_\odot/{\rm yr}}$] & $\epsilon_{\rm int}$ & ${\rm med}(\epsilon_{\rm ff})$ & $\eta$\\
		\texttt{instSNe} & $4400$ & $1.0$ & $0.0$ &  yes & yes & yes & $3.58$ & $43.4$ & $20.2\%$ & $14.4\%$ & $1.14$ \\
		\texttt{fastSNe} & $4400$ & $1.0$ & $0.5$ &  yes & yes & yes & $3.58$ & $40.9$ & $20.1\%$ & $14.1\%$ & $1.59$ \\
		\texttt{medSNe}$^*$ & $4400$ & $1.0$ & $3.0$ & yes & yes & yes & $3.52$ & $46.7$ & $19.8\%$ & $13.8\%$ & $1.39$ \\
	\rowcolor{gray!50}\multicolumn{12}{c}{Turbulence forcing series}\\
		\rowcolor{gray!30}Name & $M_{\rm cl}$ [$M_\odot$] & $b_{\rm turb}$ & $\tau_{\rm start}$ [${\rm Myr}$] & MFF? & Phot? & SN fbk? & $M_*$ [$10^9 M_\odot$] & ${\rm SFR}_{50}$ [${\rm M_\odot/{\rm yr}}$] & $\epsilon_{\rm int}$ & ${\rm med}(\epsilon_{\rm ff})$ & $\eta$\\
		\texttt{solTurb} & $4400$ & $0.3$ & $3.0$ & yes & yes & yes & $2.69$ & $32.9$ & $15.1\%$ & $6.0\%$ & $2.17$ \\
		\texttt{varTurb} & $4400$ & from $\vb{v}$ & $3.0$ & yes & yes & yes & $3.78$ & $56.4$ & $21.3\%$ & $12.6\%$ & $2.74$ \\
		\texttt{compTurb}$^*$ & $4400$ & $1.0$ & $3.0$ & yes & yes & yes & $3.52$ & $46.7$ & $19.8\%$ & $13.8\%$ & $1.39$ \\
	\hline  
	\end{tabular}
    \caption{The properties of each simulation, including the star cluster particle mass $M_{\rm cl}$; the turbulence forcing parameter $b_{\rm turb}$; and whether the simulation uses the MFF model, includes photoionization feedback, and includes SN feedback. We also report summary statistics for each simulation, including the total stellar mass $M_*$ in the galaxy at $z=9$, the average SFR over the last $50~{\rm Myr}$ at $z=9$, the integrated SFE, the median local SFE, and the outflow efficiency $\eta$ at $z=9$. The calculation of these summary statistics is described in Appendix~\ref{sec:calc_sumstat}. The simulations are separated into 4 series: a feedback series, an early feedback series, a supernova feedback series, and a turbulence forcing series. The same fiducial simulation appears in each series by a different name and marked by an asterisk.}
    \label{tab:models}
\end{table*}

We run 13 simulations from $z=100$ to $z=9$. Parameters and summary statistics for each simulation are given in Table~\ref{tab:models}. The calculation of summary statistics is described in Appendix~\ref{sec:calc_sumstat} and their values are analyzed in Section~\ref{sec:SFH}. We separate our simulations into 4 series: a feedback series, an early feedback series, a SN feedback series, and a turbulence forcing series. The same fiducial simulation appears in each series by a different name and marked by an asterisk.

We use the feedback series to investigate the effects of different feedback mechanisms. \texttt{yesFbk} includes both photoionzation and SN feedback. In \texttt{SNeOnly}, we turn off photoionization feedback. In \texttt{photOnly}, we turn off SN feedback. In \texttt{noFbk}, we turn off both forms of feedback.

We use the early feedback series to investigate the effect of photoionization feedback in detail. In \texttt{highPhot}, \texttt{medPhot}, and \texttt{lowPhot}, we progressively decrease the star particle mass by factors of 5. A smaller star particle mass corresponds to more efficient photoionization feedback (Sec.~\ref{sec:earlyfbk}). The fiducial star cluster particle mass in \texttt{medPhot} is given by the minimum baryon mass resolution $M_{\rm cl,fid} = M_{\rm b,min} = 4400 M_\odot$ close to the physically-motivated value given by Equation~\ref{eq:Mcl}.

\texttt{highPhotC}, \texttt{medPhotC}, and \texttt{lowPhotC} are identical to \texttt{highPhot}, \texttt{medPhot}, and \texttt{lowPhot}, except they do not use the MFF star formation recipe. Instead, the local SFE is set to a constant value above a density threshold, following the standard star formation recipe in galaxy simulations \citep[e.g.][]{Hopkins+2014}. The density threshold and constant local SFE are calibrated to the fiducial simulation. Specifically, the density threshold is set to the 25th-percentile star-forming density $4.24\times 10^{-21}~{\rm g/cm^3}$ and the constant local SFE is set to the median local SFE $13.8\%$.

We use the SN feedback series to investigate the effect of the SN delay time. In \texttt{medSNe}, \texttt{fastSNe}, and \texttt{instSNe}, we progressively decrease $\tau_{\rm start}$, the delay time before the first SN explosion. These simulations are motivated by \citet{Dekel+2023}'s argument that at high densities, the delay time creates a window of opportunity for rapid star formation.

We use the turbulence forcing series to investigate the effect of the turbulence forcing parameter. In \texttt{solTurb}, we set $b_{\rm turb} = 0.3$, corresponding to turbulence forced by purely solenoidal modes. In \texttt{compTurb}, we set $b_{\rm turb} = 1.0$, corresponding to turbulence forced by purely compressive modes. In \texttt{varTurb}, we determine the turbulence forcing parameter from the local velocity field using Equation~\ref{eq:bturb}.

\section{Results}
\label{sec:results}
We now analyze our simulated galaxies at $z=9$ and examine the relationships between the gas conditions of the ISM, stellar feedback processes, and star formation. First, we explain how we locate the galaxy inside the simulation box (Sec.~\ref{sec:locate}). Then, we use a snapshot of our fiducial simulation at $z=9$ to analyze the structure of our simulated MDGs (Sec.~\ref{sec:structure}). Next, we discuss the lifecycle of gas in our simulated galaxies (Sec.~\ref{sec:lifecycle}). Then, we compare the local gas properties between simulations (Sec.~\ref{sec:local_gas_prop}). Next, we discuss the star formation histories of our simulated galaxies (Sec.~\ref{sec:SFH}) and the variability in their SFRs (Sec.~\ref{sec:variability}). Finally, we discuss the effect of the turbulence forcing parameter (Sec.~\ref{sec:var_turb}).

\subsection{Zooming in}
\label{sec:locate}

To locate the galaxy inside the simulation box, we start by determining the location of every peak in the dark matter density field using \texttt{PHEW}, the built-in clump finder in the \texttt{RAMSES} code introduced by \cite{Bleuler&Teyssier2014}. Around each peak, we identify the volume bounded by the enclosing saddle surface as a clump. We iteratively merged clumps within each density isosurface $\rho = 80 \rho_{\rm crit}$ and identify the resulting objects as haloes. In a $10~{\rm kpc}$ sphere centered on the dark matter barycenter of the most massive halo, we define the center of the galaxy $\vb{r}_{\rm c}$ as the gas and star barycenter.

\begin{figure}
    \centering
    \includegraphics[width=0.9\linewidth]{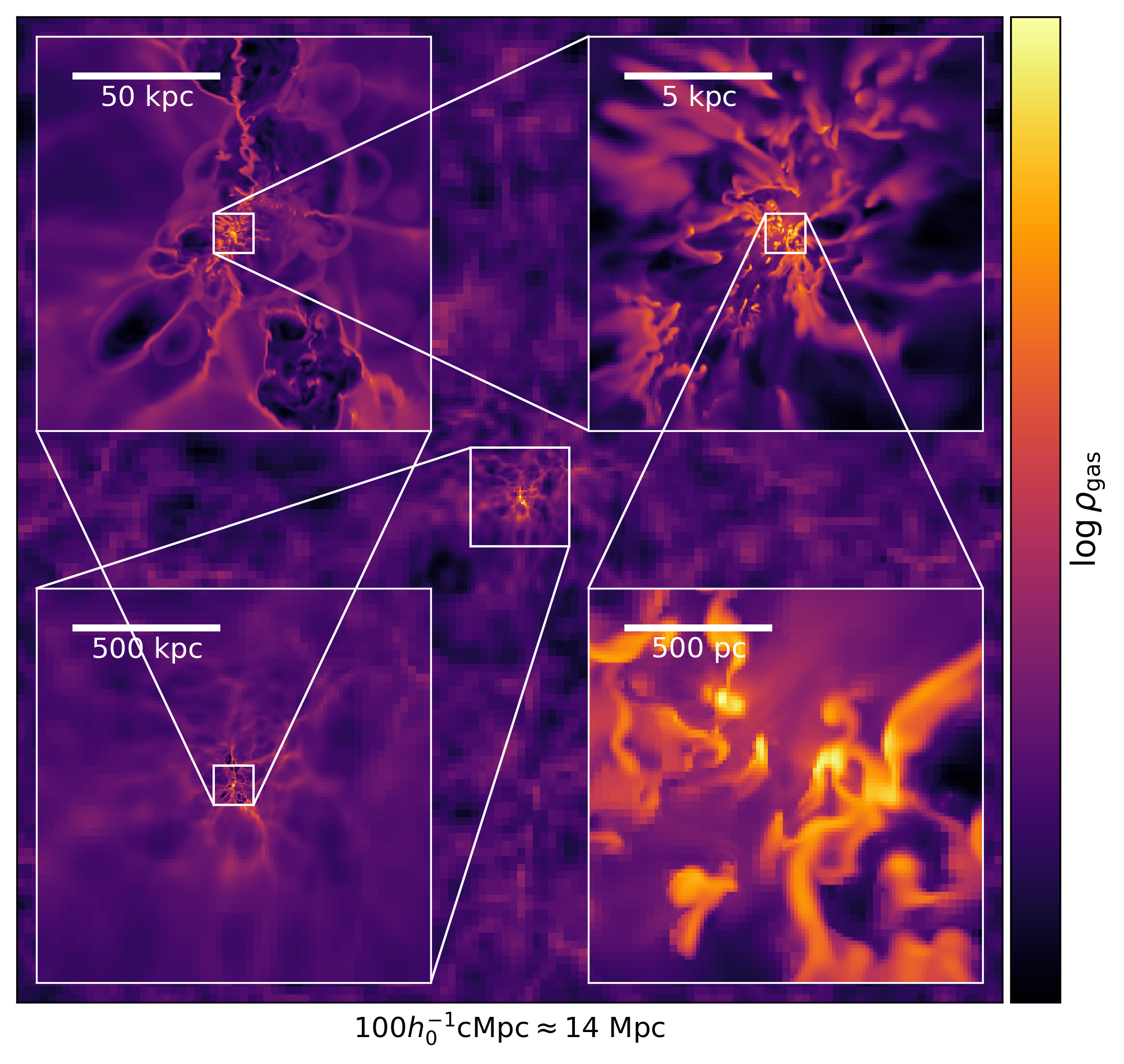}
    \caption{Logarithmic gas density in an $xy$-slice through the most massive galaxy in the fiducial simulation at $z=9$. The main plot shows the entire simulation box with side length $100~h_0^{-1}{\rm cMpc} \approx 14~{\rm Mpc}$. The inset plots show smaller scales, from $\sim 1 {\rm Mpc}$ (bottom left) to $\sim 100~{\rm kpc}$ (top right) to $\sim 10~{\rm kpc}$ (top left) to $1~{\rm kpc}$ (bottom right). The range of the density colorbar is adjusted in each inset plot to maximize contrast. At the largest scales, we identify the high resolution zoom region embedded in a cosmological box. At a scale $\sim 100~{\rm kpc}$, we identify streams accreting onto the galaxy. Zooming in further, we see the central galaxy and the individual star-forming clumps within. Movies of the simulations are available on \href{https://www.youtube.com/playlist?list=PL7YbfRC6zxzAYgFEr5oefYb5dcv0dl7Ba}{YouTube}.}
    \label{fig:zoom}
\end{figure}

In Figure~\ref{fig:zoom}, we zoom in on the location of the most massive galaxy in the fiducial simulation. At the largest scales, we identify the high resolution zoom region embedded in a cosmological box. At a scale $\sim 100~{\rm kpc}$, we identify streams accreting onto the galaxy. Zooming in further, we see the central galaxy and the individual star-forming clumps within.

Applying this algorithm to each snapshot, we can track the most massive galaxy as it moves through the simulation box. There is no guarantee that the most massive halo at $z > 9$ is the progenitor of the most massive halo at $z=9$, so we visually inspect the density field and switch haloes when necessary to enforce continuity. We fit a cubic polynomial to each spatial coordinate of the galaxy center to create a frame which smoothly follows the galaxy. We use this frame to generate movies of our simulations available on \href{https://www.youtube.com/playlist?list=PL7YbfRC6zxzAYgFEr5oefYb5dcv0dl7Ba}{YouTube}\footnote{https://tinyurl.com/53xxm3r2}.

\subsection{Galaxy structure}
\label{sec:structure}

\begin{figure*}
    \centering
    \includegraphics[width=0.667\linewidth]{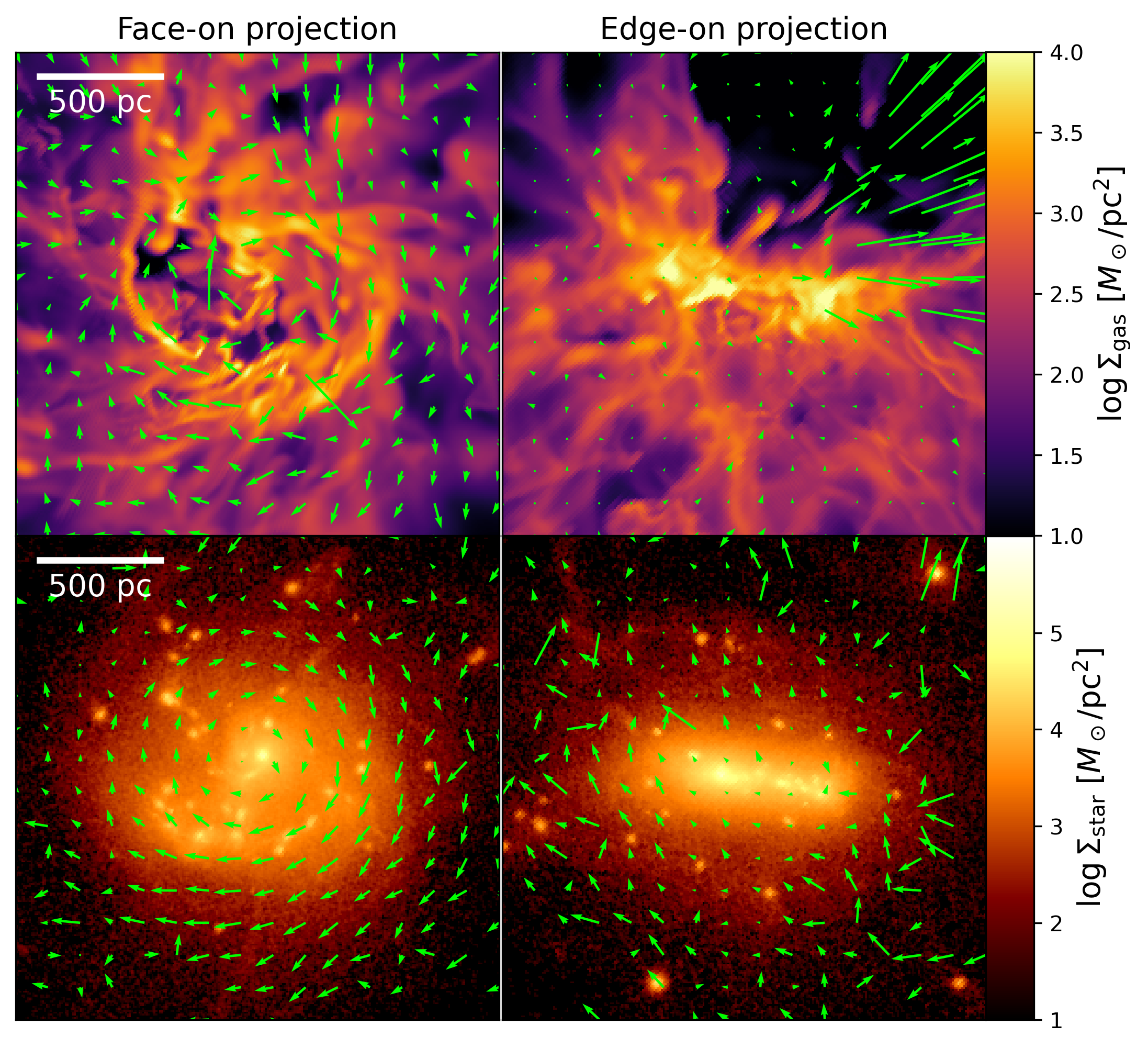}
    \caption{Logarithmic gas (top row) and star (bottom row) surface densities in a box of sidelength $2~{\rm kpc}$ around the galaxy center in the fiducial simulation at $z=9$ projected face-on (left column) and edge-on (right column). In each panel, we overlay the in-plane gas (resp. star) velocity field in green, averaged over the projection direction and weighted by gas (resp. star) mass. The galaxy mass is concentrated around a single plane with a rotational velocity field, suggesting a disc structure. The gas and star mass is organized into dense clumps of size $\sim 100~{\rm pc}$, although the clump size may be limited by our effective resolution $\simeq 10~{\rm pc}$.}
    \label{fig:proj}
\end{figure*}

In the second and third columns of Figure~\ref{fig:proj}, we show face-on and edge-on projections of the logarithmic gas and star surface densities in a box of side length $2~{\rm kpc}$. The procedure for computing the direction of net angular momentum is described in Appendix~\ref{sec:project}. We overlay the in-plane velocity fields as arrows. 

The galaxy mass is concentrated around a single plane with a rotational velocity field, suggesting a disc structure. Our estimate of the galaxy radius $720~{\rm pc}$ in Appendix~\ref{sec:surf_dens} agrees nicely with Figure~\ref{fig:proj}. The morphology of the galaxy can be quantified by the scale height $H/R$ and the ratio of the rotational velocity to the radial velocity dispersion $v_{\rm rot} / \sigma_r$. The calculation of these metrics is described in Appendix~\ref{sec:project}.

In a rotationally-supported thin disc, we expect $H/R \ll 1$ and $v_{\rm rot} / \sigma_r \gg 1$. The relative contribution of the centrifugal force to the support is \citep{Ceverino+2012}
\begin{equation}
    \mathcal{R} = \left[1 + 2 (\sigma_r / v_{\rm rot})^2 \right]^{-1}
\end{equation}
For gas with temperature $\le 10^4~{\rm K}$, we find a disc height $\approx 326~{\rm pc}$ and disc radius $\approx 696~{\rm pc}$, giving a scale height $(H/R)_{\rm gas} \approx 0.468$, although this may be artificially inflated by accreting cold gas in the polar regions. In addition, we find $v_{\rm rot} / \sigma_r\approx 2.16$, corresponding to $\mathcal{R} \simeq 0.70$. This implies that the gas disc is mostly supported by rotation (70\%) with some turbulent support (30\%).

Likewise, for stars, we find a disc height $\approx 115~{\rm pc}$ and a disc radius $\approx 431~{\rm pc}$, giving a scale height $(H/R)_{\rm star} \approx 0.266$. In addition, we find $v_{\rm rot} / \sigma_r \approx 0.919$, corresponding to $\mathcal{R} \simeq 0.30$. This implies that the stellar disc is mostly supported by turbulence (70\%) with some rotational support (30\%). The significant contributions from both rotation and turbulence explain the thick disk structure. At lower redshift, MDGs may start to resemble massive star-forming galaxies at cosmic noon, which show disc structures in observation \citep{ForsterShreiber+2006, Genzel+2006, Genzel+2008} and simulation \citep{Danovich+2015}. 

The gas and star mass is organized into dense clumps of size $\sim 100~{\rm pc}$, although the clump size may be limited by our effective resolution $\simeq 10~{\rm pc}$. Nonetheless, high-redshift disc galaxies show similar structures in observation \citep{Elmegreen+2007, Guo+2012, Guo+2018, Soto+2017} and simulation \citep{Noguchi1998, Inoue+2016, Mandelker+2017, Mandelker+2024}, strengthening the analogy. The recent simulation work of \citep{Nakazato+2024} find that similar clump structures at high redshifts $z \gtrsim 5$ might be detectable by \textit{JWST} due to their brightness in rest-frame optical emission lines.

The average face-on baryonic surface density within $1~{\rm kpc}$ of the center is $1500~M_\odot/{\rm pc^2}$, with a contribution $470~M_\odot/{\rm pc^2}$ from the gas. This is less than our estimate $\sim 10^4~M_\odot/{\rm pc^2}$ of the baryonic surface density in Appendix~\ref{sec:surf_dens}. The discrepancy is primarily due to baryons inside the halo which have not yet settled into the disk, making $f_{\rm b} M_{\rm vir}$ an overestimate of the baryon mass in the disk. The baryon mass inside the galaxy radius $720~{\rm pc}$ is only a third of the baryon mass inside the virial radius $14~{\rm kpc}$.

The surface density is higher at smaller radii, but the disc is increasingly dominated by stars. The face-on baryonic surface density within $180~{\rm pc}$ of the center is $8000~M_\odot/{\rm pc^2}$ with a contribution $880~M_\odot/{\rm pc^2}$ from the gas. Gas density is enhanced significantly within the clumps, with face-on gas surface density $\sim 10^4~M_\odot/{\rm pc^2}$ and Hydrogen number density $\sim 3 \times 10^{3}~{\rm cm^{-3}}$. Most of the star formation occurs within these clumps.

\subsection{Gas lifecycle}
\label{sec:lifecycle}

\begin{figure*}
    \centering
    \includegraphics[width=\linewidth]{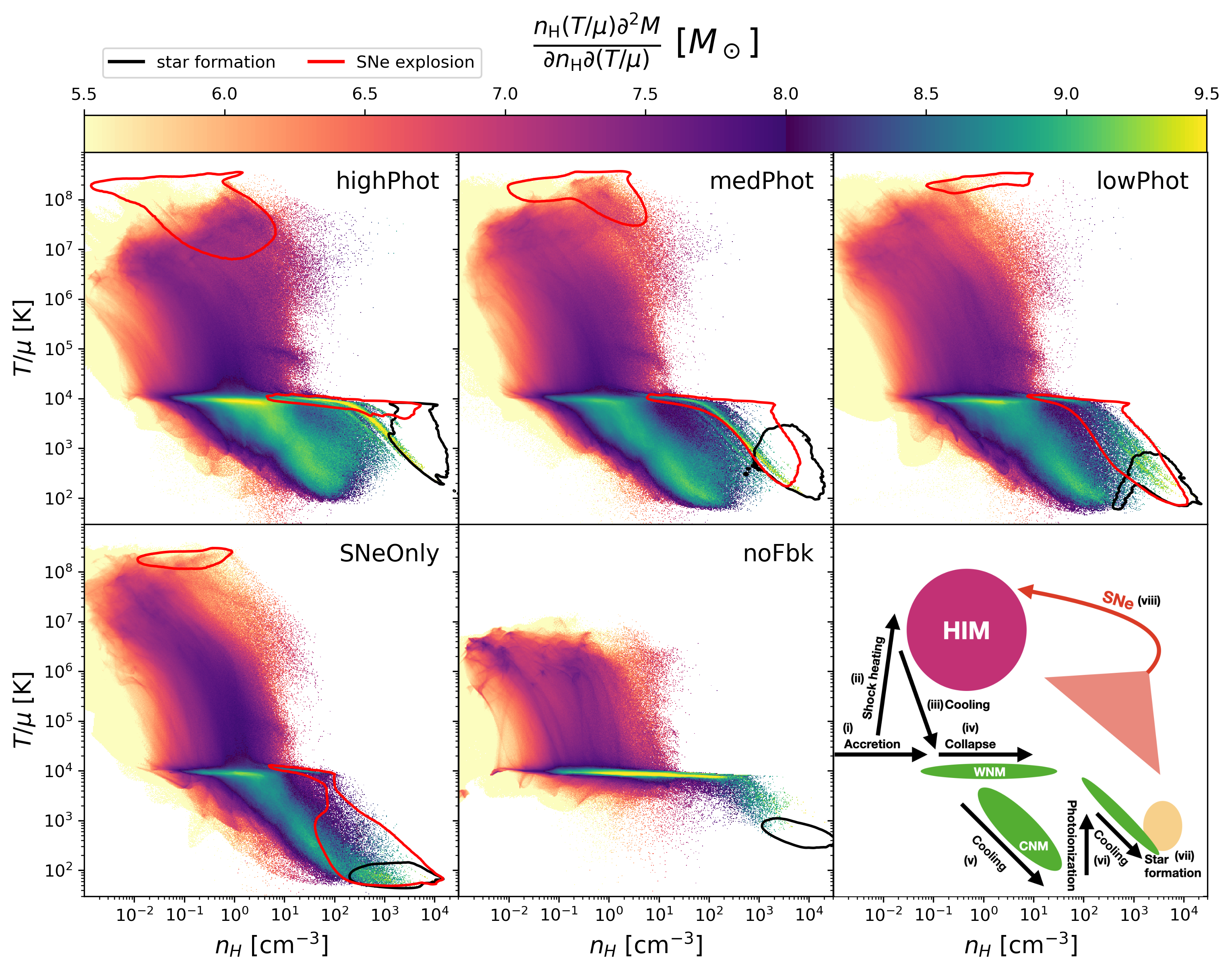}
    \caption{Two-dimensional histograms of the density and temperature in a box of sidelength $5~{\rm kpc}$ around the most massive galaxy at $z=9$ weighted by gas mass for \texttt{highPhot}, \texttt{medPhot}, \texttt{lowPhot}, \texttt{SNeOnly}, and \texttt{noFbk}. The temperature is normalized by $\mu$, the mean molecular weight. The color bar represents the phase space density i.e. the mass per bin in logarithmic density and temperature. Hotter and more diffuse gas has relatively less mass per bin and is represented by purple colors. Colder and more dense gas has relatively more mass per bin and is represented by green colors. We overplot contours outlining the smallest phase space areas containing $75\%$ of star formation events weighted by stellar mass (black) and $75\%$ of SN events (red). In the bottom right panel, we show a schematic diagram of the gas evolution, explained in Section~\ref{sec:local_gas_prop}.}
    \label{fig:phaseall}
\end{figure*}

The lifecycle of gas in our simulated galaxies can be understood in temperature-density phase space. In Figure~\ref{fig:phaseall}, we show the distribution of gas mass in temperature-density phase space for \texttt{noFbk}, \texttt{SNeOnly}, \texttt{lowPhot}, \texttt{medPhot}, and \texttt{highPhot} at $z=9$. The temperature is divided by $\mu$, the mean molecular weight. 

We identify multiple ISM gas phases, including the hot ionized medium (HIM; $n_{\rm H} \sim 0.005~{\rm cm^{-3}}$ and $T \gtrsim 10^{5.5}~{\rm K}$), the warm neutral medium (WNM; $n_{\rm H} \sim 0.6~{\rm cm^{-3}}$ and $T \sim 5000~{\rm K}$), the cold neutral medium (CNM; $n_{\rm H} \sim 30~{\rm cm^{-3}}$ and $T \sim 100~{\rm K}$), and photoionized gas ($n_{\rm H} \gtrsim 0.3~{\rm cm^{-3}}$ and $T \sim 10^4~{\rm K}$). We remind the reader that the ionization state of the gas only enters into our simulations through the prescribed cooling function.

The black contours in Figure~\ref{fig:phaseall} outline the smallest phase space area that contains 75\% of star formation events over the course of the simulation. Stars typically form at high densities $\sim 3\times 10^3~{\rm cm^{-3}}$ and low temperatures $\sim 300~{\rm K}$. The star-forming regions of phase space contain very little gas because any gas that enters this region is quickly removed by star formation or feedback. The red contours outline the smallest phase space area that contains 75\% of SN events occur. The SN events are distributed between the CNM, photoionized gas, and the HIM.

Whether gas evolves horizontally or vertically in phase space is a question of timescales. When the cooling timescale is longer than the freefall time, gas collapses, moving horizontally to higher densities. When the freefall time is longer than the cooling timescale, gas cools, moving vertically to lower temperatures. At a fixed temperature and metallicity, we have $\tau_{\rm cool} \propto \rho^{-1}$ and $\tau_{\rm ff} \propto \rho^{-1/2}$, so collapsing gas will eventually reach a density where cooling dominates. 

The gas mass in the galaxy increases over time due to accretion from the IGM. If the cooling time near the virial radius is short, this precludes the formation of a stable accretion shock, and the cold gas accretes directly onto the central galaxy (cold mode accretion). Otherwise, the shock-heated gas must cool in the circum-galactic medium (CGM) before reaching the galaxy (hot mode accretion) \citep{Birnboim&Dekel2003, Keres+2005}. Figure~\ref{fig:phaseall} probes a region deep inside the virial radius, so it does not explicitly separate these two scenarios.

Hot mode accretion typically dominates at halo masses $\gtrsim 10^{11}$ to $10^{12}~M_\odot$ for (quasi-)spherical accretion \citep{Birnboim&Dekel2003, Keres+2005}. However, at high redshift, accretion in massive haloes is increasingly dominated by dense filaments, shortening the cooling time and enhancing cold mode accretion \citep{Dekel&Birnboim2006, Keres+2009}. Therefore, we expect MDGs to accrete primarily in the cold mode despite the large halo mass and low metallicity of the accreting gas. The CGM is poorly resolved in our simulations, so we leave a detailed study of gas accretion onto MDGs to future work.

The lower right panel of Figure~\ref{fig:phaseall} is schematic diagram illustrating the processes that move gas through the phase space: (i) cold gas from cosmic filaments accrete onto the galaxy; (ii) some of the gas is shock-heated by turbulence, joining the HIM; (iii) HIM gas cools to the Hydrogen ionization temperature $\sim 10^4~{\rm K}$, where it becomes neutral and joins the WNM; (iv) the neutral gas cools less efficiently than the ionized gas, so it collapses while maintaining approximately the same temperature; (v) once the density becomes sufficiently high, cooling becomes efficient and the gas joins the CNM; (vi) eventually, stars begin to form and photoionization feedback generates photoionized gas at $\sim 10^4~{\rm K}$; (vii) the gas continues to collapse until it once again becomes cold and dense enough to form stars; (viii) SN explosions can recycle gas in any of the previous stages back into the WNM or HIM.

We can build up this picture by starting with our most simple \texttt{noFbk} simulation. In this simulation, gas accretes, cools, and collapses following steps (i)-(iv). However, without SNe to enrich the gas with metals, the gas is pristine and must reach high densities $\sim 300~{\rm cm^{-3}}$ before it can cool below $\sim 10^4~{\rm K}$. At this point, the gas is so dense that it only needs to cool to $\sim 1000~{\rm K}$ before stars form efficiently. This also describes the formation of the first generation of stars in the other simulations that include SNe.

The most striking difference between \texttt{noFbk} and \texttt{SNeOnly} is the presence of gas at temperatures $\gtrsim 10^7~{\rm K}$. Only SNe can produce gas at these extreme temperatures \citep{McCray&Snow1979, Spitzer1990}. The virial temperature reached by the accretion shock in hot mode accretion (Eq.~\ref{eq:temp_vir}) and the temperatures reached by turbulent shocks \citep{Wada&Norman2001, Ceverino&Klypin2009} are both $\sim 10^6~{\rm K}$. If a SN explodes in diffuse gas, the ejecta reaches a far greater temperature given by
\begin{equation}
    \frac{T_{\rm SN}}{\mu} = \frac{m_{\rm p}}{k_{\rm B}}\frac{E_{\rm SN}}{m_{\rm ej}} \simeq 6.1\times 10^8 E_{51} m_{{\rm ej}, 10}~{\rm K}
\end{equation}
where $E_{51} = E_{\rm SN} / 10^{51}~{\rm erg}$ and $m_{\rm ej, 10} = m_{\rm ej} / 10~M_\odot$ are the SN energy and ejecta mass respectively. SN can explode in diffuse gas when a young star particle diffuses into a diffuse region over its lifetime, or when the local gas conditions around a star cluster particle are modified by early feedback or other SNe.

Metal-enriched gas starts to cool below $\sim 10^4~{\rm K}$ at densities $\gtrsim 10~{\rm cm^{-3}}$. The cooling and collapsing gas forms a diagonal band in phase space. Eventually, gas reaches the CMB temperature $\simeq 27.25~{\rm K}$ and becomes dense enough to form stars. The metal enriched gas ends up forming stars at lower densities than pristine gas.

In simulations with photoionization feedback, star formation immediately generates photoionized gas. After a few timesteps, the temperature of the gas is set by a balance between the thermal energy injected by photoionization and the thermal energy removed by cooling. As gas collapses, the cooling rate increases and gas is able to reach lower temperatures, producing a second diagonal band in phase space. In \texttt{highPhot} where photoionization is more efficient, stars form at higher temperatures than \texttt{lowPhot}, where photoionization is less efficient. 

We do not resolve the multi-phase nature of the ISM on small scales, so the high temperatures of star formation in the simulations with photoionization does not mean that stars form out of high-temperature gas. Instead, the temperature of star formation reflects the filling factor of photoionized gas near star formation sites on unresolved scales.

\subsection{Local gas properties}
\label{sec:local_gas_prop}

\begin{figure*}
    \centering
    \includegraphics[width=\linewidth]{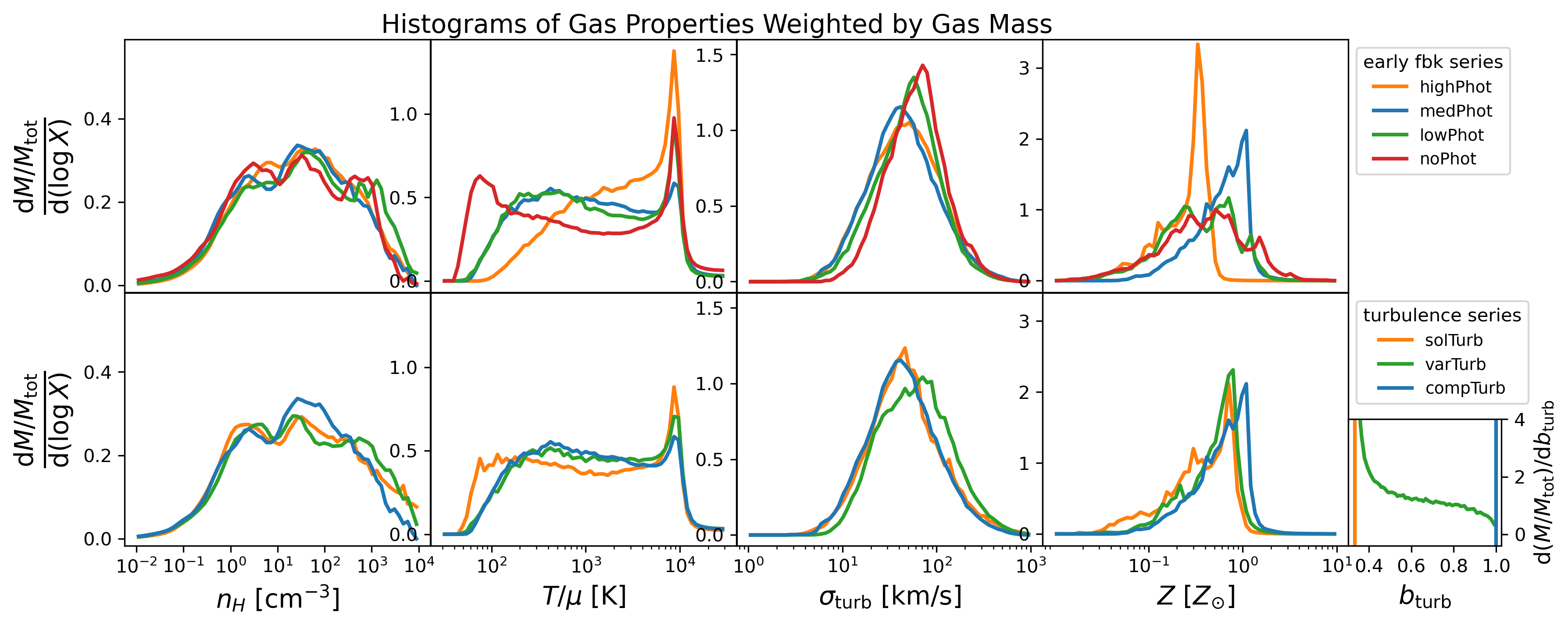}
    \caption{Histograms of the hydrogen number density (first column), modified temperature (second column), turbulent velocity dispersion (third column), and metallicity (fourth column) in a box of sidelength $5~{\rm kpc}$ around the most massive galaxy at $z=9$ in the early feedback series (top row) and turbulence forcing series (bottom row), weighted by gas mass. Each simulation is shown in a different color, with the fiducial simulation in blue. An additional histogram of the turbulence forcing parameter is included for the variable turbulence forcing simulation.}
    \label{fig:mgashist}
\end{figure*}

\begin{figure*}
    \centering
    \includegraphics[width=\linewidth]{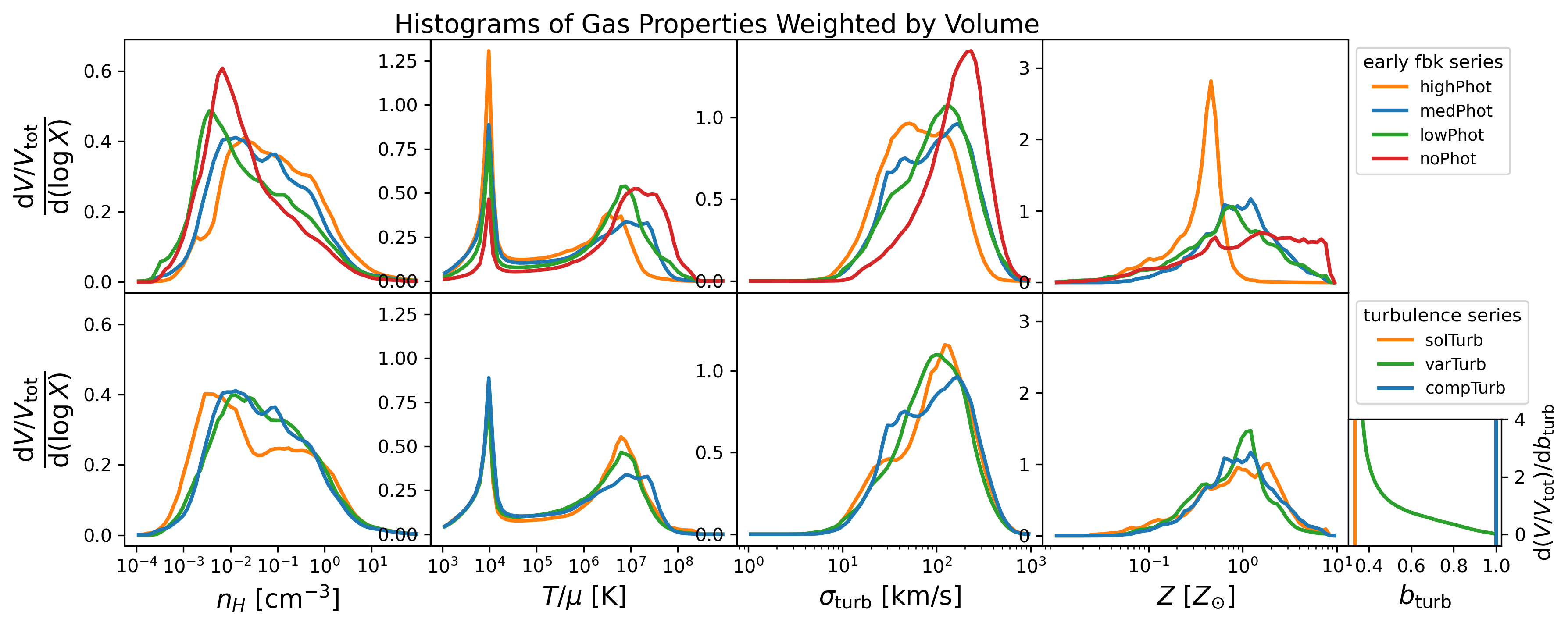}
    \caption{Histograms of the hydrogen number density (first column), modified temperature (second column), turbulent velocity dispersion (third column), and metallicity (fourth column) in a box of sidelength $5~{\rm kpc}$ around the most massive galaxy at $z=9$ in the early feedback series (top row) and turbulence forcing series (bottom row), weighted by volume. Each simulation is shown in a different color, with the fiducial simulation in blue. An additional histogram of the turbulence forcing parameter is included for the variable turbulence forcing simulation.}
    \label{fig:vgashist}
\end{figure*}

In Figures~\ref{fig:mgashist} and \ref{fig:vgashist}, we show the mass-weighted and volume-weighted distributions respectively of gas density, temperature, turbulent velocity dispersion, and metallicity for the early feedback and turbulence forcing series. We plot turbulent velocity dispersion $\sigma_{\rm turb}^2 = (2/3)\varepsilon_{\rm turb}/\rho$ rather than turbulent Mach number to isolate the differences in turbulence between simulations from the temperature-dependence $\mathcal{M}_{\rm turb} \propto T^{-0.5}$ at a constant TKE.

Compared to the lower-redshift Universe, the gas has higher densities, higher turbulent velocity dispersions, and lower metallicities. The mass-weighted density and temperature distributions in the early feedback series are consistent with the two-dimensional histograms in Figure~\ref{fig:phaseall}. 

HII regions created by photoionization feedback produce a sharp peak in the temperature distribution at $\sim 10^4~{\rm K}$. The density distribution does not change significantly across the early feedback series, suggesting that the thermal pressure from photoionization feedback is insufficient to halt the collapse of the star-forming clouds. This is expected, because for high mass star-forming clouds $\gtrsim 10^6~M_\odot$, the sound speed $\sim 10~{\rm km/s}$ in photoionized gas is less than the escape speed of the cloud \citep[e.g.][]{Krumholz&Matzner2009}.

The volume-weighted distributions tell a similar story. In this case, the peak in the temperature distribution at $T\sim 10^4~{\rm K}$ represents the volume of WNM rather photoionized gas, whose volume fraction is generally smaller. As photoionization feedback becomes more efficient, volume moves from HIM gas to less diffuse WNM gas.

Solenoidal forcing decreases the local SFE for the same gas conditions and TKE. Therefore, gas must reach higher densities to form stars efficiently, resulting in more gas mass at high densities in \texttt{solTurb}.

\begin{figure*}
    \centering
    \includegraphics[width=\linewidth]{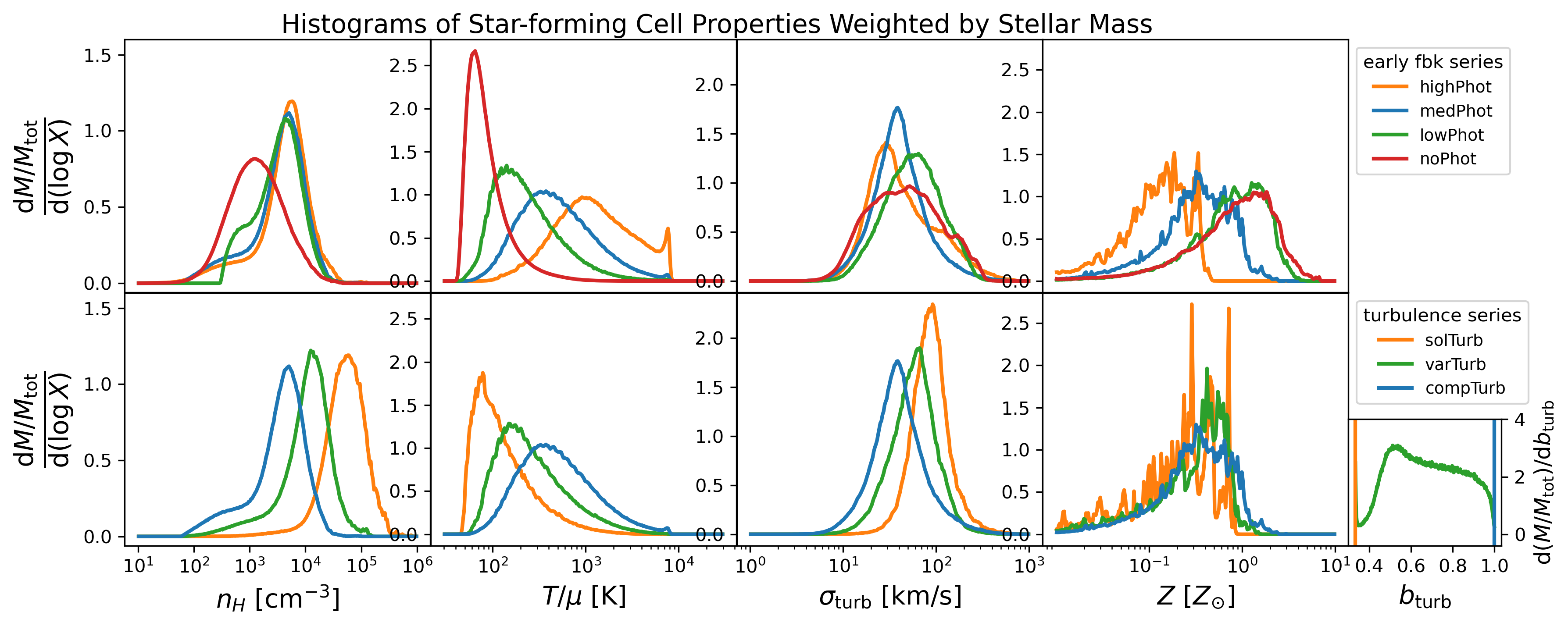}
    \caption{Histograms of the hydrogen number density (first column), modified temperature (second column), turbulent velocity dispersion (third column), and metallicity (fourth column) of star-forming cells in the early feedback series (top row) and turbulence forcing series (bottom row), weighted by stellar mass. Each simulation is shown in a different color, with the fiducial simulation in blue. The inset plot shows a histogram of the turbulence forcing parameter for the turbulence forcing series. An additional histogram of the turbulence forcing parameter is included for the variable turbulence forcing simulation.}
    \label{fig:mstarhist}
\end{figure*}

In Figure~\ref{fig:mstarhist}, we show the distributions of gas density, temperature, turbulent velocity dispersion, and metallicity in star-forming cells for the early feedback and turbulence forcing series. In the early feedback series, photoionization feedback forces stars to form at higher temperatures. Therefore, the gas must reach higher densities and lower turbulent velocity dispersions for star formation to become efficient. This explains the trends of increasing density and decreasing turbulent velocity dispersion with increasing photoionization feedback efficiency.

A similar analysis applies to the turbulence forcing series. In the solenoidal forcing limit, the SFE is lower for the same gas conditions, so gas must reach higher densities, lower temperatures, and lower turbulent velocity dispersions for star formation to become efficient. This explains the trends of increasing star-formation density and decreasing star-formation temperature with increasingly solenoidal turbulence forcing. 

However, the trend in the turbulent velocity dispersion seems to go the wrong way, increasing as the turbulence forcing becomes more solenoidal. The reason is that regions of high density, low temperature, and high turbulence are spatially coincident (App.~\ref{sec:gaspropcomp}). In \texttt{solTurb}, stars are forced to form in high density and low temperature regions, which also forces them to form in high turbulence regions, even though turbulence generally decreases the local SFE. 

\begin{figure*}
    \centering
    \includegraphics[width=\linewidth]{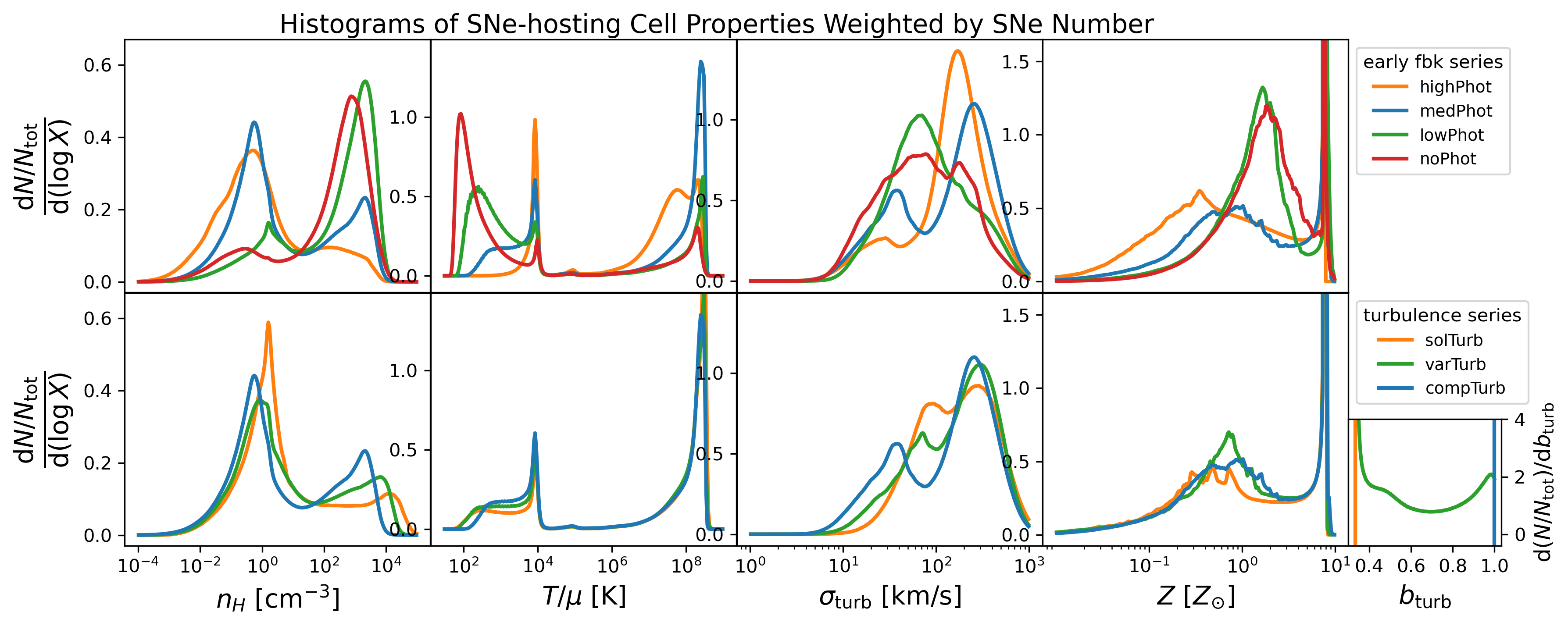}
    \caption{Histograms of the hydrogen number density (first column), modified temperature (second column), turbulent velocity dispersion (third column), and metallicity (fourth column) of cells with SN events in the early feedback series (top row) and turbulence forcing series (bottom row), weighted by SN number. Each simulation is shown in a different color, with the fiducial simulation in blue. An additional histogram of the turbulence forcing parameter is included for the variable turbulence forcing simulation.}
    \label{fig:nsnehist}
\end{figure*}

In Figure \ref{fig:nsnehist}, we show the distributions of gas density, temperature, turbulent velocity dispersion, and metallicity in cells where a SN event occurs for the early feedback and turbulence forcing series. The distribution of gas conditions near SNe is multi-modal, reflecting the multi-phase nature of the ISM. As photoionization feedback becomes more efficient, more SNe occur in photoionized gas and HIM rather than CNM. This effect can also be seen in Figure~\ref{fig:phaseall}, where the portion of the red contour in photoionized gas and HIM expands as photoionization feedback becomes more efficient. SNe occur in more dense environments as turbulence forcing becomes more solenoidal, following the trend in the density of star-forming cells. 

In the SN feedback series, we vary the SN delay time $\tau_{\rm start}$. We find that the summary statistics and gas property distributions (not shown in figures) are almost identical for all simulations in the series. This suggests that the SN delay time is not important for star formation, even in the extreme case of \texttt{instSNe} where SNe can occur immediately after star formation.

Our finding contradicts the prediction of the FFB model \citep{Dekel+2023} that the SN delay time provides a crucial window of opportunity for efficient star formation. However, our ability to test the FFB model is limited by our effective resolution $\simeq 10~{\rm pc}$. We cannot fully resolve the parsec-scale dynamics which control star-forming clouds, especially in dense regions where the SN cooling radius is unresolved (Sec.~\ref{sec:snecav}).

\subsection{Star formation histories}
\label{sec:SFH}

At a high level, we describe the SFH using summary statistics in Table~\ref{tab:models}, including the total stellar mass $M_*$ in the galaxy at $z=9$, the average SFR over the last $50~{\rm Myr}$ at $z=9$ ${\rm SFR}_{50}$, the integrated SFE $\epsilon_{\rm int} = M_* / f_{\rm b} M_{\rm halo}$, the median local SFE ${\rm med}(\epsilon_{\rm ff})$, and the outflow efficiency $\eta = \dot{M}_{\rm out} / {\rm SFR}_{\rm 50}$, where $\dot{M}_{\rm out}$ is the mass outflow rate. The calculation of these summary statistics is described in Appendix~\ref{sec:calc_sumstat}. Nearly all our simulations produce significant stellar populations ($M_* \gtrsim 10^9~M_\odot$) with high SFRs ($\dot{M}_* \gtrsim 100~M_\odot/{\rm yr}$) and high SFEs ($\epsilon_{\rm ff}, \epsilon_{\rm int} \gtrsim 10\%$), supporting the notion that star formation is intrinsically efficient at Cosmic Dawn.

\begin{figure}
    \centering
    \includegraphics[width=\linewidth]{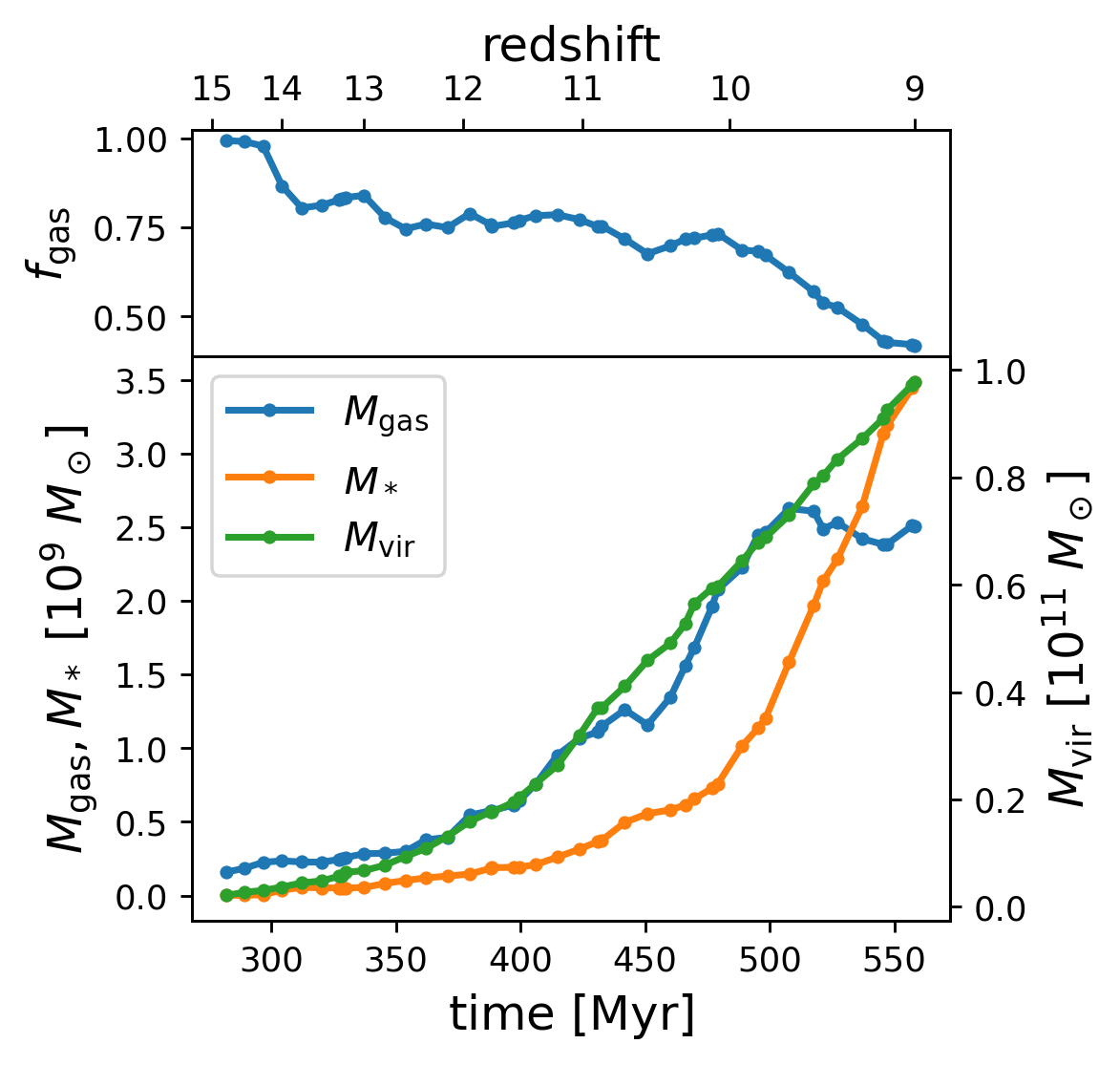}
    \caption{Gas mass (blue), stellar mass (orange), virial mass (green), and gas fraction (top panel) as a function of time for the fiducial model. These quantities are computed at 43 discrete data dumps. By design, the virial mass reaches $10^{11}~M_\odot$ at $z=9$. At $z\simeq 10$, gas accretion is balanced by star formation, causing the gas mass to plataeu and the gas fraction to drop to $40\%$ by $z=9$.}
    \label{fig:gas_frac}
\end{figure}

In Figure~\ref{fig:gas_frac}, we plot the time evolution of the gas mass, stellar mass, and virial mass for the fiducial model. The details of the calculation are described in Appendix~\ref{sec:calc_sumstat}. We also plot the gas fraction $f_{\rm gas} = M_{\rm gas} / (M_{\rm gas} + M_*)$. By design, the virial mass reaches $10^{11}~M_\odot$ at $z=9$. Initially, the galaxy is dominated by the gas supplied by accretion. As the galaxy grows, it begins to form stars efficiently. At $z\simeq 10$, gas accretion is balanced by star formation, causing the gas mass to plataeu and the gas fraction to drop to $40\%$ by $z=9$. This process will likely continue until the gas fraction reaches values $\lesssim 10\%$ similar to massive galaxies in the local Universe \citep{Wiklind+2019}.

Throughout this paper, we use three different measurements of SFE. The local SFE $\epsilon_{\rm ff}$ is the fraction of gas within a single cell that is converted into stars within one local freefall time. The global SFE $\epsilon_{\rm glob} = \dot{M}_* / f_{\rm b} \dot{M}_{\rm acc}$ is the ratio of the SFR in the galaxy to the gas accretion rate. The integrated SFE $\epsilon_{\rm int} = M_* / f_{\rm b} M_{\rm halo}$ is the ratio of the stellar mass to the total gas mass accreted onto the galaxy over its lifetime. Any measure of SFE is defined on a particular spatial and temporal scale. Measurements of the SFE on multiple scales are necessary ingredients of a holistic picture of star formation. Even simulations with a fixed local SFE $\sim 1\%$ can exhibit high integrated SFEs \citep[e.g.][]{Ceverino+2024}.

\begin{figure}
    \centering
    \includegraphics[width=\linewidth]{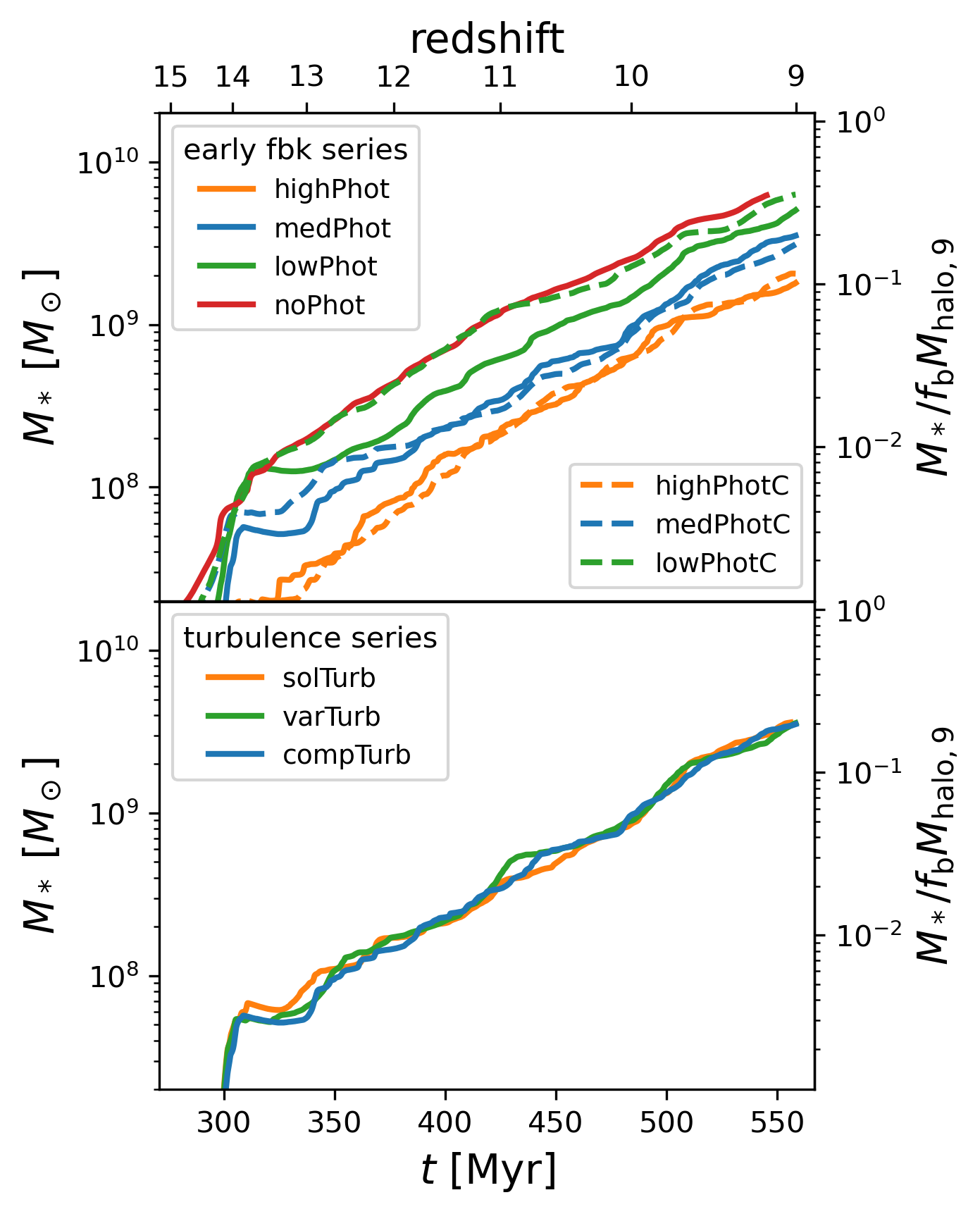}
    \caption{Stellar mass as a function of time for the early feedback series (top panel) and turbulence forcing series (bottom panel). The integrated SFE defined relative to the halo mass at $z = 9$ is shown on a secondary $y$-axis in both panels. Within each series, each simulation is shown in a different color, with the fiducial simulation in blue. Dashed lines in the upper panel indicate constant SFE simulations. The stellar mass increases almost exponentially with time, following the expected accretion rate onto a dark matter halo in the early Universe.}
    \label{fig:starmass}
\end{figure}

\begin{figure}
    \centering
    \includegraphics[width=\linewidth]{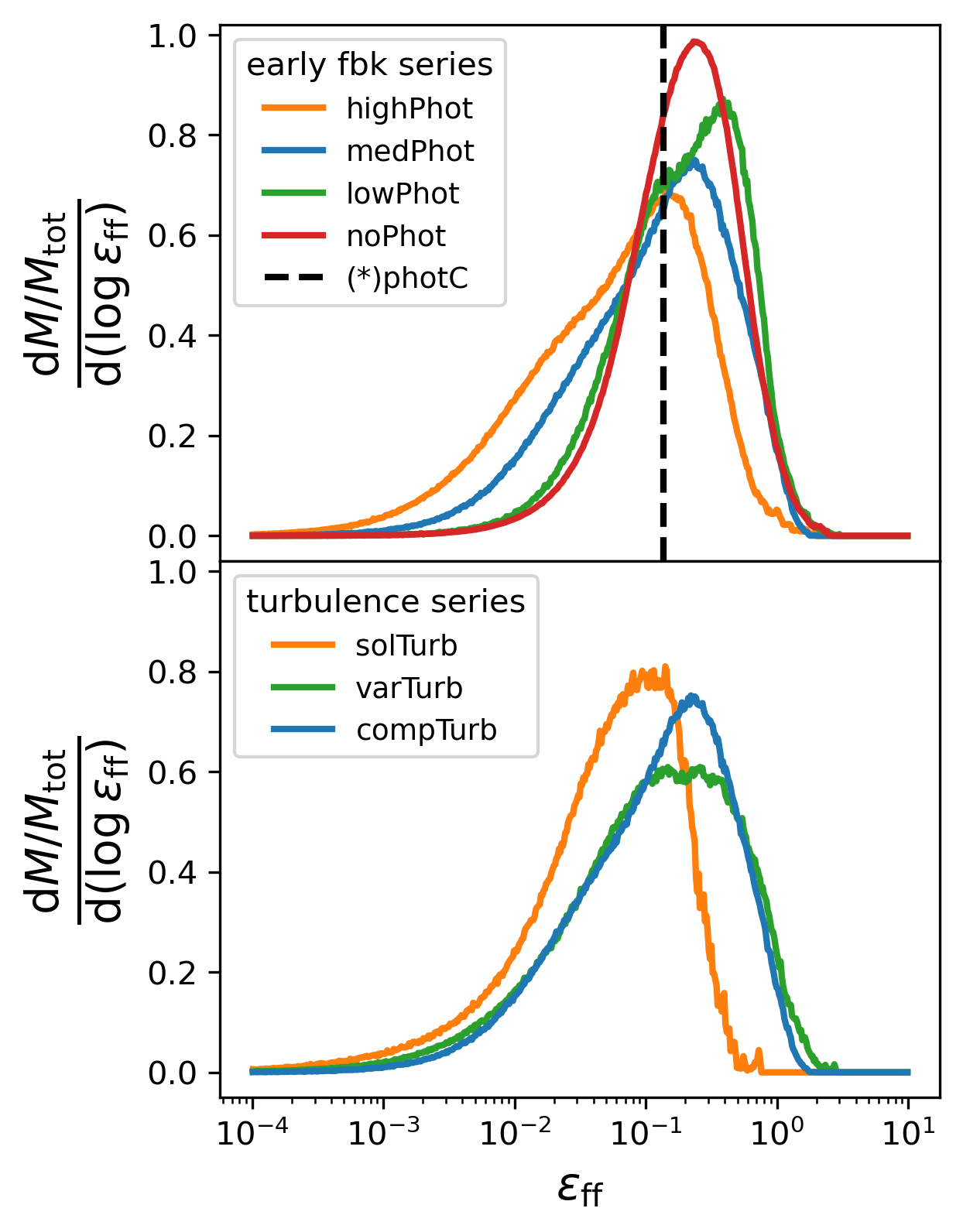}
    \caption{A histogram of local SFE weighted by stellar mass for the early feedback series (top panel) and the turbulence forcing series (bottom panel). Within each series, each simulation is shown in a different color, with the fiducial simulation in blue. The dashed line in the upper panel indicates the local SFE of the constant local SFE simulations. In all simulations, the local SFE ranges from $\sim 1\%$ to $\sim 100\%$, reflecting the large range of gas conditions in which stars can form.}
    \label{fig:sfe_loc}
\end{figure}

Figures~\ref{fig:starmass} and \ref{fig:sfe_loc} visually represent the integrated and local SFEs respectively. In Figure~\ref{fig:starmass}, we show the stellar mass as a function of time, which is proportional to the integrated SFE. In Figure~\ref{fig:sfe_loc}, we show the distribution of local SFEs in star formation events.

The relative factors between the stellar masses of different simulations remain approximately constant throughout time, implying that the differences between simulations are not redshift dependent for $z \ge 9$. This lends credence to our approach of analyzing a single snapshot in time at $z=9$. 

The stellar mass, and therefore also the mean SFR, increases almost exponentially with time, following the expected accretion rate onto a dark matter halo in the early Universe (Sec.~\ref{sec:gas_mass}). In all simulations, the local SFE ranges from $\sim 1\%$ to $\sim 100\%$, reflecting the large range of gas conditions in which stars can form (Fig.~\ref{fig:mstarhist}).

Comparing \texttt{photOnly} to \texttt{noFbk}, we isolate the effect of photoionization feedback. Both simulations have similar integrated and local SFEs, suggesting that by itself photoionization feedback is ineffective at suppressing star formation. Alternatively, comparing \texttt{SNeOnly} to \texttt{noFbk}, we isolate the effect of SN feedback. \texttt{SNeOnly} has a lower integrated SFE but a similar local SFE, suggesting that SN feedback suppresses star formation without changing the local gas conditions at star formation sites.

In \texttt{yesFbk}, both feedback mechanisms operate simultaneously. \texttt{yesFbk} has the lowest integrated and local SFE in the feedback series. The low integrated SFE suggests that photoionization is effective, but only in tandem with SN feedback. The low local SFE suggests that when both feedback mechanisms work together, they alter the local gas conditions, even though this effect is not seen for either mechanism in isolation.

These results are consistent with previous simulations, which show that early feedback pressurizes the gas in star-forming regions, driving expansion which reduces the gas density \citep[e.g.][]{Ceverino+2014, Rosdahl+2015}. When SNe occur in more diffuse gas, more of their energy makes it into the ISM rather than being absorbed locally and radiated away. Once in the ISM, this energy can suppress star formation by disrupting star-forming clouds and launching outflows. The detailed simulations of dense gas clouds in \citet{Walch&Naab2015} show that $\sim 50\%$ more SN energy goes into the ISM when SN go off in previously photo-ionized regions. 

In the early feedback series, the integrated and local SFEs change by a similar factor between simulations. For example, from \texttt{medPhot} to \texttt{highPhot}, both efficiencies decrease by a factor of 2. Therefore, it is tempting to conclude that the integrated SFE is proportional to the local SFE, which is consistent with a picture where star formation is not regulated by feedback. However, our other simulations provide evidence to the contrary. 

First, the constant efficiency simulations all have a fixed local SFE, but they show the same trend in integrated SFE as the MFF simulations. Second, in the turbulence forcing series, the local SFE decreases by a factor of 3 as turbulence forcing becomes more solenoidal. The trend in integrated SFE goes in the same direction, but by a much smaller factor. These results do not rule out a correlation between local and integrated SFE, but they suggest that the local SFE is not the main driver for the trends we see.

The trends in outflow efficiency suggest an alternative explanation. SNe can regulate star formation by launching outflows which remove gas from the galaxy. When there are no SNe (\texttt{noFbk} and \texttt{photOnly}), the galaxy is not able to launch outflows $\eta \approx 0$. When there are SNe, we see a small outflow efficiency $\eta \simeq 0.5$. 

As photoionization feedback becomes more efficient, the outflow efficiency increases, reaching $\eta \simeq 10$ in \texttt{highPhot}. This supports the idea that photoionization feedback enhances the coupling between SN explosions and the ISM. It is also consistent with the local gas properties, which show that photoionization feedback is associated with an increase in the volume fraction of the WNM (Fig.~\ref{fig:vgashist}) and more SN explosions in photoionized gas and HIM rather than CNM (Fig.~\ref{fig:nsnehist}). 

\subsection{Star formation variability}
\label{sec:variability}

In the first row of Figure~\ref{fig:variability}, we plot the SFR as a function of time. The calculation of the SFR is described in Appendix~\ref{sec:calc_sumstat}. The SFR fluctuates on multiple timescales, from small bursts lasting only a few ${\rm Myr}$ to large bursts lasting 10s of ${\rm Myr}$. Because the mean SFR increases rapidly as a function of time, measurements of the SFR on long timescales may under-represent the instantaneous SFR (Sec.~\ref{sec:indicator}). 

After major bursts, the SFR temporarily decreases, possibly because the bursts consume dense gas available to form stars. The movies associated with this paper reveal that the major bursts in the SFR coincide with significant merger events, which was also seen in the cosmological zoom-in simulations of Cosmic Dawn galaxies in the \textsc{FirstLight} project \citep{Ceverino+2018}. However, analyzing the role of mergers in detail is beyond the scope of this paper.

We quantify the variability of the SFR using the power spectral density (PSD) in the time domain. We compute the PSD following the procedure of \citet{Iyer+2020}. We start by computing the SFR using time bins of width $0.1~{\rm Myr}$, which gives a reasonable balance between shot noise and time resolution. The SFR spans a large dynamic range, so we work with the logarithm to better quantify the relative strengths of fluctuations in the SFR.

We can remove the exponential trend in the mean SFR with a linear detrending of the logarithmic SFR. We limit our analysis to times $\ge 350~{\rm Myr}$, when all simulations have a nonzero SFR such that the logarithm is well-defined. The second row of Figure~\ref{fig:variability} shows the residuals of the linear detrending.

Finally, we apply Welch's method \citep{Welch1967} as implemented by \texttt{scipy.signal.welch}. In this method, the data are divided into overlapping segments. We choose segments of 256 bins which 50\% overlap on each side with their neighbors. We compute a modified periodogram for each segment and then average the periodograms. We use the Hann window function in the computation of each periodogram to reduce edge effects.

Our techniques are similar to the recent variability analysis of \citet{Pallottini&Ferrara2023} for the \textsc{Serra} simulations with a few differences. \citet{Pallottini&Ferrara2023} fit the mean SFR trend to a polynomial in log space rather than an exponential, which is necessary because their simulations extend to lower redshifts where the mean SFR deviates from the exponential trend. They also use a \citet{Lomb1976}-\citet{Scargle1998} periodogram analysis rather than a PSD analysis. The former has several benefits including better isolation of individual frequency contributions and straight-forward quantification of peak significance \citep{Dome+2024}. In future variability analyses, we will prefer this technique.

In the third row of Figure~\ref{fig:variability}, we plot the PSD of the SFR in the third row of Figure~\ref{fig:variability}, inverting the $x$-axis to plot against fluctuations timescale rather than frequency. The PSD has three regimes. On short timescales, the PSD is flat and represents uncorrelated noise. On intermediate timescales, the PSD roughly follows a power law $\mathcal{P} \propto (\Delta t)^2$. This red noise is characteristic of the damped random walk expected from a stochastic Orstein-Uhlenbeck process \citep{Kaplar&Tacchella2019}. 

The transition between these two regimes occurs on a timescale $\sim 1~{\rm Myr}$, which can be interpreted as the timescale for star formation i.e. the free-fall time in a star-forming cloud. On shorter timescales, it is impossible for one star formation event to be correlated with another. A freefall time $\sim 1~{\rm Myr}$ is associated with a density $\sim 3\times 10^3~{\rm cm^{-3}}$, which is consistent with the typical density of star formation in Figure~\ref{fig:mstarhist}.

On longer timescales, the PSD becomes flat again. The transition between these two regimes occurs on a timescale $\sim 10~{\rm Myr}$, which can be interpreted as the lifetime of star forming clouds in our simulation \citep{Tacchella+2020}. Due to the 550~{\rm Myr} duration of our simulations, we cannot compute the PSDs on longer timescales using Welch's method. However, one might expect the PSDs to regain a positive slope at even longer timescales due to correlations caused by galaxy mergers, outflow recycling, and other processes affecting the gas reservoir \citep{Tacchella+2020}. For example, \citet{Ceverino+2018} identify bursts in the SFR on timescales of 100s of Myr.

\begin{figure*}
    \centering
    \includegraphics[width=0.667\linewidth]{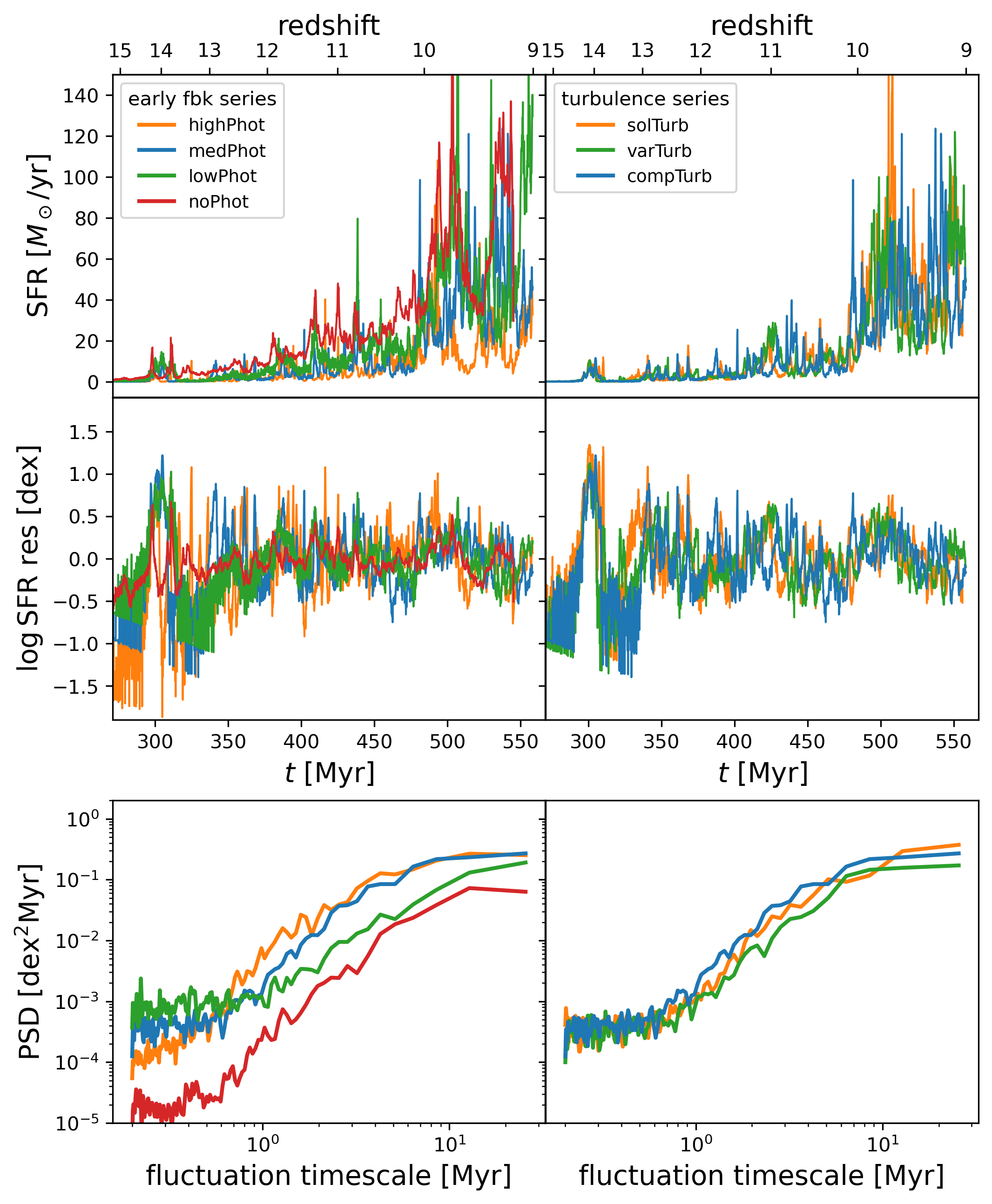}
    \caption{SFR (first row), logarithmic SFR residuals (second row), and the PSD of the SFR (third row) for the early feedback series (left column) and the turbulence forcing series (right column). Within each series, each simulation is shown in a different color, with the fiducial simulation in blue. The transition from white noise to red noise occurs on the star formation timescale $\sim 1~{\rm Myr}$. The transition from red noise back to white noise occurs on timescale representing the lifetime of star forming clouds $\sim 10~{\rm Myr}$. }
    \label{fig:variability}
\end{figure*}

For clarity of visualization, do not show our constant SFE simulations in Figure~\ref{fig:variability}. In general, we find that the SFRs in the constant SFE simulations have less power in their PSDs than the corresponding MFF simulations, consistent with previous work \citep{MartinAlvarez+2023}.

Our star formation histories show some agreement with the FFB model \citep{Dekel+2023}. The timescale of 10s of ${\rm Myr}$ is similar to the timescale predicted for generations of FFBs \citep{Li+2024}. In addition, the highest star formation rates are achieved during a time window of $\sim 80~{\rm Myr}$ between $z=10$ and $z=9$, similar to the halo-crossing timescale predicted by \citet{Dekel+2023} over which FFBs are active. However, a longer simulation is required to see if rapid star formation is suppressed by feedback from earlier generations of star clusters.

\subsection{Variable turbulence forcing parameter}
\label{sec:var_turb}

In \texttt{varTurb}, we determine the turbulence forcing parameter from the local velocity field using Equation~\ref{eq:bturb}. \texttt{varTurb} behaves almost like a simulation with constant forcing parameter in between the solenoidal and compressive forcing limits. In Figure~\ref{fig:mstarhist}, the density, temperature, and turbulent velocity dispersion distributions of star-forming cells for \texttt{varTurb} are somewhere in-between the distributions for \texttt{solTurb} and \texttt{compTurb}. 

However, in some metrics, \texttt{varTurb} deviates from this behavior. The median local SFE in \texttt{varTurb} is nearly the same as \texttt{compTurb} and the turbulent velocity dispersion of the gas is higher in \texttt{varTurb} than in \texttt{solTurb} or \texttt{compTurb}.

These effects can be explained because stars preferentially form in compressively forced regions, which have a higher local SFE for the same gas conditions and TKE. This is apparent in the forcing parameter distribution for star-forming cells (Fig.~\ref{fig:mstarhist}), which is shifted more towards the compressive end than the forcing parameter distribution for the gas (Fig.~\ref{fig:mgashist}). Because a disproportionate number of stars form in compressively forced regions, the local SFE distributions are similar between \texttt{varTurb} and \texttt{compTurb}. Our fiducial model, which assumes the compressively forced limit, is therefore a reasonable approximation for model \texttt{varTurb}, where the forcing parameter is determined more self-consistently.

In Appendix~\ref{sec:gaspropcomp}, we show the turbulence forcing parameter in an $xy$-slice through the galaxy in the fiducial simulation at $z=9$. Close to the galaxy, most of the volume is dominated by solenoidal forcing, but there are small pockets of compressive forcing. The turbulence forcing parameter distribution of SN-hosting cells (Fig.~\ref{fig:nsnehist}) has a peak at the compressive forcing limit $b_{\rm turb} = 1.0$, hinting that these pockets may be formed by SN explosions.

We leave a more rigorous analysis of self-consistent turbulence forcing to future work which simulates the forcing parameter more carefully. \citet{Mandelker+2024} describe a more rigorous approach to implement a self-consistent forcing parameter which involves decomposing the shear tensor into three components representing compression, solid body rotation and shear flow. Their method will be described in an upcoming paper (Ginzburg et. al., in prep.).

\section{Discussion}
\label{sec:discussion}
In this section, we interpret our results with toy models, compare to similar works, and discuss observational implications. First, we show that when the gas depletion time is sufficiently short, star formation and gas metallicity are regulated by ejective stellar feedback (Sec.~\ref{sec:gas_mass} and \ref{sec:metal_enrichment}). Then, we show that the previous arguments are predicated on the high local SFE in MDGs (Sec.~\ref{sec:gas_deplete}). Next, we describe the relationship between turbulence and star formation (Sec.~\ref{sec:turbdiscuss}). Then, we discuss the role of early feedback processes not included in our models (Sec.~\ref{sec:otherfbk}). Next, we discuss the limitations of our star formation recipe (Sec.~\ref{sec:beyond_mff}), SN feedback recipe (Sec.~\ref{sec:snecav}), and other aspects of our modeling (Sec.~\ref{sec:misc}). Finally, we discuss the observational implications of our results (Sec.~\ref{sec:obs}).

\subsection{Self-regulation of global SFE}
\label{sec:gas_mass}

We can use a simple one-zone model to explain the global SFE in our simulations and the exponential growth of the stellar mass, inspired by the ``bathtub'' toy model of \citet{Dekel&Mandelker2014} and similar models from earlier works \citep[e.g.][]{Tinsley1968}. The gas mass $M_{\rm gas}$ in a galaxy is depleted by star formation and outflows. Simultaneously, the gas mass is augmented by accretion and SN explosions. These processes can be described by a differential equation:
\begin{equation}\begin{split}
	\dot{M}_{\rm gas} =\ & f_{\rm b} \dot{M}_{\rm acc} - \dot{M}_* + \dot{M}_{\rm SNe} - \dot{M}_{\rm out}\\
	 =\ & f_{\rm b} \dot{M}_{\rm acc} - (1 - \chi + \eta) \dot{M}_*
	\label{eq:bathtub1}
\end{split}\end{equation}
where $\chi$ is the mass fraction of massive stars and $\eta = \dot{M}_{\rm out} / \dot{M}_*$ is the outflow efficiency. The model can account for the recycling of gas by adding negative contributions to $\eta$. 

The outflow efficiency characterizes the strength of ejective feedback, the component of stellar feedback that removes gas from the galaxy. Preventative feedback, the component of stellar feedback which does not remove gas from the galaxy, can still suppress star formation by moving gas out of the star-forming state, but its effect on the global SFE is more subtle (App.~\ref{sec:star_formation}).

If $\dot{M}_* \propto M_{\rm gas}$ and $\dot{M}_{\rm acc}$ is constant, then Equation~\ref{eq:bathtub1} asymptotically approaches a steady state solution where $\dot{M}_{\rm gas} = 0$ \citep{Dekel&Mandelker2014}. In practice, $\dot{M}_{\rm acc}$ changes as a function of time. However, if the gas depletion time $\tau_{\rm dep} = M_{\rm gas} / \dot{M}_*$ is short compared to the timescale on which the accretion rate changes $\tau_{\rm acc} = \dot{M}_{\rm acc} / \ddot{M}_{\rm acc}$, then we can treat $\dot{M}_{\rm acc}$ as constant. We estimate values for $\tau_{\rm dep}$ and $\tau_{\rm acc}$ relevant to our simulations in Sec.~\ref{sec:gas_deplete}. 

Setting $\dot{M}_{\rm gas} = 0$ in Equation~\ref{eq:bathtub1}, we find
\begin{equation}
	\dot{M}_* = \frac{f _{\rm b} \dot{M}_{\rm acc}}{1-\chi+\eta}
	\label{eq:SFR_simple}
\end{equation}
Dividing by the accretion rate to get the global SFE, we find
\begin{equation}
	\epsilon_{\rm glob} = \frac{1}{1 - \chi + \eta}
	\label{eq:eps_glob_ss}
\end{equation}

The global SFE is independent of the local SFE and decreases with stronger ejective feedback. We call this behavior self-regulation, because the SFR is regulated by stellar feedback. Self-regulation is consistent with the trend in decreasing integrated SFE with increasing outflow efficiency in our simulations (Sec.~\ref{sec:SFH}).

Equation~\ref{eq:SFR_simple} also demonstrates that the SFR is proportional to the gas accretion rate onto the galaxy. At Cosmic Dawn, the Universe is well-approximated by an Einstein-deSitter (EdS) cosmology. The specific accretion rate onto a halo of mass $M_{\rm halo}$ in an EdS universe is approximately \citep{Dekel+2013}
\begin{equation}
	\frac{\dot{M}_{\rm acc}}{M_{\rm halo}} \simeq s \left( \frac{M_{\rm halo}}{10^{12}~M_\odot} \right)^\beta (1 + z)^{5/2}
	\label{eq:tau_acc}
\end{equation}
where $\beta$ is the power law exponent of the fluctuation power spectrum and $s$ is the specific accretion rate into a halo of $M_{\rm halo} = 10^{12}~{\rm M_\odot}$ at $z=0$. \citet{Dekel+2013} find that values $\beta \simeq 0.14$ and $s \simeq 0.030~{\rm Gyr^{-1}}$ are consistent with the Millennium cosmological simulation \citep{Springel+2005}.

Ignoring the weak $M_{\rm halo}^\beta$ dependence and integrating, we find \citep[][Eq.~9]{Dekel+2013}
\begin{equation}
	M_{\rm halo} \simeq M_{\rm halo, 0} e^{-\alpha (z - z_0)}
\end{equation}
where $M_{\rm halo, 0}$ is the halo mass at redshift $z_0$, $\alpha = (3/2) s t_{\rm H, 0}$, $t_{\rm H, 0} = (2/3) \Omega_{\rm m, 0}^{-1/2} H_0^{-1} \approx 17.5~{\rm Gyr}$ is the age of an EdS universe at redshift $z=0$, and we have used
\begin{equation}
	a = (1 + z)^{-1} = \left( \frac{ t }{ t_{\rm H, 0} } \right)^{2/3}
\end{equation}

This implies that the accretion rate onto a given halo as it grows is \citep[][Eq.~10]{Dekel+2013}
\begin{equation}
	\dot{M}_{\rm acc}(z) \simeq s M_{\rm halo, 0} e^{-\alpha (z - z_0)} (1 + z)^{5/2}
	\label{eq:Mdot_acc_grow}
\end{equation}
At high redshift, the accretion rate is well-described as an exponential function of redshift \citep[see also][]{Correa+2015}. Therefore, the exponential growth of stellar mass in our simulated galaxies is simply a reflection of the exponential growth of the halo accretion rate. 

\subsection{Metal enrichment}
\label{sec:metal_enrichment}

Using a similar approach, we can describe the process of metal enrichment. The metal mass in a galaxy is governed by
\begin{equation}\begin{split}
	\dot{M}_{\rm metal}=\ & -Z (\dot{M}_* + \dot{M}_{\rm out})+ y \dot{M}_{\rm SNe}\\
	=\ & -[ Z (1+ \eta) - y \chi ] \dot{M}_*
	\label{eq:bathtub2}
\end{split}\end{equation}
where $y$ is the SN metal yield and we have assumed that the metal content of the primordial, accreting gas in negligible. 

This model assumes that metals produced in SN explosions are rapidly mixed throughout the ISM rather than remaining localized to the SN site. The turbulent environment at Cosmic Dawn facilitates that mixing. Our simulated galaxies have typical velocity dispersions $\simeq 50~{\rm km/s}$ and radii $\simeq 700~{\rm pc}$, so the metal mixing time is approximately $\tau_{\rm mix} \sim R_{\rm gal} / \sigma_{\rm turb, 1D} \simeq 14~{\rm Myr}$, far shorter than the lifetime of the galaxy.

Again assuming a steady state $\dot{M}_{\rm metal} = 0$, we find that
\begin{equation}
    Z = y\chi/(1 + \eta)
\end{equation}
In our simulations, we have $y = 0.1$ and $\chi = 0.2$. If the outflow efficiency is negligible $\eta \approx 0$, then we have $Z \approx 0.02 \approx 1~Z_\odot$, which is similar to the typical gas metallicity in \texttt{noPhot} (Fig.~\ref{fig:mgashist}). 

If, on the other hand, outflows become efficient at removing gas, then the gas metallicity will be reduced accordingly. For example, $1 + \eta$ is larger by a factor of 5 in \texttt{highPhot} compared to \texttt{medPhot}, and the typical gas metallicity is reduced by the same factor.

\subsection{Condition for self-regulation}
\label{sec:gas_deplete}

Section~\ref{sec:gas_mass}, we assume that the depletion timescale is short compared to the accretion timescale, and therefore the gas mass reaches a steady state. To define the condition for steady state more precisely, we apply a bathtub-like model to the star-forming gas mass $M_{\rm sf}$, following the arguments of \citet{Semenov+2017} and \citet{Semenov+2018}. 

Star-forming gas is depleted by star formation, stellar feedback, and dynamical processes (e.g. turbulent shear) operating on a timescale $\tau_{\rm dyn}$. Simultaneously, star-forming gas is augmented by cooling and collapse on a timescale $\tau_{\rm cool}$. These processes can be described by a differential equation
\begin{equation}
	\dot{M}_{\rm sf} = \frac{1 - f_{\rm sf}}{\tau_{\rm cool}} M_{\rm gas} - \frac{f_{\rm sf}}{\tau_{\rm dyn}} M_{\rm gas} - (1 + \mu) \dot{M}_*
\end{equation}
where $f_{\rm sf} = M_{\rm sf} / M_{\rm gas}$ is the star-forming mass fraction and $\mu \dot{M}_*$ is the rate at which gas mass is removed from the star-forming state by stellar feedback. $\mu$ characterizes the strength of preventative feedback.

The timescales $\tau_{\rm cool}$, $\tau_{\rm ff}$, and $\tau_{\rm dyn}$ which set the star-forming gas mass are shorter than the accretion timescale, so we can assume a steady state where $\dot{M}_{\rm sf} = 0$. Then, we can solve for $f_{\rm sf}$, which yields
\begin{equation}
	f_{\rm sf} = \frac{\tau_{\rm ff}}{\epsilon_{\rm ff}} \left[ f_{\rm dyn} \frac{\tau_{\rm ff}}{\epsilon_{\rm ff}} + (1 + \mu) \tau_{\rm cool} \right]^{-1}
	\label{eq:f_sf}
\end{equation}
where $f_{\rm dyn} = 1 + \tau_{\rm cool} / \tau_{\rm dyn}$ is a factor of a few. We have assumed that the SFR is related to the gas mass by the Schmidt Law
\begin{equation}
	\dot{M}_* = \frac{\epsilon_{\rm ff}}{\tau_{\rm ff}} f_{\rm sf} M_{\rm gas}
	\label{eq:schmidt}
\end{equation}

Equation~\ref{eq:schmidt} implies that $\tau_{\rm dep} = \tau_{\rm ff} / ( f_{\rm sf} \epsilon_{\rm ff} )$. Therefore, the gas depletion time is
\begin{equation}
	\tau_{\rm dep} = f_{\rm dyn}\frac{\tau_{\rm ff}}{\epsilon_{\rm ff}} + (1 + \mu) \tau_{\rm cool}
	\label{eq:semenov}
\end{equation}

We can combine Equation~\ref{eq:semenov} and our simulations to estimate the depletion time in MDGs. In our simulations, the typical local SFE is $\epsilon_{\rm ff} \sim 10\%$ and the typical density of star-forming regions is $\sim 3\times 10^3~{\rm cm^{-3}}$, corresponding to a freefall time $\tau_{\rm ff} \sim 1~{\rm Myr}$. Feedback is weak, so we can approximate $\mu \approx 0$. $f_{\rm dyn}$ and $\tau_{\rm cool} / \tau_{\rm ff}$ are both factors of a few. These values imply that the depletion time is on the order of 10s of ${\rm Myr}$. 

Differentiating Equation.~\ref{eq:Mdot_acc_grow}, we find the rate of change of the accretion rate
\begin{equation}
	\ddot{M}_{\rm acc} \simeq s^2 M_{\rm halo, 0} e^{-\alpha (z - z_0)} (1 + z)^4 \left[ (1 + z) - \frac{5}{2 \alpha} \right]
\end{equation}
Dividing $\dot{M}_{\rm acc}$ by $\ddot{M}_{\rm acc}$ to get the accretion timescale, we find
\begin{equation}
	\tau_{\rm acc} \simeq \frac{\alpha (1 + z)^{-5/2} / s}{\alpha - (5/2) (1 + z)^{-1} }
\end{equation}
For a halo of mass $10^{11}~{\rm M_\odot}$ at redshift $z = 9$, we find $\tau_{\rm acc} \simeq 154~{\rm Myr}$. This is longer than the depletion time, so the assumption of steady state in Section~\ref{sec:gas_mass} and the resulting self-regulation behavior are valid. 

If the local SFE were on the order of a few percent, similar to present-day galaxies, than the accretion timescale would be shorter than the depletion timescale and the assumption of steady state would break down. In Appendix~\ref{sec:star_formation}, we find that in this case, self-regulation is not guaranteed, especially for low local SFE.

This simple calculation emphasizes an important point. One might think that because feedback is weak in MDGs, that star formation should not be regulated by feedback. However, we have shown that this is not the case, because the high local SFE compensates for the weak feedback.

\subsection{Turbulence}
\label{sec:turbdiscuss}

Turbulence broadens the subgrid density PDF (Eq.~\ref{eq:sigs}). In relatively hotter and more diffuse gas, the broadening effect extends the high-density tail of the distribution into the Jeans unstable region, enhancing local star formation. Physically, this represents star formation in high-density regions created by turbulent compression. 

In relatively cooler and more dense gas, the broadening effect extends the low-density tail of the distribution into the Jeans stable region, suppressing local star formation. However, in this case the broadening effect does not significantly change the local SFE, because the low-density tail of the distribution is only associated with a small amount of mass.

Making the turbulence more compressive also has a broadening effect, because $\mathcal{M}_{\rm turb}$ and $b_{\rm turb}$ play the same role in Equation~\ref{eq:sigs}. This explains the trend of increasing local SFE as turbulence becomes more compressive. However, changing the TKE has an additional effect which does not occur when changing the turbulence forcing parameter.

Turbulent pressure assists thermal pressure in preventing gravitational collapse (Eq.~\ref{eq:alphavir}). This results in a larger critical density in the gravitational stability criterion (Eq.~\ref{eq:scrit}), reducing the local SFE for the same gas conditions. Local star formation is suppressed by turbulent pressure support for any gas condition, but the effect is most potent when the mean density is close to the critical density for collapse. 

Turbulent pressure support becomes more important as turbulence forcing becomes more solenoidal because the PDF becomes narrower. In the solenoidal forcing limit, a small change in the critical density for collapse has a larger effect on the portion of the PDF which is gravitationally unstable.

\begin{figure*}
    \centering
    \includegraphics[width=\linewidth]{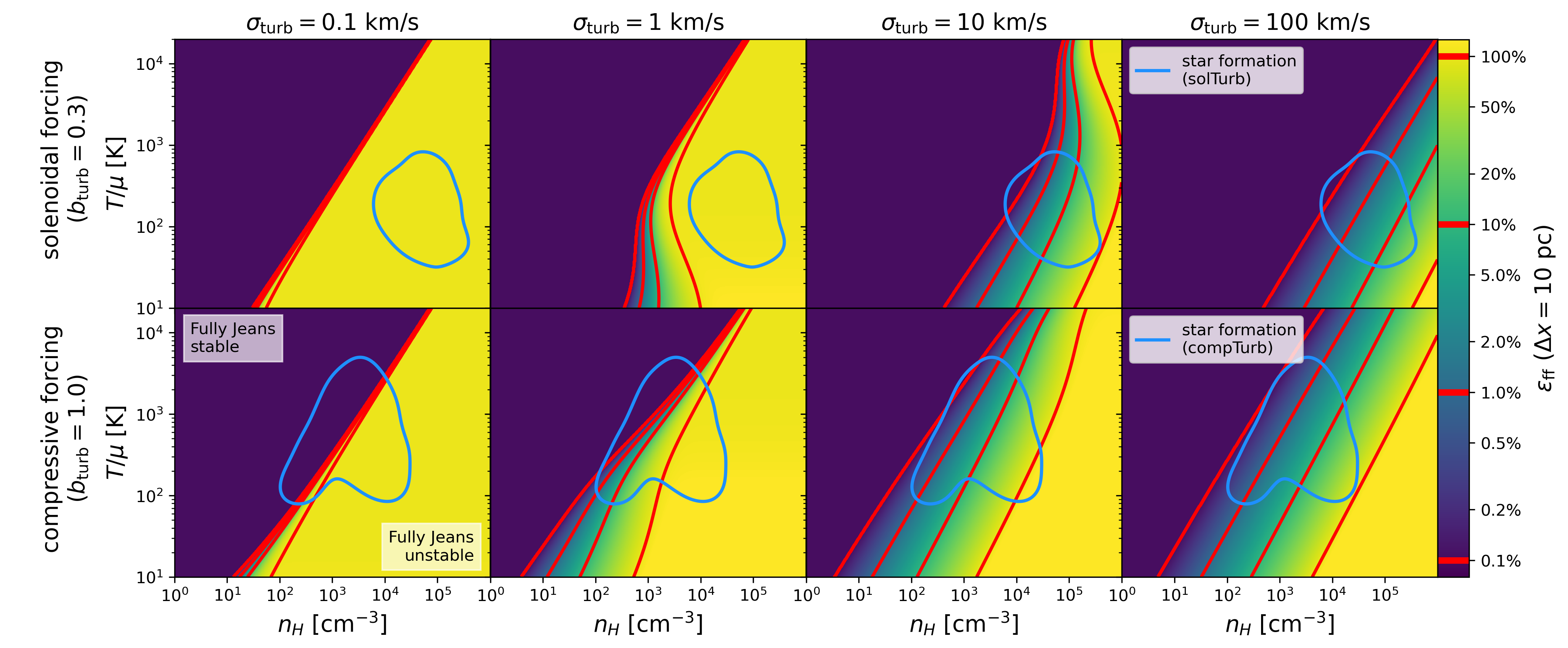}
    \caption{Local SFE as a function of temperature and Hydrogen number density in the MFF model (Eq.~\ref{eq:mff}). Each column shows a different turbulent velocity dispersion. The top row is the solenoidal forcing limit ($b_{\rm turb} = 0.3$) and the bottom row is the compressive forcing limit ($b_{\rm turb} = 1.0$). The resolution scale is set to its minimum value $\Delta x = 10~{\rm pc}$ in our simulations, and we emphasize that the local SFE is a resolution-dependent quantity by definition. Iso-efficiency contours are shown in red for $\epsilon_{\rm ff} \in \{0.1\%, 1\%, 10\%, 100\%\}$. When $\epsilon_{\rm ff} \sim 0\%$, the gas is Jeans stable. When $\epsilon_{\rm ff} = 100\%$, the gas is fully Jeans unstable. In each panel, we overplot a blue contour outlining the region of phase space containing 90\% of star formation events weighted by stellar mass in \texttt{solTurb} (top row) and \texttt{compTurb} (bottom row). The contours are smoothed by a $10~{\rm px}$ Gaussian filter for ease of viewing. In general, the local SFE decreases with increasing turbulent velocity dispersion for the same gas conditions, except for gas conditions which are marginally Jeans stable in the compressive forcing limit. }
    \label{fig:mff}
\end{figure*}

In Figure~\ref{fig:mff}, we show how the local SFE in the MFF model varies as a function of temperature, density, and turbulent velocity dispersion $\sigma_{\rm turb} = c_{\rm s} \mathcal{M}_{\rm turb}$ in the solenoidal and compressive forcing limits at a fixed smoothing scale $10~{\rm pc}$. Physically, the smoothing scale can be understood as a mixing length. In numerical contexts, it can be understood as the size of a resolution element. 

The local SFE is easy to understand in the extreme regimes of fully Jeans stable ($\epsilon_{\rm ff} \sim 0\%$) or fully Jeans unstable ($\epsilon_{\rm ff} \sim 100\%$). In the transitional regime, where the local SFE takes on intermediate values, the result becomes sensitive to turbulence. In our simulated MDGs, most star formation occurs in the transitional regime (Fig.~\ref{fig:mff}) where the effect of turbulence is important. 

In general, the local SFE decreases with increasing turbulent velocity dispersion for the same gas conditions, except for gas conditions which are marginally Jeans stable in the compressive forcing limit. These marginally stable, supersonic gas clouds are common in low redshift galaxies such as the Milky Way. 

In real galaxies, the relationship between the forcing of small-scale turbulence and large-scale galactic dynamics is complex. \texttt{varTurb} provides a first insight into how this could play out. Solenoidal forcing modes in the turbulence may be driven by the differential rotation within a disc, suppressing star formation in discs. Meanwhile, compressive forcing modes may be driven by SN explosions, galaxy-galaxy mergers \citep{Renaud+2014}, or accretion \citep{Mandelker+2024}, enhancing star formation in those environments.

Due to self-regulation (Sec.~\ref{sec:gas_mass}), the decrease in local SFE due to the high turbulence of Cosmic Dawn only has a weak effect on the stellar mass which ultimately forms. For turbulent velocity dispersions even larger than the typical values $\sim 50~{\rm km/s}$ in our simulated galaxies, the local SFE may drop enough to take the galaxy out of the self-regulated regime. In this case, the integrated SFE would begin to decrease in proportion to the decrease in local SFE.

\subsection{Other early feedback processes}
\label{sec:otherfbk}

Our early feedback model only includes thermal pressure from photoionized gas (Sec.~\ref{sec:earlyfbk}). However, there are other early feedback processes we do not model, including radiation pressure from FUV and multiply-scattered IR photons and stellar winds. In this section, we discuss the effect of these mechanisms on star formation in MDGs.

\subsubsection{Radiation pressure feedback}

Massive stars produce a significant flux of UV photons with energies below the Lyman limit i.e. FUV. In dense environments like giant molecular clouds, these photons are readily absorbed by dust grains, imparting their momentum onto the surrounding gas \citep{Krumholz&Matzner2009, Fall+2010, Murray+2010, Sales+2014}. This radiation pressure is augmented by multiply-scattered Ly-$\alpha$ \citep{Kimm+2018} and reprocessed IR \citep{Murray+2010, Skinner&Ostriker2015} photons. The former is not as well-studied, so we will focus our discussion on FUV and IR radiaton pressure feedback.

FUV radiation pressure feedback is diminished in high surface densities environments because the radiation pressure force cannot overcome the gravity of the cloud \citep{Fall+2010, Murray+2010}. In addition, radiation pressure forces partially cancel out when FUV is absorbed on scales smaller than the typical separation between radiation sources \citep{Menon+2023}. On the other hand, IR radiation feedback is enhanced in high surface density environments because each IR photon undergoes many scatterings \citep{Thompson+2005, Krumholz&Matzner2009, Murray+2010}. When $\Sigma_{\rm b} \gtrsim 3000~M_\odot/{\rm pc}^2$, IR radiation pressure feedback dominates over FUV radiation pressure feedback \citep{Menon+2023}, even though dust opacities are $\sim 100$ times greater in the FUV than in the IR \citep{Semenov+2003}. Therefore, we expect IR radiation feedback to dominate over FUV radiation feedback in MDGs. 

There are several additional considerations which may affect the strength of IR radiation feedback in MDGs. First, IR wavelengths are much larger than interstellar grain radii, so the dust opacity varies as $\kappa \propto \nu^2$ following the electric dipole limit \citep{Draine2011}. This causes the mean photon frequency to decrease as stellar radiation diffuses through the dust, weakening the dust-radiation coupling \citep{Reissl+2018, Krumholz+2019}. Second, the low metallicity at Cosmic Dawn leads to a low dust-to-gas ratio, which reduces the effective IR opacity \citep{Menon+2024}.

Third, in a turbulent medium, radiation preferentially escapes through low column density channels \citep{Skinner&Ostriker2015, Tsang&Milosavljevic2018, Menon+2022}. Radiation pressure feedback may even self-regulate by driving turbulence \citep{Krumholz&Thompson2012}. This argument only applies to the IR photons, which are produced diffusely by dust and thus sample the turbulent density structure. The FUV photons are produced by the massive stars themselves which are preferentially surrounded by high-density regions. Matter-radiation anti-correlation is especially important in the star-forming regions of MDGs, where the turbulent velocity dispersions are high $\sim 50~{\rm km/s}$ (Fig.~\ref{fig:mstarhist}).

\begin{figure}
    \centering
    \includegraphics[width=\linewidth]{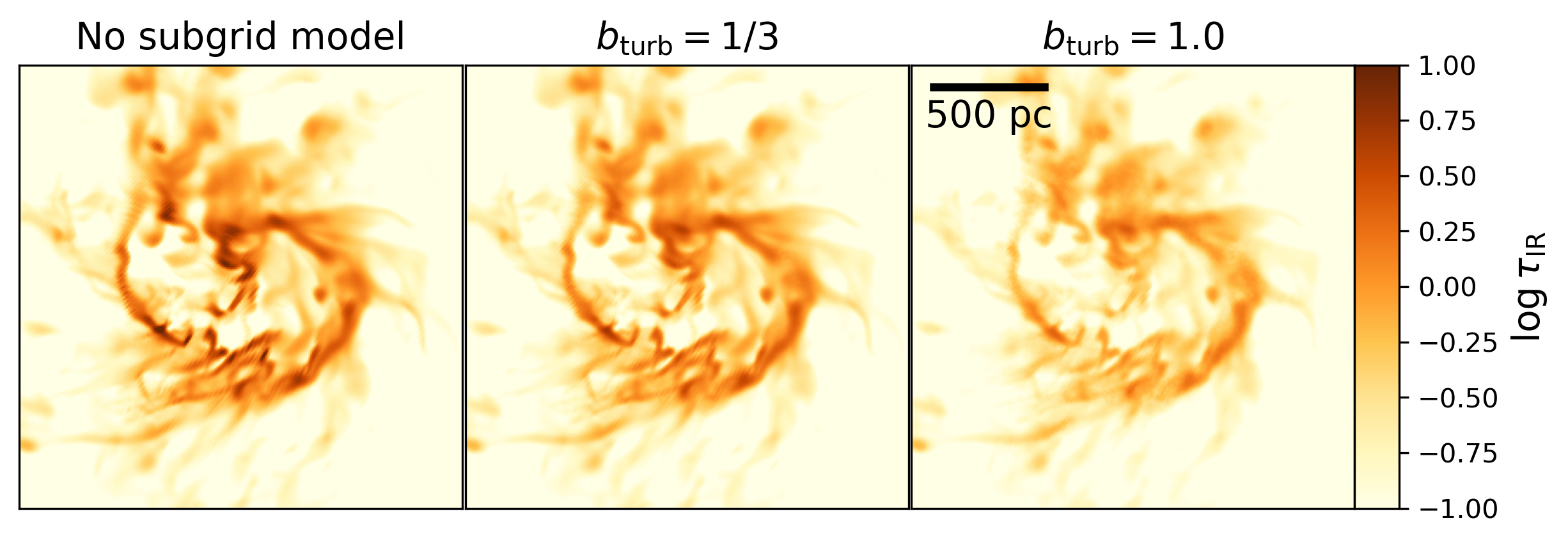}
    \caption{Logarithmic IR dust optical depth in a box of sidelength $2~{\rm kpc}$ around the most massive galaxy in the fiducial simulation, projected in the face-on direction. In the left panel, we compute the contribution of each cell as $\rho \kappa_{\rm IR}$. In the other panels, we compute the contribution of each cell using an effective optical depth accounting for the matter-radiation anti-correlation (App.~\ref{sec:coldenspdf}) for solenoidal (middle panel) and compressive (right panel) turbulence forcing models. We find that despite the low metallicity, the star-forming regions of the galaxy are optically thick to IR. However, accounting for the matter-radiation anti-correlation significantly reduces the optical depth.}
    \label{fig:opacityIR}
\end{figure}

In Figure~\ref{fig:opacityIR}, we estimate the face-on IR dust optical depth in our fiducial simulation given by integrating the contributions from each cell along the projection direction. In the left panel, we compute the contribution of each cell as $\rho \kappa_{\rm IR}$, where $\kappa_{\rm IR} = \widetilde{\kappa}_{\rm IR} (Z / Z_\odot)$ is the IR dust opacity at solar metallicity assuming that the dust-to-gas ratio scales linearly with metallicity. This may overestimate the opacity at low metallicity \citep{Feldman2015, Choban+2022}. 

$\widetilde{\kappa}_{\rm IR}$ is roughly constant $\sim 5~{\rm g/cm^2}$ in the dust temperature range $100~{\rm K} \lesssim T_{\rm dust} \lesssim 1200~{\rm K}$, and takes on smaller values outside of that range \citep{Semenov+2003}. We adopt this constant value and ignore the temperature dependence. In the other panels, we compute the contribution of each cell using an effective optical depth accounting for the matter-radiation anti-correlation (App.~\ref{sec:coldenspdf}) for solenoidal and compressive turbulence forcing models.

We find that despite the low metallicity, the star-forming regions of the galaxy are optically thick to IR. However, accounting for the matter-radiation anti-correlation significantly reduces the optical depth. Turbulent Mach number is positively correlated with density in our simulations (Fig.~\ref{fig:gaspropcomp}), so the regions that contribute most to the optical depth are most strongly affected by matter-radiation anti-correlation. This estimate suggests that IR radiation pressure feedback may still be important in MDGs, but matter-radiation anti-correlation needs to be accounted for. This should be investigated by future work.

\subsubsection{Stellar winds}

Line photons from massive stars drive stellar winds which serve as an additional early feedback mechanism \citep{Castor+1975, Weaver+1977, Lancaster+2021}. The integrated kinetic energy carried by stellar winds over a massive star lifetime is comparable to that delivered by a SN explosion \citep{Castor+1975}. The potential of stellar winds to rival SNe as a feedback mechanism has been shown in simulations \citep{Ceverino&Klypin2009, Fierlinger+2016}. 

The low metallicity in MDGs makes launching line-driven winds more difficult. Even when winds are launched, their impact may be weakened by the leakage of hot gas through low-density channels and energy losses due to turbulent mixing \citep{Rogers&Pittard2013, Rosen+2014, Lancaster+2021}. These effects are enhanced in the turbulent environment of MDGs. For these reasons, we expect that stellar winds are a subdominant feedback mechanism in MDGs. However, detailed simulations of stellar winds in turbulent, low-metallicity environments are needed to test this hypothesis. Given an appropriate subgrid model, stellar wind feedback could be easily included in our simulations.

\subsection{Beyond the multi-freefall model}
\label{sec:beyond_mff}

Our star formation recipe is based on the MFF model of \citet{Federrath&Klessen2012} as implemented in \texttt{RAMSES} by \citet{Kretschmer&Teyssier2020}. In this section, we discuss the shortcomings of this approach and compare our recipe to other works with similar prescriptions.

\subsubsection{Other star formation recipes}

\citet{Federrath&Klessen2012} discuss three variations of the MFF model with different choices for the critical density threshold $s_{\rm crit}$ from \citep{Krumholz&McKee2005} (KM), \citep{Padoan&Nordlund2011} (PN), and \citep{Hennebelle&Chabrier2011} (HC). The KM and PN versions simultaneously fit all of their detailed simulations to within a factor of two over two magnitudes in SFR. 

Our MFF model is based on the KM version with a modification by \citet{Kretschmer&Teyssier2020}. Other works \citep{Kimm+2017, Trebitsch+2017, Mitchell+2018, Rosdahl+2018} use a MFF model based on the PN version, except that they only allow stars to form in cells larger than the turbulent Jeans length. This thermo-turbulent model differs from our model because it does not allow star formation in unresolved locally Jeans-unstable clouds if they reside in a cell which is globally Jeans-stable. The thermo-turbulent model has been extended to include magnetic pressure support \citep{MartinAlvarez+2020} and has been incorporated into multi-physics simulations including on-the-fly radiative transfer and cosmic rays \citep{MartinAlvarez+2023, Dome+2024}. 

There are a plethora of star formation recipes outside the MFF model framework. Based on detailed MHD simulations, \citet{Padoan+2012} propose an exponential SFE law of the form $\epsilon_{\rm ff} \propto \exp ( \alpha_{\rm vir}^{1/2} )$, noting that the local SFE is nearly independent of turbulent Mach number. More recent simulations by \citet{Kim+2021} also favor the exponential model, although with a different numerical coefficient in the argument of the exponential. This model has successfully been applied to measurements of the SFE in the Milky Way \citep{Evans+2022, Elia+2025}. Future work is needed to develop a theoretical foundation for the exponential model, and to describe under what conditions it is favored over MFF-type models.

\subsubsection{Prestellar core efficiency}

We assume that all of the gas in a gravitationally unstable gas cloud at the sonic scale forms stars. However, the process of star formation is not 100\% efficient even at the sonic scale. Previous observational \citep{Andre+2010} and theoretical \citep{Matzner&McKee2000} work suggest that only $\epsilon_{\rm core} \sim 30-50\%$ of the gas in a collapsing prestellar core ends up in stars. 

The MFF model can easily be modified by a prefactor $\epsilon_{\rm core}$ in Equation~\ref{eq:mff} which accounts for the efficiency of star formation at the sonic scale. Any value $\epsilon_{\rm core} < 1$ would result in a lower local SFE $\epsilon_{\rm ff}$ for the same gas conditions. This is essentially the same effect as when we go from compressive to solenoidal turbulence forcing. \texttt{solTurb} indeed has a smaller local SFE than \texttt{compTurb}, but the integrated SFEs only differ by $\sim 5\%$ because our MDGs are in the self-regulated regime where their integrated SFEs are primarily determined by the strength of stellar feedback. Therefore, we expect that simulations with lower values of $\epsilon_{\rm core}$ will also have lower integrated SFEs, but only by a small factor. 

\subsubsection{Subgrid density PDF}

We assume that the subgrid density PDF is lognormal (Eq.~\ref{eq:mffpdf}) with width given by Equation~\ref{eq:sigs}. However, the true small-scale density field may differ significantly from this description. First, Equation~\ref{eq:sigs} comes from fitting a functional form to the small-scale turbulence simulations of \citet{Padoan&Nordlund2011}. Those simulations consider the steady state in a gas forced by the idealized Ornstein-Uhlenbeck process. However, ISM turbulence is not a steady state. Forcing from the bulk velocity field and stellar feedback is multi-scale and time-variable. 

Second, the lognormal form of the subgrid density PDF is only an approximation \citep{Federrath+2010, Beattie+2022}. Deviations from a lognormal shape are more pronounced at high Mach numbers \citep{Federrath&Klessen2012}, such as those in MDGs. After a star-formation event, there may be a delay before the lognormal PDF can be reestablished by gravitational collapse which the model does not account for.

Third, the MFF model is fundamentally limited by its use of the density PDF alone to characterize the unresolved density field. The density PDF is a one-point statistical description that ignores spatial correlation. The same density PDF could be produced by numerous dense subsolar gas clumps or a single mass gas clump, which should physically correspond to different SFEs.

Recently, \citet{Hennebelle+2024} presented the turbulent support (TS) model to address some of these shortcomings. The TS model substitutes the lognormal PDF in favor of the more complex Castaing-Hopkins PDF, which has been shown to produce better agreement in the high-Mach regime \citep{Castaing1996, Hopkins2013}. Compared to the lognormal PDF, the Castaing-Hopkins PDF predicts less gas at high densities, a consequence of small-scale turbulent pressure support. The TS model also accounts for unresolved spatial correlations parameterized by the power-law index of the mass function of self-gravitating clumps.

\citet{Brucy+2024} compare the predictions of the TS model to detailed turbulence simulations. Surprisingly, they find that the lognormal PDF reproduces the SFE in the high-Mach regime better than the Castaing-Hopkins PDF. It is unclear whether this discrepancy is physical or related to numerical dissipation in the turbulence simulations. Regardless, they show that the MFF tends to overpredict the SFE in the high-Mach regime (their Figure~14) in agreement with \citet{Hennebelle+2024}. At high Mach number, the turbulent Jeans length becomes larger than the turbulence injection scale. This implies that turbulence no longer generates gravitationally unstable structures. 

These arguments imply that that MFF may overestimate the SFR in MDGs. However, it's worth noting that our simulations are at an intermediate resolution $\sim 10~{\rm pc}$ relative to the resolution of the detailed turbulence simulations of \citet{Brucy+2024} at $\sim 2~{\rm pc}$ and the size of the simulation domain $\sim 1000~{\rm pc}$ on which they apply the TS model. It's possible that although we use a more simple star formation recipe, we still capture some of the relevant behavior simply due to our high resolution. Future work should investigate whether the TS model converges with the MFF model at smaller scales.

\subsubsection{Subgrid equation of state}

Another aspect of star formation we have neglected is the subgrid EOS. The detailed turbulence simulations used to calibrate the MFF model assume an isothermal EOS \citep{Padoan&Nordlund2011}. This is generally appropriate in dense star-forming regions where cooling is efficient. This also matches the behavior of our simulations in dense regions due to the cooling term in the energy equation.

Variations in the EOS lead to changes in subgrid density statistics \citep{Passot&VazquezSemadeni1998, Li+2003, Audit&Hennebelle2010}. The simulations of \citet{Glover&MacLow2007a, Glover&MacLow2007b} indicate that the effective polytropic index may be sub-unity (soft EOS) in the density range $\sim 10-10^4~{\rm cm^{-3}}$ and super-unity (stiff EOS) at much higher densities $\gtrsim 10^9~{\rm cm^{-3}}$ where the gas starts to become optically thick \citep{Masunaga&Inutsuka2000}. In our simulations, the high-density tail of the subgrid density PDF enters this range (Fig.~\ref{fig:rho_mff}).

\citet{Federrath&Banerjee2015} derive density variance vs. Mach number relations for multiple EOSs, analogous to Equation~\ref{eq:sigs}. At low Mach numbers, the density variance only depends weakly on the EOS. However, at high Mach numbers relevant to the conditions in MDGs, the density variance can increase significantly as the EOS becomes softer. Future detailed turbulence simulations should account for how the EOS changes at different density scales and produce a more accurate parameterization of the density distribution.

\subsection{Feedback modeling caveats}
\label{sec:snecav}

\subsubsection{Supernova clustering}

Each star cluster particle in our simulations represents a bound collection of individual stars. The star particle mass is tied to the efficiency of photoionization feedback (Sec.~\ref{sec:earlyfbk}). However, the star particle mass also affects the efficiency of SN feedback through the clustering of SN explosions. A larger star particle mass produces more clustered SN explosions which enhances the efficiency of SN feedback through the formation of superbubbles with delayed radiative cooling \citep[e.g.][]{Sharma+2014, Gentry+2017}. However, a smaller star particle mass samples more diffuse environments where the efficiency of SN feedback is higher. Ideally, one could decouple the photoionization feedback model from the star particle mass and sample star particle masses from a cluster mass function parameterized by the local gas conditions. However, although the cluster mass function has some constraints in local Universe \citep{Fall+2012, Adamo+2015}, it could differ substantially in MDGs. This is further complicated by the dependence of cluster sizes on the efficiency of early feedback \citep{Smith+2021}.

SN clustering is also impacted by the clustering of the star clusters themselves. In our simulations, star particles inherit the velocity of the gas in their parent cell. However, the velocity of the star cluster could deviate from this value on the scale of the turbulent velocity dispersion. This implies that our simulations underestimate the dispersion of star clusters.

Dynamical interactions can also remove stars from the cluster. Weak two-body interactions produce ``walk-away'' stars which slowly diffuse out of the cluster. Strong two-body interactions produce ``run-away'' stars ejected at rapid velocities. Run-away stars can travel a significant distance from the cluster center before going SN. The fraction of massive stars ejected this way is enhanced by their tendency to concentrate at the centers of clusters where two-body encounters are more common \citep{Oh&Kroupa2016}. Hydrodynamics simulations demonstrate that the ejection of massive stars enhances the efficiency of SNe feedback by producing more SN explosions in diffuse regions \citep{Ceverino&Klypin2009, Kimm&Cen2014}.

Finally, only a fraction of stars form in bound clusters, described by the cluster formation efficiency (CFE) \citep{Bastian2008}. Milky-Way-like galaxies in the local Universe have typical CFEs $\lesssim 15\%$ \citep[][Tab.~3]{Kruijssen2012}. Empirically, the CFE increases in higher gas surface density environments, both within and between galaxies \citep{Adamo+2020a}, reaching $37\pm 7\%$ in the Milky Way Galactic center \citep{Ginsburg&Kruijssen2018} and $30-100\%$ in nearby mergers \citep{Adamo+2020b}. Therefore, the CFE is likely higher in MDGs but still $<100\%$. The deviation from $100\%$ may reduce SN clustering and thus the efficiency of SN feedback \citep[][Sec.~7.3.4]{Kruijssen2012}.

SN clustering also depends on the distribution of SN in time. We determine the SN rate at each timestep assuming that SNe are uniformly distributed within the time interval between $\tau_{\rm start}$ and $\tau_{\rm end}$. In reality, SN explosions are more clustered towards the end of that interval due to the negative slope of the IMF. Previous work has accounted for this by adjusting the SN rate over the star particle lifetime \citep{Kimm+2015}.

\subsubsection{Supernova cooling radius}

In dense regions when the SN cooling radius is unresolved, we model the SN explosion using momentum feedback. Without momentum feedback, the thermal energy injected by the SN would be unphysically radiated away before the snowplow phase of the SN blastwave. However, by skipping the earlier phases of the SN explosion, we do not account for the effect of this heating on the star-forming cloud. This small-scale heating could allow SNe to more rapidly shut off star formation within a cloud. If we could resolve these scales, the SN delay time might become important to star formation as predicted in the FFB model \citep{Dekel+2023}.

The supernova cooling radius should also be affected by the unresolved density field. \citet{Martizzi+2015} simulate SN explosions in inhomogeneous media with a variety of turbulent Mach numbers. They find that as the Mach number increases, the SN explosion deposits its thermal energy and momentum at progressively larger radii (their Fig.~7). This effect increases the effectiveness of SN feedback. Future modeling should produce power law scalings similar to those in \citet{Martizzi+2015} which incorporate the turbulent Mach number as a continuous parameter.

\subsubsection{Secondary star formation}

In the dense galactic environments of Cosmic Dawn, it may be possible for the dense shells of a SN blastwave to themselves be gravitationally unstable and trigger additional star formation. Using simple one-dimensional models, \citet{Nagakura+2009} found that a triggered star formation can occur from a typical core-collapse SN in gas with density $\gtrsim 10^3~{\rm cm^{-3}}$. In our simulations, a significant fraction of SNe occur in this regime (Fig~\ref{fig:nsnehist}). Therefore, this effect may increase the SFR relative to what we predict. Secondary star formation might even be enhanced when accounting for small scale turbulence, because the conditions inside a SN blastwave are dominated by compressive forcing modes.

\subsection{Miscellaneous considerations}
\label{sec:misc}

\subsubsection{Pop III stars}

We do not include a model for Pop III stars. At $z = 9$, this is a good approximation, but our simulations begin at $z\sim 100$ and therefore they include the time period where Pop III stars dominate the cosmic star formation history $z\sim 15-20$ \citep[see][for a review]{Klessen&Glover2023}. Pop III stars with masses $140~M_\odot \lesssim m_* \lesssim 260~M_\odot$ are expected to undergo pair-instability SNe, which have energies $100\times$ that of more standard core-collapse SN, leading to more effective SN feedback. 

Even after Pop III star formation ceases, stars forming in low metallicity $Z \lesssim 3\%~Z_\odot$ environments could form directly from atomic gas due to the discrepancy between the thermal and chemical equilibrium timescales \citep{Glover&Clark2012, Krumholz2012}. The formation timescale for ${\rm H}_2$ is slow at low metallicities due to the lack of dust grain surfaces the catalyze the reaction, so the molecular fraction remains small until the density rises high enough to form $H_2$ via 3-body reactions. If the gas surrounding a young massive star is primarily atomic Hydrogen, the effect of photoionization feedback may be enhanced due to the lower ionization energy relative to molecular Hydrogen. More detailed studies are needed to explore these effects, especially in less massive haloes at Cosmic Dawn, which may be less rapidly enriched by metals.

\subsubsection{Active galactic nuclei}

As discussed in Section~\ref{sec:intro}, some MDGs show evidence of AGN activity. AGN might provide an additional source of feedback which we do not consider. From a theoretical perspective, the formation mechanisms of supermassive BHs (SMBHs) is an open question. It is typically assumed that SMBHs form from massive BH seeds, which grow by accretion and mergers. As pointed out by \citet{Dekel+2023}, the enhanced SFE in massive $z\sim 10$ galaxies might suppress accretion onto the BH by consuming most of the gas in star formation. On the other hand, the inefficiency of feedback may enhance accretion of the residual gas. Future simulations could address the question of BH growth and AGN feedback by using the sink particle functionality already available in \texttt{RAMSES}.

\subsubsection{Chaotic variance}

The multiple non-linear coupled PDEs in our galaxy simulations form a chaotic system, and therefore our results are subject to chaotic variance, seeded by truncation errors and random algorithmic elements. \citet{Genel+2019} found that global properties of a single galaxy can vary on a level of $\sim 2-25\%$ due to chaotic variance, depending on the property in question. In a similar analysis, \citet{Keller+2019} concluded that differences $\lesssim 20\%$ in time-average SFRs between different galaxy simulations require statistical evidence to show that it is not due to stochasticity, with mergers being a particularly large source of chaos. Ideally, we could run multiple copies of each simulation to quantify the range of chaotic variance, but this was impossible due to computational cost. Most of our conclusions are based on integrated quantities or statistical analyses, so we believe that our results are robust to chaotic variance, but we do not have data to confirm this.

\subsection{Observational implications}
\label{sec:obs}

\subsubsection{SFR indicators}
\label{sec:indicator}

\citet{Kaplar&Tacchella2019} point out that when the SFR is determined from observations, one must rely on tracers such as IR, UV, and emission lines. These tracers do not reflect the instantaneous SFR because they have different contributions from stars of different ages. To predict what the star formation history looks like as seen by a given indicator, we convolve the intrinsic SFR with the response function of that indicator.

Using the Flexible Stellar Population Synthesis (FSPS) code \citep[\texttt{FSPS}, ][]{Conroy+2009}, we estimate the response function for two star formation indicators: the H$\alpha$ emission line and the \textit{JWST} f200w filter. At $z=9$, the H$\alpha$ line is redshifted to $\lambda = 5.9~{\rm \mu m}$, which is outside the spectral range of \textit{JWST}'s \texttt{NIRSpec} instrument. However, we found that viable emission line indicators such as H$\beta$ and the OIII lines produce similar response functions to H$\alpha$. At $z=9$, \textit{JWST}'s f200w and f277w filters probes mid-UV and near-UV frequencies respectively. We find that the f277w filter produces a similar response function to the f200w filter.

We assume a redshift $z=9$, a Chabrier IMF \citep{Chabrier2003}, and a metallicity $Z = 0.331~Z_\odot$, equal to the median metallicity of stars in our fiducial simulation. \texttt{FSPS} also includes dust and nebular emission models. \texttt{FSPS} has been extensively calibrated against observations in the low redshift Universe \citep{Conroy&Gunn2010}. There may be large systematic errors at $z\sim 10$, which we ignore for the sake of qualitative argument.

\begin{figure}
    \centering
    \includegraphics[width=\linewidth]{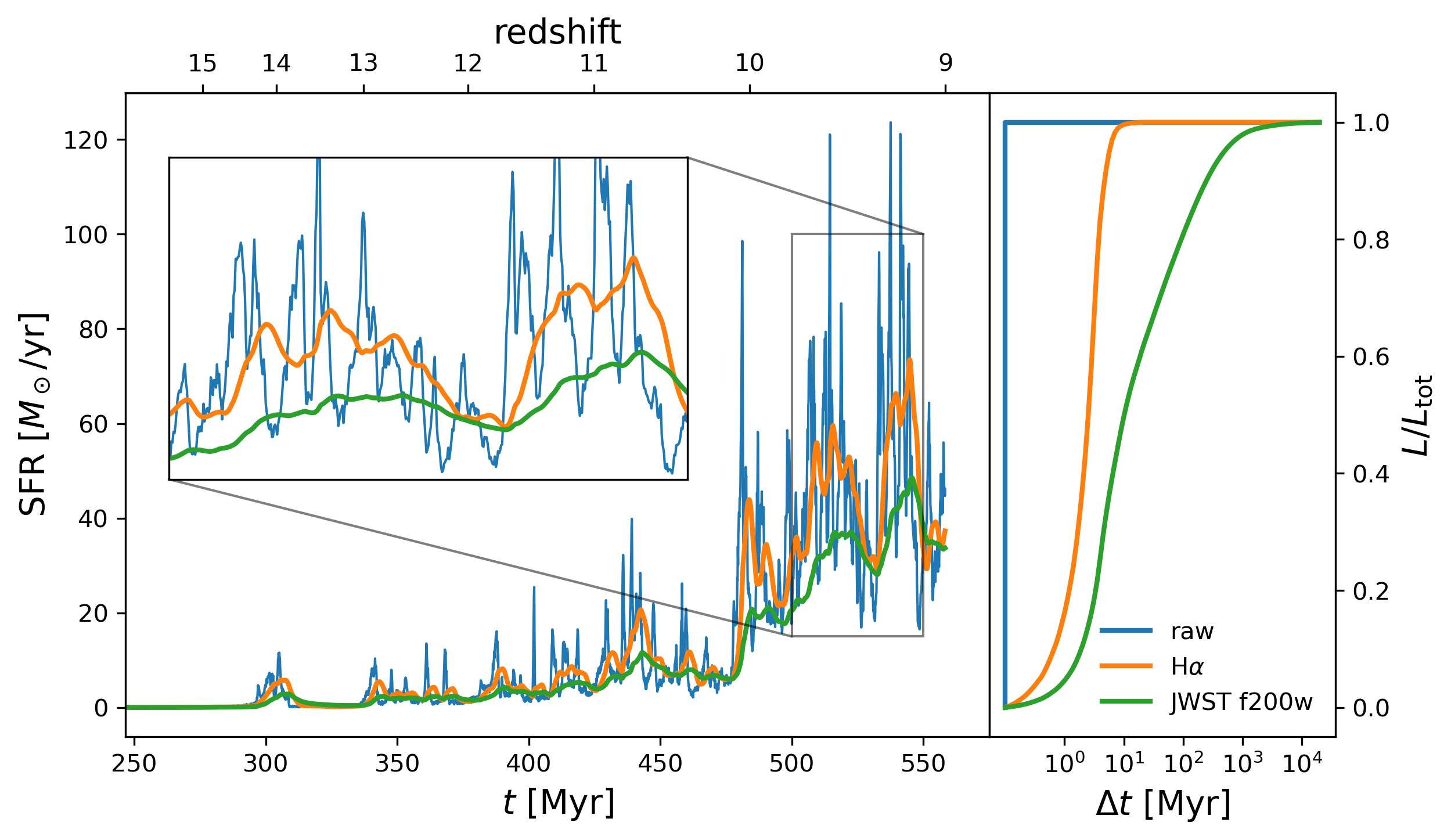}
    \caption{SFR as a function of time (left panel) for the fiducial simulation, including both the intrinsic SFR (blue) and the SFR inferred by H$\alpha$ (orange) and the \textit{JWST} f200W filter (green). An inset plot shows a zoom to highlight how the indicators respond to short-timescale variability. In the right panel, we show the cumulative luminosity for a given indicator as a function of time $\Delta t$ since star birth. The H$\alpha$ SFR is similar in magnitude to the intrinsic SFR, but the \textit{JWST} f200w SFR is systematically smaller, resulting from the long-timescale contributions which sample earlier times when the SFR was lower. }
    \label{fig:SFR_conv}
\end{figure}

We plot the response functions in the right panel of Figure~\ref{fig:SFR_conv}. H$\alpha$ responds rapidly on a timescale of $\sim 3~{\rm Myr}$, while the \textit{JWST} f200w filter responds on longer timescales of $\sim 10~{\rm Myr}$, but with small contributions on timescales as long as $\sim 100~{\rm Myr}$. Note that our response function for the \textit{JWST} f200w filter is only valid at $z=9$. At higher redshifts, the \textit{JWST} f200w filter responds slightly faster e.g. on timescales $\sim 5~{\rm Myr}$ for $z = 19$. The OIII$\lambda$4960 line, another plausible emission line indicator, has a similar response function to H$\alpha$. Likewise, the \textit{JWST} f277w filter, another plausible photometric indicator, has a similar response function to the \textit{JWST} f200w filter. 

In the left panel of Figure~\ref{fig:SFR_conv}, we show the SFR as a function of time for the fiducial simulation, including both the intrinsic SFR and the SFR inferred by H$\alpha$ and \textit{JWST} f200w star formation indicators. The H$\alpha$ SFR is similar in magnitude to the intrinsic SFR, but the \textit{JWST} f200w SFR is systematically smaller, resulting from the long-timescale contributions which sample earlier times when the SFR was lower. This effect is a consequence of the rapid increase in the mean SFR as a function as time in this epoch.

Both SFR indicators hide the intrinsic variability in the SFR on short timescales, reducing the observed variability as pointed out by \citet{Furlanetto&Mirocha2022}. Following \citet{Pallottini&Ferrara2023}, we quantify this effect using the standard deviation of fluctuations in the log SFR $\sigma_{\rm SFR}$, computed using the procedure in Section~\ref{sec:variability}. For the raw and \textit{JWST} f200w SFRs in our fiducial simulation, the standard deviations are $0.28$ and $0.12$. The variability in UV magnitudes is directly related to the variability in the log SFR by $\sigma_{\rm UV} = 2.5 \sigma_{\rm SFR}$. For the raw and \textit{JWST} f200w SFRs in our fiducial simulation, this implies UV variabilities $0.83$ and $0.35$ respectively. The latter more accurately characterizes the variability that would actually be observed. This value is significantly smaller than the variability $\sigma_{\rm UV} \gtrsim 1.5$ required to explain the discrepancies between models and data from burstiness alone \citep{Mason+2023, Shen+2023}. Our conclusion is similar to \citet{Pallottini&Ferrara2023}, expect that we account for the effect of the SFR indicator. However, our simulations do not capture enhanced burstiness due to coupling with radiation and cosmic rays as demonstrated by \citet{Dome+2024}.

\subsubsection{Stellar masses and SFRs}

\begin{figure}
    \centering
    \includegraphics[width=\linewidth]{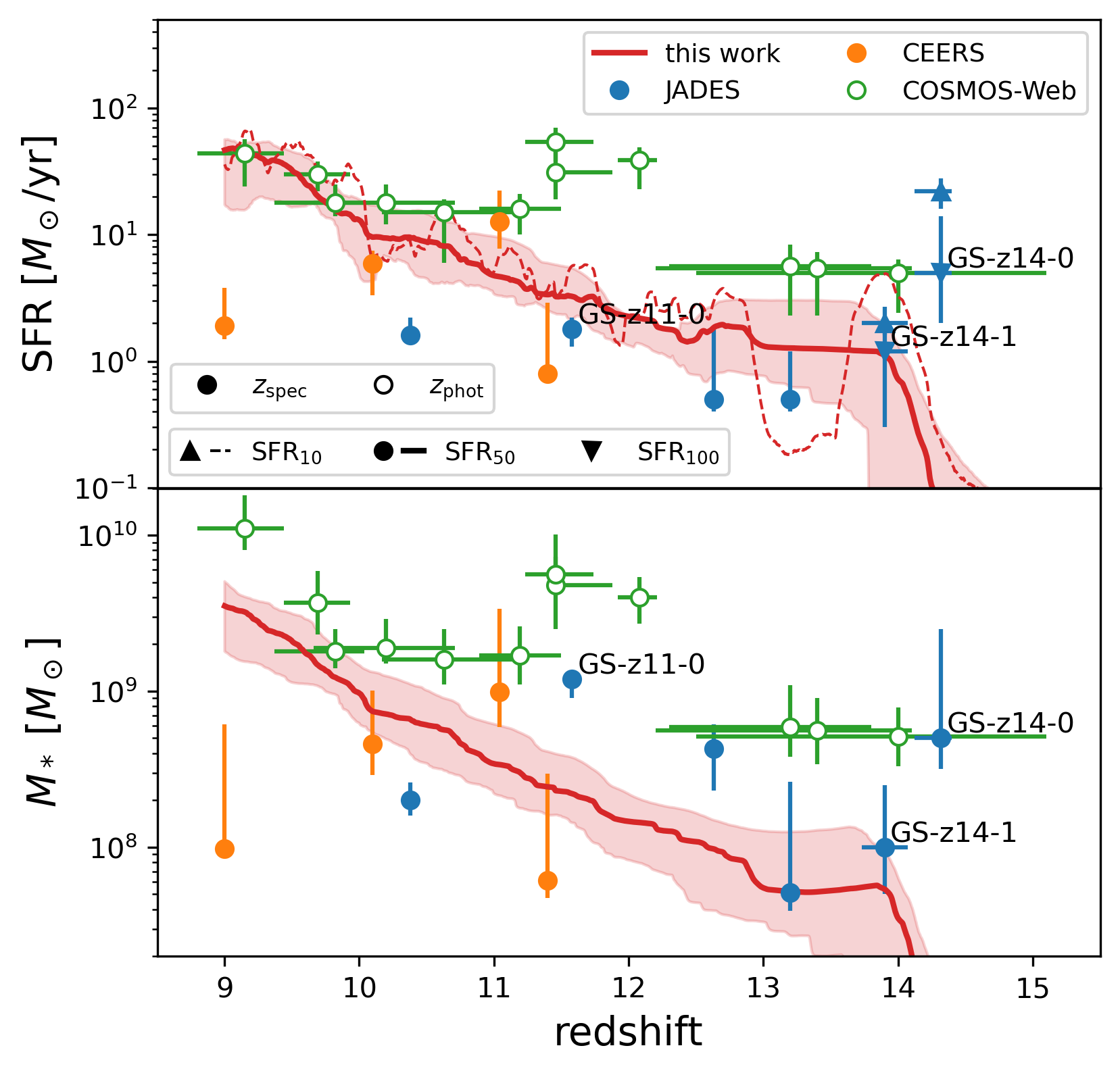}
    \caption{SFR (top panel) and stellar mass (bottom panel) as a function of time in our simulations (red) and a selection of the most luminous galaxy candidates at Cosmic Dawn in various surveys, including JADES (blue), CEERS (orange), and COSMOS-Web (green). The solid red line shows data from our fiducial simulation and the shaded area includes the range of values from \texttt{lowPhot}, \texttt{highPhot}, \texttt{solTurb}, and \texttt{varTurb}. Galaxy candidates which are not spectroscopically confirmed are marked with an open circle. SFR is averaged over $10~{\rm Myr}$ (up triangle, solid line), $50~{\rm Myr}$ (circle), and $100~{\rm Myr}$ (down triangle, dashed line). The stellar mass and SFR data for each simulation are available on the Princeton Research Data Service \citep{Andalman&Teyssier2024}.}
    \label{fig:obs_comp}
\end{figure}

In Figure~\ref{fig:obs_comp}, we plot the stellar mass and SFR in our fiducial simulation as a function of redshift in red. We also plot our epistemic uncertainty, the range of stellar masses and SFRs across simulations in the early feedback and turbulence forcing series. The stellar mass and SFR data for each simulation are available on the Princeton Research Data Service\footnote{\href{https://doi.org/10.34770/v56h-ps15}{https://doi.org/10.34770/v56h-ps15}} \citep{Andalman&Teyssier2024}.

In blue, green, and orange, we show estimates of stellar mass and SFR from \textit{JWST} \texttt{NIRSpec} spectropscopy and \texttt{NIRCam} photometry for a selection of luminous galaxies from the \textit{JWST} Advanced Deep Extragalactic Survey \citep[JADES][]{Eisenstein+2023}, Cosmic Evolution Early Release Science \citep[CEERS][ERS-1345]{Finkelstein+2023b}, and \citep[COSMOS-Web][GO\#1727]{Casey+2023}. We source our data for each survey from \citet{Harikane+2024}, \citet{Carniani+2024}, and \citet{Casey+2024} respectively. This selection includes the most luminous and high redshift galaxies spectroscopically confirmed at the time of writing, such as GS-z11-0, GS-z14-0, and GS-z14-1. 

Ideally, we should only compare our simulated galaxy to observed galaxies whose halo mass matches that of our simulated galaxy at the appropriate redshift. However, the star-to-halo mass ratio at Cosmic Dawn is not well-constrained, since it depends on the integrated SFE. As argued in Section~\ref{sec:intro}, the most extreme objects discovered at Cosmic Dawn in current surveys should have halo masses in around the value $M_{\rm halo}\sim 10^{11}~M_\odot$ for our simulated galaxy. However, surveys which sample a larger volume, like COSMOS-Webb, probably include more massive haloes $M_{\rm halo} \ge 10^{11}~M_\odot$ which naturally have larger stellar masses and SFRs. 

The COSMOS-Webb candidates are not spectroscopically confirmed, indicated by open circles in the plot, so their redshifts should be taken with caution. Previous work suggests that photometric estimates of redshift systematically overestimate redshift for galaxies at Cosmic Dawn \citep[e.g.][Fig. 4]{Fujimoto+2023}. Even when spectra are available, stellar masses and SFRs are typically estimated using spectral fitting codes like \textsc{Bagpipes} \citep{Carnall+2018} and \textsc{Prospector-$\alpha$} \citep{Leja+2019}. However, at Cosmic Dawn, \texttt{NIRSpec} data provides very little constraints on dust-obscured star formation or the presence of an AGN, which would artificially decrease and increase stellar population estimates respectively.

The crude comparison in Figure~\ref{fig:obs_comp} is reasonable because the stellar masses and SFRs inferred from observations are already highly model-dependent. We are only suggesting that our simulated galaxies are in the ballpark of stellar masses and SFRs of the MDGs seen by \textit{JWST}. Future work could make this comparison more quantitative using abundance matching and accounting for the effects of cosmic variance \citep{Jespersen+2024}, but a more detailed approach may not be justified until the observational data are better understood.

More observational evidence for MDGs are found at Cosmic Noon ($z\sim 3$), near the peak of the cosmic SFR density, where a population of massive quiescent galaxies (MQGs) have emerged \citep{Glazebrook+2017, Valentino+2020, Antwi-Danso+2023, Carnall+2023, Carnall+2024, deGraaff+2024, Glazebrook+2024}. These galaxies typically have large stellar masses $M_* \gtrsim 10^{10}~M_\odot$ and low specific SFRs $\lesssim 0.2$. Recently, \citet{Weibel+2024} reported the discovery of a MQG at $z = 7.3$ confirmed spectroscopically with \textit{JWST} \texttt{NIRSpec} as part of the Cycle 2 program of RUBIES (GO\#4233) with a stellar mass $M_* \sim 10^{10}~M_\odot$.

Detailed modeling of MQG spectra suggests that MQGs must have formed the bulk of their stellar populations within the first billion years of cosmic history in short, extreme starbursts lasting hundreds of ${\rm Myr}$ at most \citet[e.g.][]{Weibel+2024}. This is qualitatively consistent with our simulated MDGs, hinting that MDGs could be the progenitors of MQGs. Future work should extend run simulations out to lower redshift to investigate this possibility. Cosmological simulations tend to underproduce MQGs at high redshift \citep{AlcaldePampliega+2019, Merlin+2019}, so they have been understudied in the theoretical galaxy formation community.

\section{Conclusions and Future Work}
\label{sec:conclusion}
In this work, we run zoom-in simulations of a MDG ($M_{\rm halo} = 10^{11}~M_\odot$, $z = 9$) using the cosmological hydrodynamics code \texttt{RAMSES} at an effective resolution $\simeq 10~{\rm pc}$. Our simulations capture a unique regime where star formation is expected to be enhanced by high gas densities. Our suite of simulations vary in the details of their star formation and feedback recipes, allowing us to investigate the connection between global SFE and local gas conditions. Our fiducial simulation uses a physically-motivated, turbulence-based, multi-freefall model of star formation, avoiding \textit{ad hoc} extrapolation from lower redshifts.

We summarize our main findings in bullet points:
\begin{itemize}

\item
\textit{Star formation efficiencies}: The high gas densities $\sim 10^4~{\rm M_\odot/pc^2}$, $\sim 3 \times 10^3~{\rm cm^{-3}}$ in star-forming regions of MDGs naturally result in efficient star formation, with a local SFE $\epsilon_{\rm ff} \sim 10 - 20\%$ and an integrated SFE $\epsilon_{\rm int} \sim 10 - 30\%$.

\item
\textit{Galaxy structure}: Our simulated galaxies form a thick disc with support from both rotation and turbulence. For cold gas, the disc radius is $R_{\rm gas}\sim 700~{\rm pc}$ and the scale height is $H/R\sim 0.5$. For stars the disc radius is $R_{\rm gas} \sim 400~{\rm pc}$ and scale height is $H/R\sim 0.25$. The baryonic mass is organized into dense, cold, and turbulent clumps of diameter $\sim 100~{\rm pc}$ where most of the star formation occurs, although the size of these clumps may be an artifact of our $\simeq 10~{\rm pc}$ effective resolution. These discs may have analogues in the discs of star-forming galaxies at Cosmic Noon.

\item
\textit{Star formation history and variability}: The stellar mass in the MDG increases in proportion to the accretion rate onto the halo i.e. nearly exponential growth for $z \ge 9$. The SFR fluctuates on various timescales, producing an intrinsic UV variability $\sigma_{\rm UV} \simeq 0.83$ which is reduced to $\sigma_{\rm UV} \simeq 0.35$ when measured with a wide-band filter. This is insufficient to explain the discrepancies between models and data from bursty star formation alone. The Fourier transform of the SFR encodes information about the typical density of star formation $\sim 3\times 10^3~{\rm cm^{-3}}$ and the typical lifetime of star-forming clouds $\sim 10~{\rm Myr}$.

\item
\textit{Stellar feedback}: Our implementation of stellar feedback is crude, but our preliminary results suggest that stellar feedback in MDGs is weak due to the high density conditions. This is consistent with the general prediction of the FFB scenario, despite the fact that our current simulations don't resolve the SN cooling radius or the star-forming clusters where FFBs would occur. Weak feedback can be partially mitigated by photoionization from massive stars, which decreases the density around SN explosions. However, even in our strongest photoionization feedback model the integrated SFE is still high $\ge 10\%$. 

\item
\textit{Self-regulation}: One might think that the weak feedback in MDGs cannot regulate star formation. However, we show that star formation is still regulated by feedback due to the high local SFE. As a result, the global SFE is nearly independent of the local SFE and depends inversely on the strength of ejective feedback.

\item
\textit{Metal enrichment}: Similar to the global SFE, gas metallicity depends inversely on the strength of ejective feedback. By $z=9$, SN feedback enriches our simulated galaxies to a significant fraction of solar metallicity.

\item 
\textit{Turbulence}: Turbulence generally decreases the local SFE for the same gas conditions due to turbulent pressure support. Turbulence in MDGs is strong $\sim 50~{\rm km/s}$ due to intense accretion, but not strong enough to take MDGs out of the self-regulated regime, so it has only a weak effect on global and integrated SFEs.

\item
\textit{Comparison to observations}: Our simulated galaxies are in the ballpark of the estimated stellar masses and SFRs of the MDGs observed by \textit{JWST}.

\end{itemize}

Our study is the first step towards understanding efficient star formation in MDGs. Our treatment of stellar feedback is crude due to the limitations of our effective resolution $\simeq 10~{\rm pc}$ in a pure hydrodynamics simulation. Although these factors affect the accuracy of our results, they also make the simulations easier to interpret by simplifying the physics. We will use the insights from this work to make informed choices about more computationally expensive future simulations which capture more of the physics.

The interplay between photoionization and SN feedback in our simulations suggests that early feedback is important in MDGs, even when it cannot suppress star formation directly. In future simulations, we will model early feedback in more detail using \texttt{RAMSES-RT}, including explicit models for photoionization, radiation pressure, and stellar winds. These models should account for the effect of strong turbulence, which can decrease the effective opacity (App.~\ref{sec:coldenspdf}). Using \texttt{RAMSES-RT} will also enable a more self-consistent treatment of the extragalactic radiation field.

The high densities in our simulations mean that we do not resolve the scale of star-forming clouds, which can be roughly approximated by the Jeans radius $R_{\rm J} = c_{\rm s} / \sqrt{G \rho}$. For a cloud at the Hydrogen ionization temperature $\sim 10^4~{\rm K}$ and a density given by the median star-forming density of our fiducial simulation $\sim 3\times 10^{3}~{\rm cm^{-3}}$, the Jeans radius is $\sim 20~{\rm pc}$, and the cloud will only be resolved by eight cells. For a lower temperature given by the median star-forming temperature of our fiducial simulation $\sim 300~{\rm K}$, the Jeans radius is $\sim 3~{\rm pc}$, and the cloud will not be resolved at all. The addition of 3 AMR levels would give an effective resolution $\simeq 1~{\rm pc}$, resulting in fully resolved star-forming clouds. A high resolution would also solve the problem of unresolved HII regions, revealing the multi-phase structure of the ISM.

There are many other avenues for future work, including (i) exploring the growth of black holes and impact of AGN on star formation in MDGs and (ii) running simulations beyond Cosmic Dawn to investigate whether our MDGs could be the progenitors of massive quiescent galaxies at lower redshifts.

\section*{Acknowledgements}
Z.A. would like to acknowledge helpful conversations with David Setton, Jenny Greene, Eve Ostriker, Ronan Hix, Nicholas Choustikov, Matthew Sampson, Diederek Kruijssen, Mike Boylan-Kolchin, Tibor Dome, Robert Feldman, and work by Diego Solorio on the simulation initial conditions. We are grateful to our referee Daniel Ceverino, whose time and attention significantly improved the quality of this manuscript.

This material is based upon work supported by the National Science Foundation (NSF) and the U.S.-Israel Binational Science Foundation (BSF) under Award Number 2406558 and Award Title ``The Origin of the Excess of Bright Galaxies at Cosmic Dawn''.

This material is based upon work supported by the U.S. Department of Energy, Office of Science, Office of Advanced Scientific Computing Research, Department of Energy Computational Science Graduate Fellowship under Award Number DE-SC0024386. 

A.D. has been partly supported by the Israel Science Foundation grant ISF 861/20 and by NSF-BSF grant 2023723.

This report was prepared as an account of work sponsored by an agency of the United States Government. Neither the United States Government nor any agency thereof, nor any of their employees, makes any warranty, express or implied, or assumes any legal liability or responsibility for the accuracy, completeness, or usefulness of any information, apparatus, product, or process disclosed, or represents that its use would not infringe privately owned rights. Reference herein to any specific commercial product, process, or service by trade name, trademark, manufacturer, or otherwise does not necessarily constitute or imply its endorsement, recommendation, or favoring by the United States Government or any agency thereof. The views and opinions of authors expressed herein do not necessarily state or reflect those of the United States Government or any agency thereof.

The simulations presented in this article were performed on computational resources managed and supported by Princeton Research Computing, a consortium of groups including the Princeton Institute for Computational Science and Engineering (PICSciE) and the Office of Information Technology’s High Performance Computing Center and Visualization Laboratory at Princeton University.

This research made use of \texttt{SciPy} \citep{Virtanen+2020}, \texttt{NumPy} \citep{Harris+2020}, and \texttt{Matplotlib} \citep{Hunter2007}. 

\section*{Data Availability}
The simulations in this work were run on Stellar cluster at Princeton University. Each simulation contains approximately $1.3~{\rm Tb}$ of data. Due to data storage limitations, we have only maintained data from our fiducial simulation, which can be provided upon reasonable request to the corresponding author. Likewise, the exact \texttt{RAMSES} patch and initial conditions used for our simulations can be provided upon reasonable request to the corresponding author.

A dataset containing the stellar mass and SFR as a function of proper time and redshift for the fiducial model and models \texttt{lowPhot}, \texttt{highPhot}, \texttt{solTurb}, and \texttt{varTurb} are provided on the Princeton Research Data Service \citep{Andalman&Teyssier2024} under the Creative Commons Attribution 4.0 International (\href{https://choosealicense.com/licenses/cc-by-4.0/}{CC-BY-4.0}) license. This dataset may be useful for comparing our simulated MDGs results to other simulated or observed MDGs.

Movies of the simulations are publicly available on \href{https://www.youtube.com/playlist?list=PL7YbfRC6zxzAYgFEr5oefYb5dcv0dl7Ba}{YouTube} (https://tinyurl.com/53xxm3r2).

\bibliographystyle{mnras}
\bibliography{main}

\appendix
\section{Simple estimate of baryon surface density}
\label{sec:surf_dens}

We define a dark matter halo as a sphere about a density peak with mean overdensity $\Delta \simeq 200$ above the cosmological background $\rho_{\rm crit}$, following the standard definition from the spherical-collapse model \citep{Gunn&Gott1972}. This implies a matter density inside the halo of
\begin{equation}
	\rho_{\rm halo} = \Delta \rho_{\rm crit, 0} \Omega_{\rm m, 0} (1 + z)^3
    \label{eq:rho_halo}
\end{equation}
and a halo virial radius
\begin{equation}\begin{split}
	R_{\rm vir} =\ & \left( \frac{M_{\rm vir}}{(4\pi/3) \rho_{\rm halo}} \right)^{1/3} = \left( \frac{M_{\rm vir}}{(4\pi/3) \Delta \rho_{\rm crit, 0} \Omega_{\rm m, 0}} \right)^{1/3} (1 + z)^{-1}\\
    \simeq\ & 14 M_{11}^{1/3} (1 + z)_{10}^{-1}~{\rm kpc}
\end{split}\end{equation}
where $M_{11} = M_{\rm vir} / 10^{11}~M_\odot$, $(1 + z)_{10} = (1 + z) / 10$, and $\rho_{\rm crit, 0} = 3 H_0^2 / ( 8 \pi G )$ is the critical density of the Universe at redshift $z = 0$. The corresponding virial temperature is
\begin{equation}
    \frac{T_{\rm vir}}{\mu} = \frac{m_{\rm p}}{k_{\rm B}} \frac{G M_{\rm vir}}{2 R_{\rm vir}} \simeq 3.6 \times 10^6 M_{11}^{2/3} (1 + z)_{10}~{\rm K}
    \label{eq:temp_vir}
\end{equation}

Following \citet{Dekel+2013}, we assume that the radius of the galactic disc scales with the halo radius via a contraction factor $\lambda$
\begin{equation}
	R_{\rm gal} = \lambda R_{\rm vir} \simeq 720\lambda_{0.05} M_{11}^{1/3} (1+z)_{10}^{-1}~{\rm pc}
\end{equation}
where $\lambda_{0.05} = \lambda / 0.05$. Based on tidal-torque theory, \citet{Dekel+2013} argue that the average $\lambda$ is similar to the halo spin parameter, with typical value $\lambda \simeq 0.05$ \citep[e.g.][]{Bullock+2001}. We further assume that the baryon mass in the galactic disc is approximately equal to the total accreted baryon mass,
\begin{equation}
    M_{\rm b} = f_{\rm b} M_{\rm vir}, \quad f_{\rm b} = \Omega_{\rm b, 0} / \Omega_{\rm m, 0}
\end{equation}

Under these assumptions, the typical baryon surface density in the galaxy is
\begin{equation}\begin{split}
    \Sigma_{\rm b} =\ & \frac{M_{\rm b}}{\pi R_{\rm gal}^2} = \frac{f_{\rm b} M_{\rm vir}^{1/3}}{\pi \lambda^2} \left( \frac{4\pi}{3} \Delta \rho_{\rm crit, 0} \Omega_{\rm m, 0} \right)^{2/3} (1 + z)^2\\
    \simeq\ & 9.6\times 10^3\lambda_{0.05}^{-2} M_{11}^{1/3} (1 + z)_{10}^2~M_\odot/{\rm pc^2}
    \label{eq:sigma_b}
\end{split}\end{equation}
As a corollary, we see that $M_{\rm vir} \propto (1 + z)^6$, which is similar to a result from \citet{Boylan-Kolchin2024}. We use Equation~\ref{eq:sigma_b} to estimate the baryon surface density in a galaxy as a function of redshift and the mass of its host halo in Figure~\ref{fig:haloregime}. The gas and baryon surface densities are related by $\Sigma_{\rm gas} = (1 - \epsilon_{\rm int}) \Sigma_{\rm b}$, where $\epsilon_{\rm int} = M_* / M_{\rm b}$ is the integrated SFE. Given the disk scale height $(H/R)_{\rm gal}$, the typical number density is
\begin{equation}\begin{split}
    n_{\rm H} =\ & \frac{M_{\rm gas} / m_{\rm H}}{\pi R_{\rm gal}^3 (H/R)_{\rm gal}} = \frac{(1 - \epsilon_{\rm int}) f_{\rm b}}{\pi \lambda^3 (H/R)_{\rm gal}} \frac{4\pi}{3} \Delta \rho_{\rm crit, 0} \Omega_{\rm m, 0} (1 + z)^3\\
    \simeq\ & 540~(1 - \epsilon_{\rm int}) \lambda_{0.05}^{-3} (H/R) (1 + z)_{10}^3~{\rm cm^{-3}}
    \label{eq:num_dens}
\end{split}\end{equation}

\section{Contribution of massive stars to photoionization}
\label{sec:QvsM}

The average rate of Hydrogen-ionizing photons emitted per stellar mass depends on both the IMF $\xi(m)$ and the rate of ionizing photon emission as a function of stellar mass $Q(m)$. The latter has empirical constraints from the local Universe. 

\begin{figure}
    \centering
    \includegraphics[width=\linewidth]{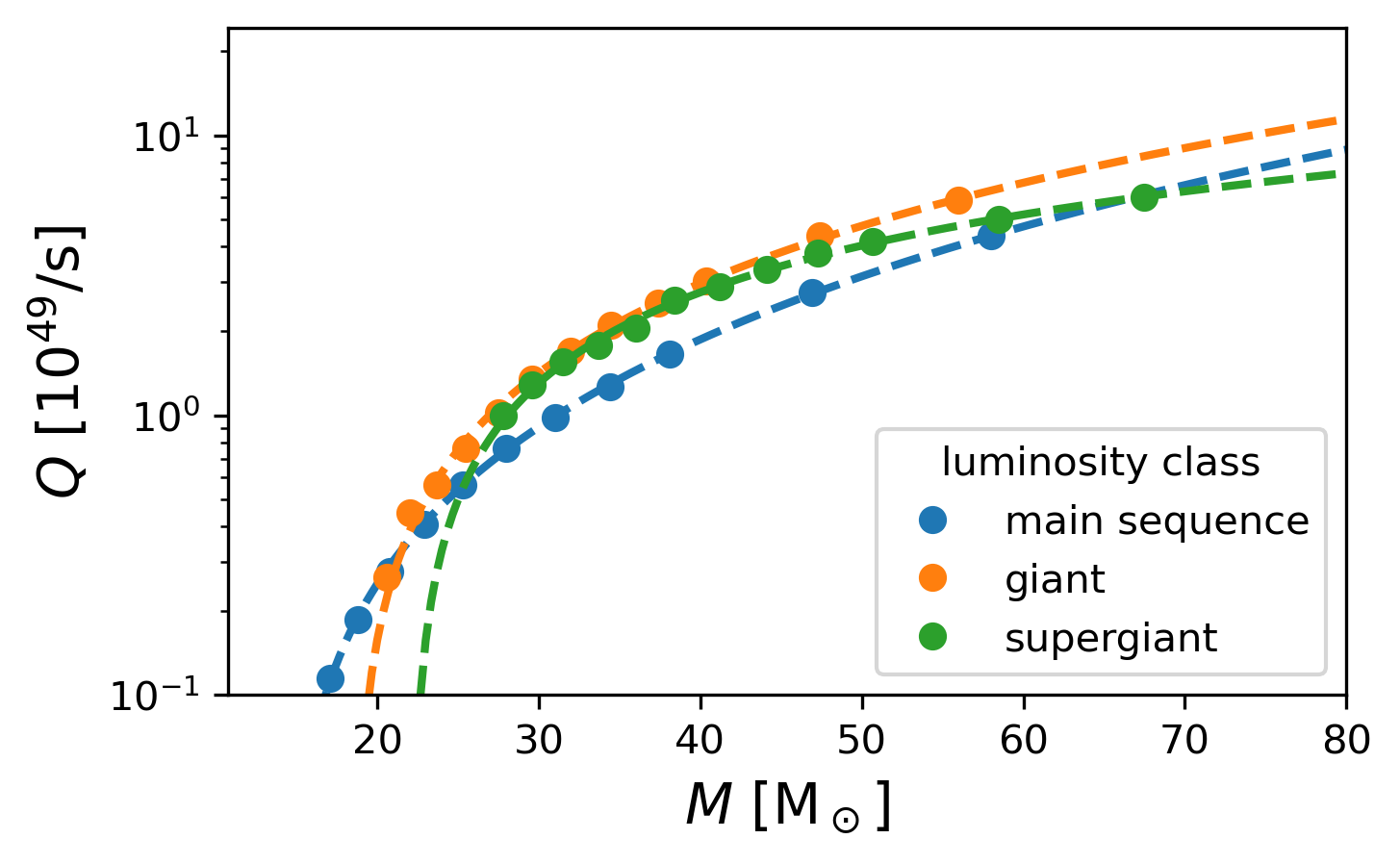}
    \caption{The rate of ionizing photon emission as a function stellar mass for massive stars on the main sequence (blue), the giant branch (orange), and the supergiant branch (green). The dots are data from \citet{Draine2011}, Table~15.1 and the dashed lines are fits to the functional $Q(M) \sim a M^b + c$.}
    \label{fig:QvsM}
\end{figure}

In Fig.~\ref{fig:QvsM}, we plot data from Table~15.1 of \citet{Draine2011} and fits to the functional form $Q(m) \sim a m^b + c$ using the least-squares method. Across luminosity classes, only stars with masses $m \gtrsim 20~M_\odot$ make significant contributions to the ionizing photon production, making photoionization feedback sensitive to the high-mass end of the IMF. 

This uncertainty motivates the large range of values for $M_{\rm cl}$ in our early feedback series. The crudeness of our early feedback recipe is a reflection of the limitations of a pure hydrodynamics simulation and large uncertainties at Cosmic Dawn.

\section{Calculation of net angular momentum}
\label{sec:project}

In Figure~\ref{fig:proj}, we show face-on and edge-on projections of the galaxy. In this section, we describe the procedure for computing the direction of net angular momentum required to generate these projections.

First, we estimate the bulk velocity $\vb{v}_{\rm bulk}$ of the galaxy as the mass-weighted average velocity in a $3~{\rm kpc}$ sphere with center $\vb{r}_{\rm c}$, the center of the galaxy. Contributions from stars and gas are included in the mass weight. The typical bulk velocity of the galaxy relative to our grid is on the order of hundreds of ${\rm km/s}$. 

Next, we compute the direction of the net angular momentum of the galaxy $\vu{\boldsymbol \Omega}$, which is in general not aligned with the Cartesian simulation grid. We estimate the direction of the net angular momentum as the mass-weighted sum of the specific angular momentum in a $3~{\rm kpc}$ sphere with center $\vb{r}_{\rm c}$, where the specific angular momentum is given by $\vb{l} = (\vb{r} - \vb{r}_{\rm c}) \times (\vb{v} - \vb{v}_{\rm bulk})$. Contributions from stars and gas are included in the mass weight. 

We define a primed coordinate system $\vb{r}'$ with origin $\vb{r}_{\rm c}$ and $z$-axis parallel to $\vu{\boldsymbol \Omega}$. By projecting parallel to $\vu{z}'$, we can view the galaxy face-on. Likewise, by projecting parallel to $\vu{x}'$ or $\vu{y}'$, we can view the galaxy edge-on.

In Section~\ref{sec:structure}, we used the scale height to quantify the disciness of the galaxy. We compute the scale height using an iterative method based on Appendix~B of \citet{Mandelker+2014}. 

Let $M_{\rm cyl}(H, R)$ denote the mass contained inside a cylinder of height $H$ and radius $R$ with axis along $\vu{z}'$ and center $\vb{r}_{\rm c}$. We start with an initial guess for the disc height $H$ and an upper bound on the disc radius $R_{\rm max}$. Then, we iterate the following steps until convergence to a less than 1\% change between iterations in $H$ and $R$:
\begin{enumerate}
	\item Determine the radius $R$ such that $M_{\rm cyl}(H, R) = f M_{\rm cyl}(H, R_{\rm max})$;
	\item Determine the height $H$ such that $M_{\rm cyl}(H, R) = f M_{\rm cyl}(R, R)$.
\end{enumerate}

We set the initial disc height $H = 500~{\rm pc}$ and the upper bound on the disc radius $R_{\rm max} = 1~{\rm kpc}$. However, our results are not sensitive to changes in these parameters by factors of a few. We chose a mass fraction $f = 85\%$ consistent with \citet{Mandelker+2014}. The calculation can be repeated separately for cold gas and stars. We only include gas with temperatures $\le 10^4~{\rm K}$ to exclude outflows and accreting gas.

To compute the ratio of the rotation velocity to the radial velocity dispersion, we first compute the relative velocity components in the primed coordinate system
\begin{align}
	\Delta v_{x'} =\ & (\vb{v} - \vb{v}_{\rm bulk}) \vdot \vu{x}', \quad \Delta v_{y'} = (\vb{v} - \vb{v}_{\rm bulk}) \vdot \vu{y}'
\end{align}
In cylindrical polar coordinates $(s, \varphi, z)$, this becomes
\begin{align}
	\Delta v_{s'} = s^{-1} ( x' \Delta v_{y'} - y' \Delta v_{x'} ), \quad \Delta v_{\varphi'} = s^{-1} ( x' \Delta v_{x'} + y' \Delta v_{y'} )
\end{align}

The ratio of the rotation velocity to the radial velocity dispersion at cylindrical radius $s'$ is
\begin{equation}
	\frac{v_{\rm rot}}{\sigma_r}(s') = \expval{\Delta v_{\varphi'}}_\rho(s') \left( \expval{(\Delta v_{s'})^2 }_\rho (s') - \expval{ \Delta v_{s'} }_\rho^2(s') \right)^{-1/2}
\end{equation}
where $\expval{\cdot}_\rho(s')$ denotes a density-weighted average over a cylindrical shell at cylindrical radius $s'$. 

We use cylindrical shells of width $\Delta s' = 25~{\rm pc}$ for $s \le R$ and height $H$. The overall ratio of the rotation velocity to the radial velocity dispersion is the average over cylindrical shells, weighted by the mass of each shell. The ratio can be computed separately for gas and stars using the corresponding disc dimensions.

We adopt the notation $v_{\rm rot} / \sigma_r$ to be consistent with previous literature in the main body of the paper. However, we emphasize that the radial velocity dispersion is computed in cylindrical polar coordinates, not spherical polar coordinates.

\section{Calculation of summary statistics}
\label{sec:calc_sumstat}

We use various summary statistics to compare our simulations in Section~\ref{sec:SFH}. In this appendix, we explain the calculation details for each statistic. For these calculations, we restrict the domain to a box $\mathcal{B}$ of side length $5~{\rm kpc}$ centered on the galaxy to avoid contributions from other galaxies in the simulation box. At $z=9$, $\mathcal{B}$ contains $\sim 1/3$ of the stellar mass in the simulation box.

Our simulations record the time of birth for each star particle. To compute the SFR, we start by binning the star particles in $\mathcal{B}$ at $z=9$ by time of birth and dividing by the bin width:
\begin{equation}
	\widetilde{\dot{M}}_*(t) = \frac{1}{\Delta t} \sum_{\substack{\vb{r}_k(t_9) \in \mathcal{B}\\ t_{{\rm b}, k} \in [t, t+\Delta t]}} m_{k}(t_9)
	\label{eq:SFR_tilde}
\end{equation}
where $\vb{r}_k$ is the position of star particle $k$, $t_{{\rm b}, k}$ is the time of birth of star particle $k$, $\Delta t$ is the bin width, and $t_9$ is the time at $z=9$. We use 5500 time bins of width $100~{\rm kyr}$ in the range $[0, 550]~{\rm Myr}$.

Equation~\ref{eq:SFR_tilde} underestimates the SFR because the star particle mass decreases over its lifetime due to SN mass loss. This is accounted for by
\begin{equation}
	\dot{M}_*(t) = \frac{\widetilde{\dot{M}}_*(t)}{f(t_9 - t)}
	\label{eq:SFR}
\end{equation}
where $f(\Delta t)$ is given by
\begin{equation}
	f(\Delta t) = \begin{cases}
		1 & \Delta t \le \tau_{\rm start}\\
		1 - \chi \left( \frac{\Delta t - \tau_{\rm start}}{\tau_{\rm end} - \tau_{\rm start}}\right) & \tau_{\rm start} \le \Delta t \le \tau_{\rm end}\\
		1 - \chi & \Delta t \ge \tau_{\rm end}
	 \end{cases}
	 \label{eq:fdeltat}
\end{equation}
The SFR averaged over a time interval $\Delta t$ is 
\begin{equation}
	{\rm SFR}_{\Delta t}(t) = \frac{1}{\Delta t} \int_{0}^{\Delta t} \dd (\Delta t') \dot{M}_*(t - \Delta t')
\end{equation}

We compute the stellar mass by integrating the SFR over time, accounting for SN mass loss:
\begin{equation}
    M_*(t) = \int_0^t \dd t' \dot{M}_*(t') f(t - t')
    \label{eq:Mstar_calc}
\end{equation}
Equivalently, we can compute the stellar mass by summing the masses of the star particles in $\mathcal{B}$:
\begin{equation}
	M_*(t) = \sum_{\vb{r}_k(t) \in \mathcal{B}} m_k(t)
    \label{eq:Mstar_calc2}
\end{equation}
In the fiducial simulation, the two methods always agree to within 1\%. However, we prefer Equation~\ref{eq:Mstar_calc} because it can be used to compute the stellar mass at arbitrary times rather than only times associated with data dumps. We compute the gas mass by summing the mass $\rho \Delta V$ in each grid cell,  similar to Equation~\ref{eq:Mstar_calc2}.

To compute the virial mass, we start by computing the enclosed mass $M(<r)$ as a function of radius, including contributions from gas, stars, and dark matter. The radial coordinate is computed with respect to the center of the halo. The virial radius is the radius where $M(< R_{\rm vir}) = (4/3) \pi R_{\rm vir}^3 \rho_{\rm halo}$, where $\rho_{\rm halo}$ is given by Equation~\ref{eq:rho_halo}. The virial mass is $M_{\rm vir} = M(< R_{\rm vir})$.

The local SFE is a function of the local gas conditions by the MFF model (Eq.~\ref{eq:mff}). We compute the median local SFE from the distribution of local SFEs in star birth events, weighted by stellar mass. The integrated SFE is
\begin{equation}
	\epsilon_{\rm int} = \frac{M_*}{f_{\rm b} M_{\rm halo}}
\end{equation}

The outflow efficiency is given by $\eta = \dot{M}_{\rm out} / {\rm SFR}_{\rm 50}$, where $\dot{M}_{\rm out}$ is the mass outflow rate and ${\rm SFR}_{\rm 50}$ is the SFR averaged over the last $50~{\rm Myr}$. We use ${\rm SFR}_{50}$ rather than $\dot{M}_*$ because the raw SFR is highly stochastic, making the mass outflow rate better-correlated with ${\rm SFR}_{\rm 50}$. We compute the mass outflow rate using the mass flux of gas with positive radial velocities in a spherical shell $\partial V$ of radius $R_{\rm out} = 2~{\rm kpc}$ and thickness $\Delta R_{\rm out} = 0.1 R_{\rm out} = 200~{\rm pc}$:
\begin{equation}
	\dot{M}_{\rm out} = \frac{4 \pi R_{\rm out}^2}{\Delta V_{\rm out}} \int_{\partial V} \dd^3 \vb{r} \rho(\vb{r}) v_r(\vb{r}) H(v_r(\vb{r}))
\end{equation}
where $H$ is the Heaviside step function, $v_r$ is the radial velocity of the gas relative to the center of the galaxy, and $\Delta V_{\rm out}$ is the volume of the spherical shell, given by
\begin{equation}
	\Delta V_{\rm out} = \frac{4 \pi}{3} \left[ (R_{\rm out} + \Delta R_{\rm out})^3 - R_{\rm out}^3 \right]
\end{equation}
For a given simulation, the value of $\eta$ varies for different choices of $R_{\rm out}$, but the ratio of outflow efficiencies between simulations is robust for values from $1$ to $4~{\rm kpc}$.

\section{The density PDF on different scales}
\label{sec:pdf}

\begin{figure}
    \centering
    \includegraphics[width=\linewidth]{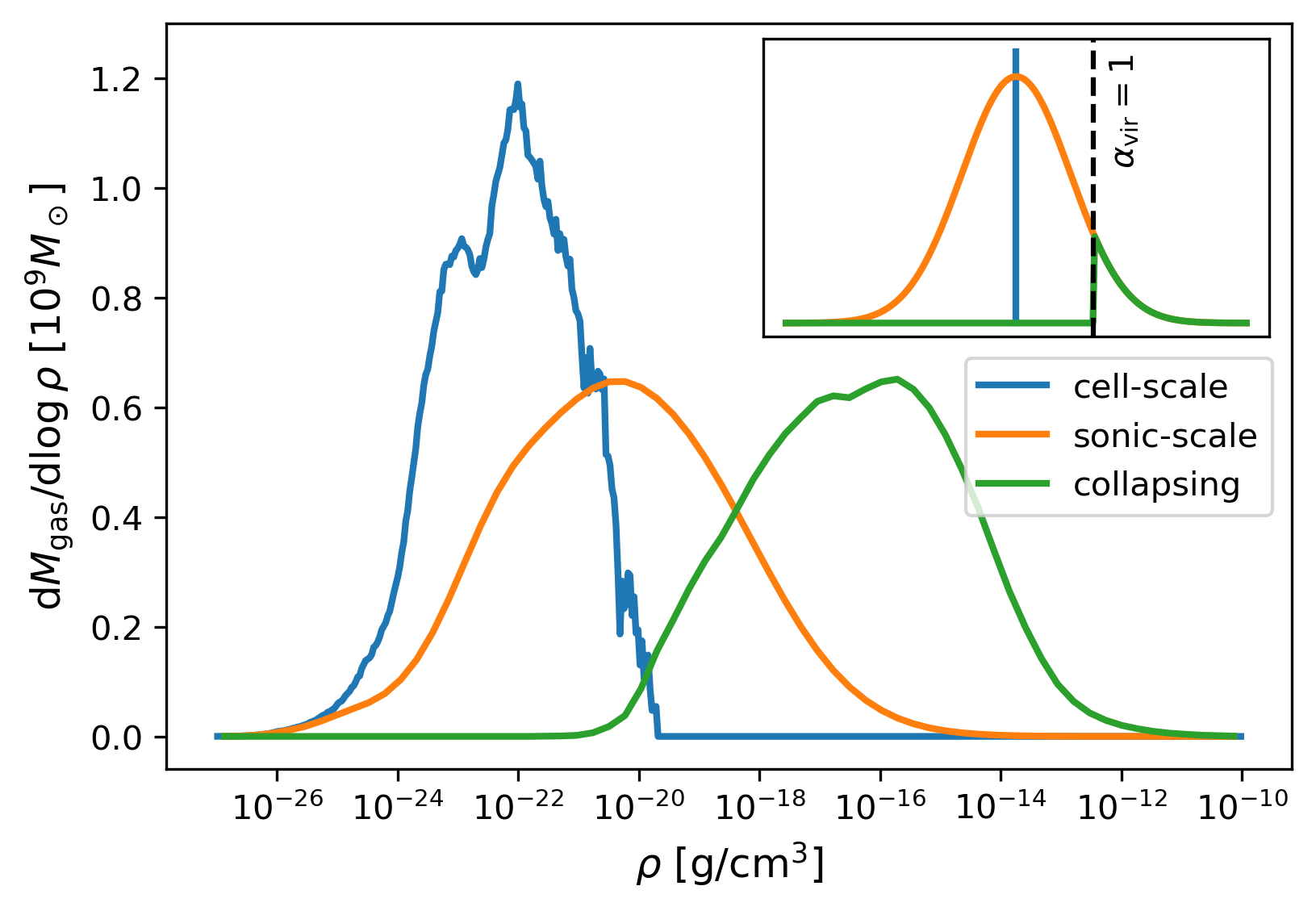}
    \caption{Histogram of the gas density in a box of sidelength $5~{\rm kpc}$ around the most massive galaxy in the fiducial simulation at $z=9$, weighted by gas mass. We compute the gas density using the raw cell value (blue), the subgrid density PDF given by the MFF model (orange), and the portion of the subgrid density PDF that is gravitationally unstable i.e. $\alpha_{\rm vir} < 1$ (green). These histograms represent the cell-scale density distribution, the sonic-scale density distribution, and the density distribution of collapsing clouds respectively. The inset plot shows the convolution functions for clarity: a Dirac delta function for the cell-scale histogram (blue), a lognormal function for the sonic-scale histogram (orange), and a truncated lognormal function for the collapsing sonic-scale histogram (green).}
    \label{fig:rho_mff}
\end{figure}

The density distribution of gas in the galaxy changes depends on the scale under consideration. In Figure \ref{fig:rho_mff}, we show the mass-weighted gas density distribution of the galaxy in our fiducial simulation at $z=9$. On the same plot, we convolve the density distribution with the subgrid density PDF given by the MFF model (Eq.~\ref{eq:mffpdf}), effectively computing the gas density distribution on the sonic scale. 

Because the mean density of a cell represents a volume-weighted average, the mass-weighted distribution shifts to higher densities in the sonic-scale PDF. This shows that the mass-weighted density PDF is scale-dependent, and therefore it can produce misleading results when there are multiple scales of interest. 

We also convolve the density distribution with the portion of the PDF that is gravitationally unstable i.e. $\alpha_{\rm vir} < 1$, effectively computing the density distribution of collapsing clouds. The tail of this distribution crosses the opacity limit $\rho \sim 10^{-13}~{\rm g/cm^3}$ \citep{Masunaga&Inutsuka1999}, where the assumption of a cold, isothermal gas begins to break down.

\section{Spatial correlations of gas properties}
\label{sec:gaspropcomp}

In this section, we use 2-dimensional projections and slices to gain an intuition for the spatial correlation of gas properties. Spatial correlations are not captured by the PDFs analyzed in Section~\ref{sec:local_gas_prop}.

\begin{figure*}
    \centering
    \includegraphics[width=\linewidth]{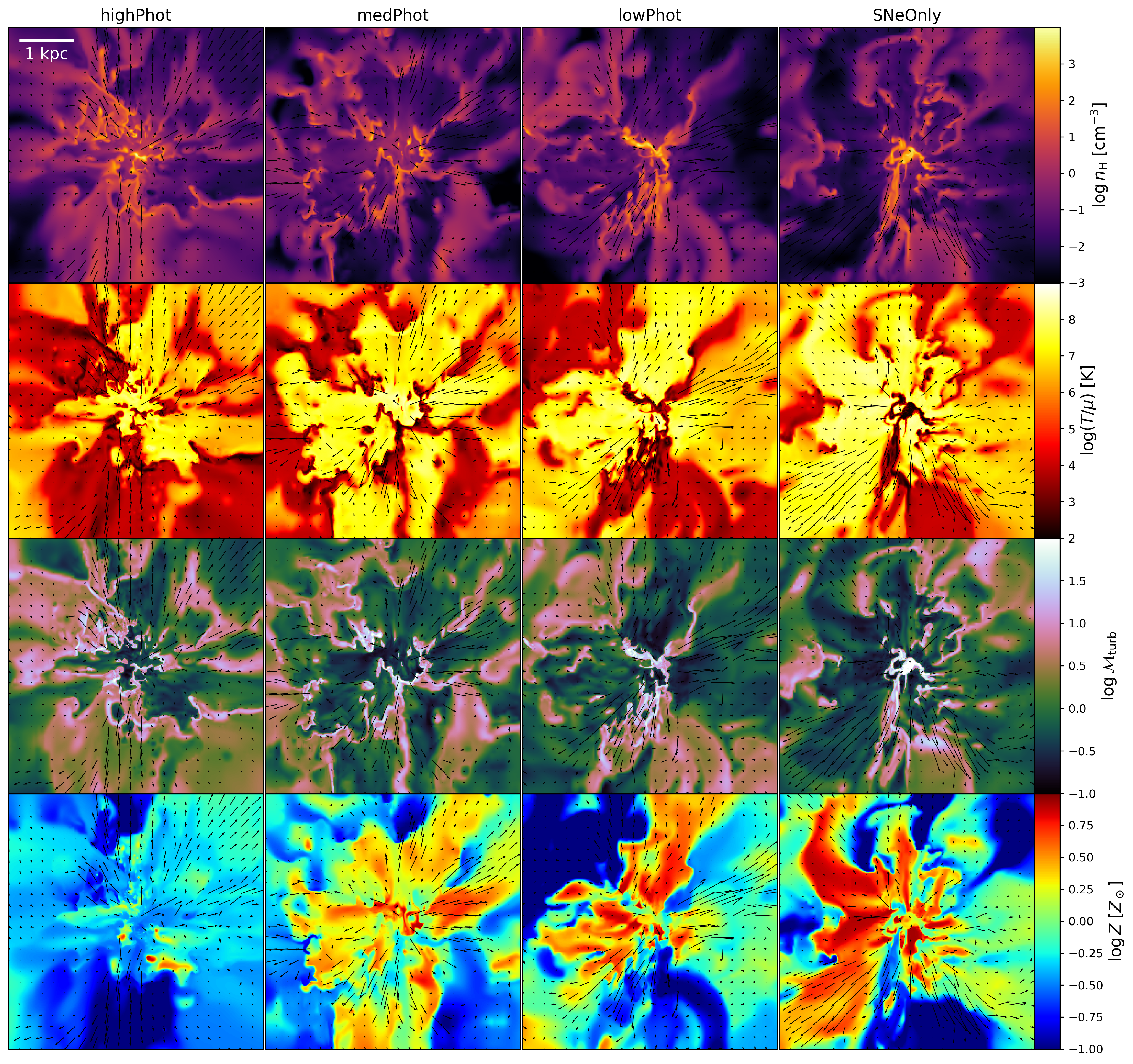}
    \caption{Logarithmic gas density (first row), temperature (second row), turbulent Mach number (third row), and metallicity (fourth row) in an $xy$ slice through the most massive galaxy in \texttt{highPhot} (first column), \texttt{medPhot} (second column), \texttt{lowPhot} (third column), and \texttt{Nophot} (fourth column) at $z=9$. We overlay arrows showing the direction and magnitude of the velocity field in each panel, with the bulk motion subtracted out.}
    \label{fig:gaspropcomp}
\end{figure*}

In Figure~\ref{fig:gaspropcomp}, we show slice plots of gas properties in the early feedback series. Regions of high density, low temperature, high Mach number, and low metallicity are spatially coincident. Likewise, regions of low density, high temperature, low Mach number, and high metallicity are spatially coincident. The low density regions tend to flow away from the galaxy. 

As photoionization becomes more effective, the density field becomes more homogeneous, confirming our interpretation of the volume-weighted PDFs (Fig.~\ref{fig:vgashist}). In \texttt{noPhot}, nearly all the gas mass is located in one central clump. In \texttt{highPhot}, the gas is distributed into multiple dense clumps and throughout the diffuse regions of the ISM. This helps explain why photoionization feedback leads to more SNe occurring in diffuse environments.

It is difficult to predict \textit{a priori} whether compressive or solenoidal forcing modes should dominate. On the one hand, the collapsing gas clouds and SN explosions associated with star formation drive compressive modes. This creates a positive feedback loop, because compressive forcing conditions enhance star formation. On the other hand, large-scale processes like accretion, mergers, and the differential rotation of a disc drive solenoidal modes, which suppress star formation. In \texttt{varTurb}, we capture these effects by computing the turbulence forcing parameter from the local velocity field. 

\begin{figure}
    \centering
    \includegraphics[width=\linewidth]{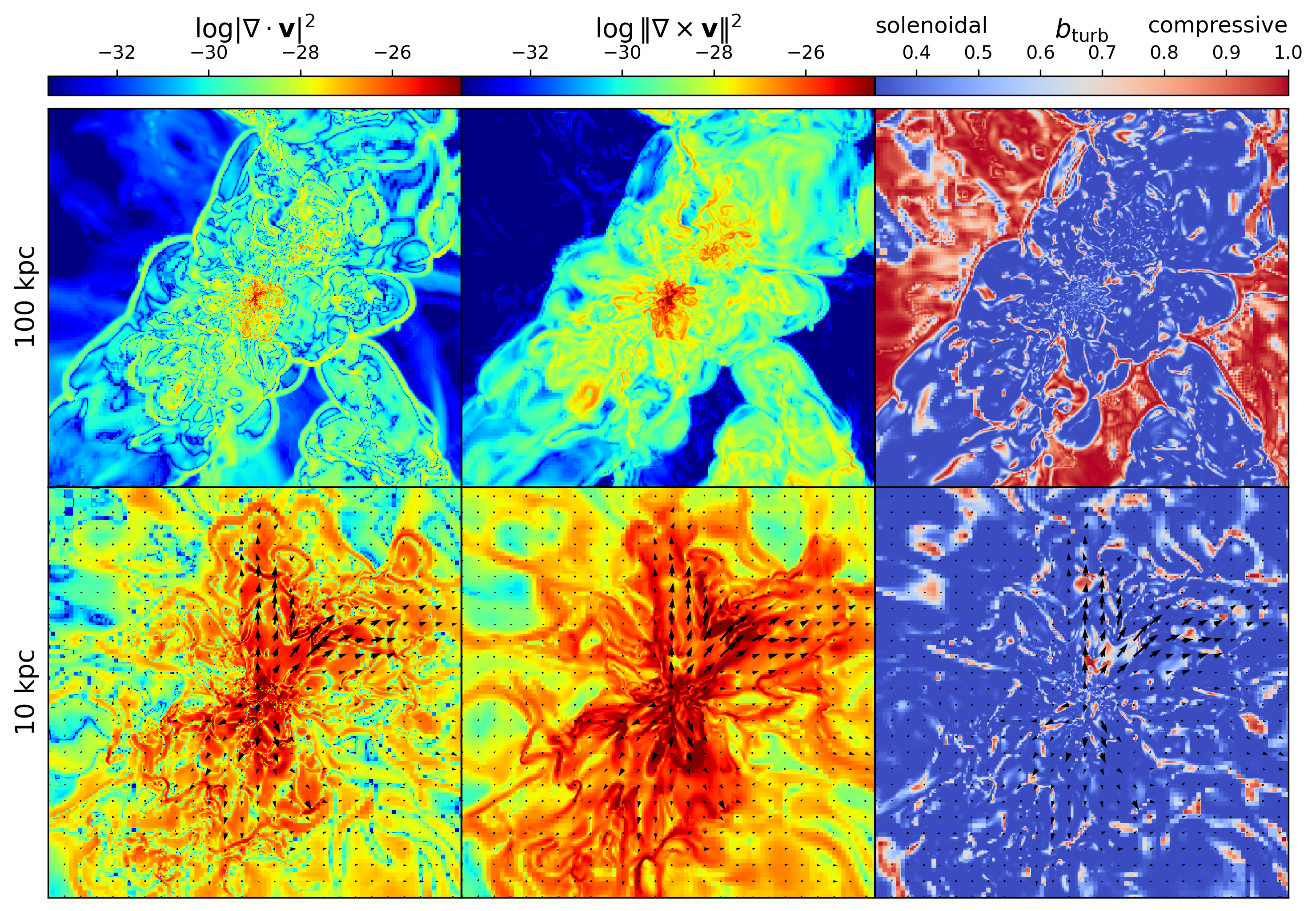}
    \caption{Logarithmic square of the velocity divergence (left column), square modulus of the velocity curl (middle column), and turbulence forcing parameter (right column) in an $xy$ slice through the most massive galaxy in the fiducial simulation at $z=9$ at a scale of $100~{\rm kpc}$ (top row) and $10~{\rm kpc}$ (bottom row). We overlay arrows showing the direction and magnitude of the velocity field in the bottom row. The square of the velocity divergence and velocity curl modulus are proxies for power in compressive and solenoidal forcing modes respectively.}
    \label{fig:bturb}
\end{figure}

In Figure~\ref{fig:bturb}, we show the velocity divergence and curl, proxies for the power in compressive and solenoidal forcing modes respectively, and the resulting forcing parameter in \texttt{varTurb}. The velocity field in regions near the galaxy is dominated by solenoidal motion, which should suppress star formation. However, the star formation history in \texttt{varTurb} is closer to \texttt{compTurb} than \texttt{solTurb}. This indicates that even though solenoidal forcing modes dominate the turbulence, a disproportionate fraction of star formation occurs in small pockets of compressive forcing. 

\section{Column density PDF}
\label{sec:coldenspdf}

The volume density PDF (Eq.~\ref{eq:mffpdf}) described by the MFF model is related to the column density PDF, which is a direct observable. Numerical simulations show that the column density PDF of supersonic turbulent gas is well-described by a log-normal distribution, similar to the volume density PDF \citep{Federrath+2010}
\begin{equation}
	p_A(\eta) = \dv{A}{\eta} = \frac{1}{\sqrt{2 \pi \sigma_\eta^2}} \exp \left[ - \frac{(\eta - \overline{\eta})^2}{2 \sigma_\eta^2} \right]
	\label{eq:pdf_col}
\end{equation}
where $\eta = \ln(\Sigma / \overline{\Sigma})$ is the logarithmic surface density, $\sigma_\eta$ is the standard deviation of the logarithmic surface density, and $\overline{\eta} = -1/2\sigma_\eta^2$ is its mean in analogy to $s$, $\sigma_s$, and $\overline{s}$ for the volume density PDF (Eq.~\ref{eq:mffpdf}). We use $\eta$ to denote the logarithmic surface density for notational consistency with other work and rely on context to disambiguate the logarithmic surface density and outflow efficiency.

Equation~\ref{eq:pdf_col} is supported observationally by measurements of column densities in the Perseus molecular cloud \citep{Goodman+2009}. In general, the dispersion in the logarithmic column density distribution $\sigma_\eta$ is less than the dispersion in the logarithmic volume density distribution $\sigma_s$, because fluctuations are averaged out by integration along the line-of-sight.

The ratio $\sigma_\eta / \sigma_s$ decreases as the spectral slope of the density field decreases because the density field has less spatial correlation, so integration averages over more fluctuations. Previous work has shown that the spectral slope can be approximated by a power law in the Mach number \citep[e.g.][]{Seon2009}.

Using that relationship, \citet{Seon2012} show that the relationship between the two dispersions is well-approximated by a constant of proportionality $\sigma_\eta^2 = A \sigma_s^2$, where $A$ depends only on the forcing parameter. For $b_{\rm turb} = 1/3, 0.5, 1$, they find best fit values $A = 0.2, 0.24, 0.38$ respectively. We fit these data with a linear function using a least-squares regression, which combined with Equation~\ref{eq:sigs}, yields
\begin{equation}
	\sigma_\eta^2 \approx (0.107 b_{\rm turb} + 0.272) \ln ( 1 + b_{\rm turb}^2 \mathcal{M}_{\rm turb}^2 )
\end{equation}

Using the column density PDF, we can compute the effective optical depth through a cell. We assume a constant opacity $\kappa$ such that $\tau = \kappa \Sigma$. Then the optical depth PDF is $p_A(\tau) = \tau^{-1} p_A(\eta)$. Consider a radiation field with intensity $I_0$ incident on the face of a cell with area $\Delta x^2$. Then the incident flux $F_0 = I_0 \Delta x^2$. The flux out the other side of the cell has contributions from each column
\begin{equation}\begin{split}
    F =\ & \int \dd A I_0 e^{-\tau} = F_0 \int_0^\infty \dd \tau p_A(\tau) e^{-\tau}\\
    =\ & F_0 \int_{-\infty}^\infty \dd \eta p_A(\eta) \exp( -\overline{\tau} e^\eta )
\end{split}\end{equation}
where $\overline{\tau} = \kappa \overline{\Sigma}$ is the mean optical depth. Let $F = F_0 e^{-\tau_{\rm eff}}$ define the effective optical depth. Then we have
\begin{equation}\begin{split}
    \tau_{\rm eff} =\ & -\ln \int_{-\infty}^\infty \dd \eta p_A(\eta) \exp( -\overline{\tau} e^\eta )
    \label{eq:tau_eff}
\end{split}\end{equation}
In the limit $\mathcal{M}_{\rm turb} \ll 1$, the logarithmic surface density PDF becomes a delta function and the effective optical depth approaches the mean optical depth as expected.

\begin{figure}
    \centering
    \includegraphics[width=\linewidth]{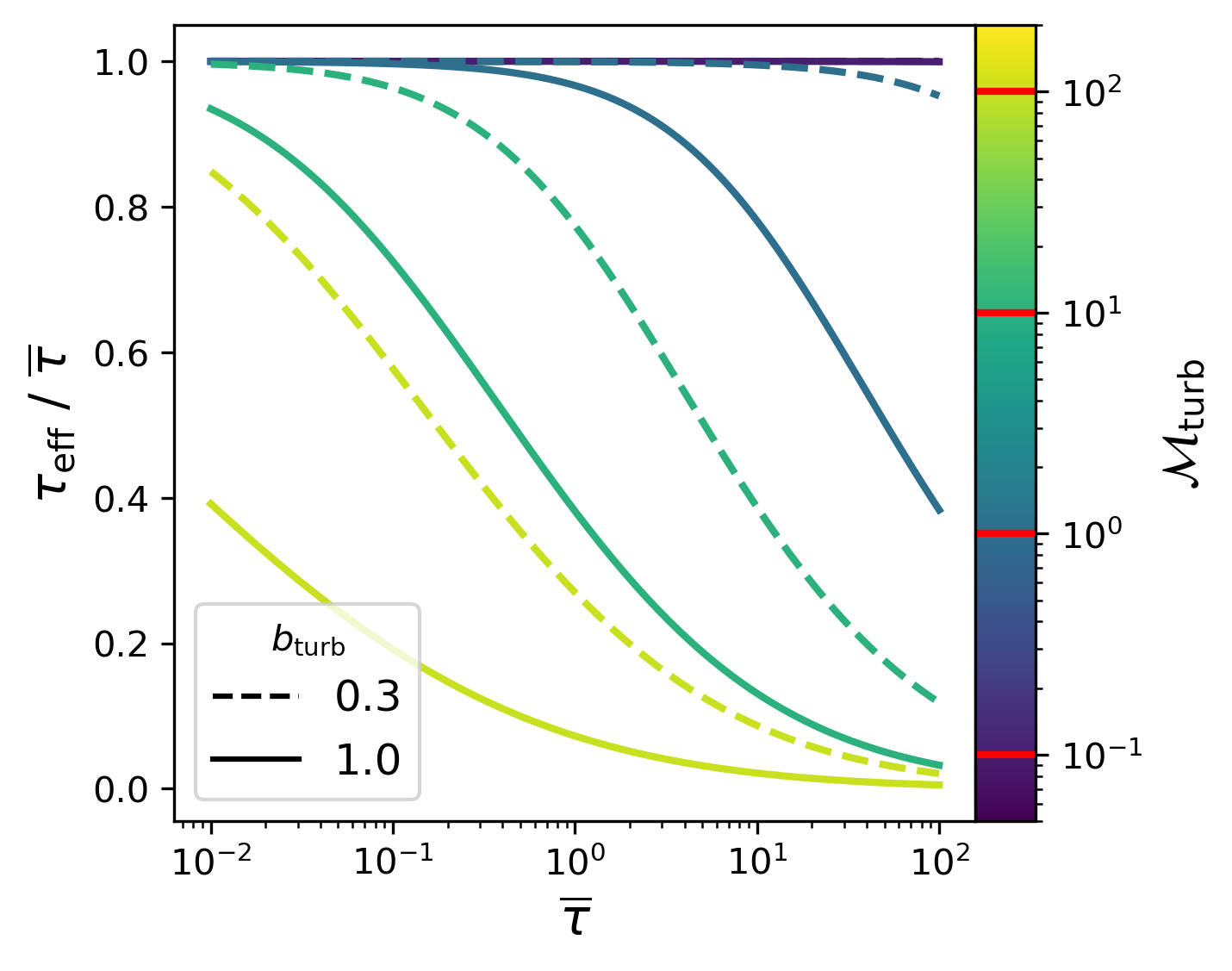}
    \caption{Effective optical depth (Eq.~\ref{eq:tau_eff}) in a turbulent medium relative to mean optical depth as a function of mean optical depth. Curves for different turbulent Mach numbers are shown in different colors indicated by the colorbar. Curves for compressive (resp. solenoidal) forcing are shown in solid (resp. dashed) lines. At high Mach numbers and high mean optical depths, the effective optical depth is significantly reduced relative to the mean.}
    \label{fig:tau_eff}
\end{figure}

The integral can be evaluated numerically using Gaussian quadrature as implemented in \texttt{scipy.integrate.quad}. For multiple simultaneous evaluations, we use the vectorized version implemented in \texttt{scipy.integrate.quad\_vec}. In Figure~\ref{fig:tau_eff}, we show the effective optical depth relative to the mean optical depth as a function of mean optical depth for various turbulent Mach numbers and turbulence forcing parameters. At high Mach numbers and high mean optical depths, the effective optical depth is significantly reduced relative to the mean.

\section{Breakdown of steady state}
\label{sec:star_formation}

If the depletion time is long compared to the accretion timescale, then the SFR responds slowly to changes in the gas mass. In this case, the SFR passively reflects the current distribution of gas and is given by Equation~\ref{eq:schmidt}. Dividing Equation~\ref{eq:schmidt} by the gas accretion rate to get the global SFE, we find
\begin{equation}
	\epsilon_{\rm glob} = \frac{M_{\rm g}}{f_{\rm b} \dot{M}_{\rm acc}} \frac{f_{\rm sf} \epsilon_{\rm ff}}{\tau_{\rm ff}}
	\label{eq:eps_glob_nss}
\end{equation}

Consider the case where the local SFE is small or preventative feedback is weak such that the first term dominates in Equation~\ref{eq:f_sf}. Then
\begin{equation}
	f_{\rm sf} \approx \frac{1}{f_{\rm dyn}}, \quad \tau_{\rm dep} \approx f_{\rm dyn} \frac{\tau_{\rm ff}}{\epsilon_{\rm ff}}
	\label{eq:f_sf_low_eps}
\end{equation}
and
\begin{equation}
	\epsilon_{\rm glob} \approx \frac{M_{\rm gas}}{f_{\rm b} \dot{M}_{\rm acc} \tau_{\rm ff}} \frac{\epsilon_{\rm ff}}{f_{\rm dyn}}
\end{equation}
In this case, the global SFE is proportional to the local SFE, so the galaxy is not self-regulated. \citep{Semenov+2018} call this the dynamics-regulation regime.

Now consider the case where the local SFE is large or preventative feedback is weak such that the second term dominates in Equation~\ref{eq:f_sf}. Then 
\begin{align}
	f_{\rm sf} \approx \frac{\tau_{\rm ff}}{\tau_{\rm cool}} \frac{1}{(1 + \mu) \epsilon_{\rm ff}}, \quad \tau_{\rm dep} \approx (1 + \mu) \tau_{\rm cool}
	\label{eq:f_sf_high_eps}
\end{align}
and
\begin{equation}
	\epsilon_{\rm glob} \approx \frac{M_{\rm gas}}{f_{\rm b} \dot{M}_{\rm acc} \tau_{\rm cool}} \frac{1}{1 + \mu}
\end{equation}
Similar to when steady state applies, the global SFE is independent of the local SFE, so the galaxy is self-regulated. However, rather than ejective feedback, the global SFE is regulated by preventative feedback. \citep{Semenov+2018} call this the feedback-regulation regime.

When steady state applies, the galaxy is always self-regulated. When steady state does not apply, the galaxy is only self-regulated if the local SFE satisfies
\begin{equation}
	\epsilon_{\rm ff} \gg \frac{\tau_{\rm ff}}{\tau_{\rm cool}} \frac{f_{\rm dyn}}{1 + \mu}
\end{equation}
Essentially, the local SFE must be large enough for preventative feedback to remove gas from the star-forming state faster than it is supplied by cooling and collapse.

\bsp	
\label{lastpage}
\end{document}